\documentclass[12pt]{article}
\usepackage{epsfig,amssymb,psfrag}

\catcode`\@=11
\def \thesection {\arabic{section}.}

\def \be  {\begin{equation}}
\def \ee  {\end{equation}}
\def \ba  {\begin{eqnarray}}
\def \ea  {\end{eqnarray}}
\def \baa {\begin{eqnarray*}}
\def \eaa {\end{eqnarray*}}
\def \bb  {\begin {thebibliography} }
\def \eb  {\end{thebibliography}}
\def \lab #1 {\label{#1}}

\newcommand \bi [1] {\bibitem{#1}}
\newcommand\re[1]{(\ref{#1})}

\def \matrix #1 {\left(\begin{array}{cc} #1 \end{array}\right)}
\def \Tr {\mathop{\rm Tr}\nolimits}

\def \Im {\mathop{\rm Im}\nolimits}
\def \Re {\mathop{\rm Re}\nolimits}

\def \e  {\mathop{\rm e}\nolimits}
\newcommand\lr[1]{{\left({#1}\right)}}
\newcommand \widebar [1] {\overline{#1}}

\newcommand \vev [1] {\langle{#1}\rangle}

\newcommand \ket [1] {|{#1}\rangle}

\newcommand{\as}{\ifmmode\alpha_{\rm s}\else{$\alpha_{\rm s}$}\fi}
\newcommand{\asbar}{\ifmmode\bar{\alpha}_{\rm s}\else{$\bar{\alpha}_{\rm s}$}\fi}

\font\cmss=cmss12 
\def\inbar{\,\vrule height1.5ex width.4pt depth0pt}
\def\IC{\relax\hbox{$\inbar\kern-.3em{\rm C}$}}
\def\IZ{\relax{\hbox{\cmss Z\kern-.4em Z}}}
\def\IR{{\hbox{{\rm I}\kern-.2em\hbox{\rm R}}}}

\def\IP{{\hbox{{\rm I}\kern-.2em\hbox{\rm P}}}}
\def\II{\hbox{{1}\kern-.25em\hbox{l}}}

\def\numberbysection{\@addtoreset{equation}{section}
                     \def\theequation{\thesection\arabic{equation}}}
\numberbysection

\newcommand \mybf[1] {\mbox{\boldmath$ {#1} $}}

\psfrag{nu_h}[cc][bc]{$\mbox{\large$\nu_h$}$}
\psfrag{-E_3/4}[cc][cc]{$-E_3/4$}
\psfrag{-E_4/4}[cc][cc]{$-E_4/4$}
\psfrag{-E_N/4}[cc][cc]{$-E_N/4$}
\psfrag{q4/q2}[cc][cc]{$q_{_{\scriptstyle 4}}/q_{_{\scriptstyle 2}}$}
\psfrag{N}[cc][bc]{$N$}
\psfrag{qN/q2}[cc][cc]{$|q_{_{\scriptstyle N}}|/q_{_{\scriptstyle 2}}$}

\psfrag{Re(q4^(1/4))}[cc][bc]{$\Re[q_4^{1/4}]$}
\psfrag{Im(q4^(1/4))}[cc][cc]{$\Im[q_4^{1/4}]$}

\psfrag{Im(q3^(1/3))}[cc][cc]{$\Im[q_3^{1/3}]$}
\psfrag{Re(q3^(1/3))}[cc][bc]{$\Re[q_3^{1/3}]$}

\psfrag{Im[q3^(1/3)]}[cc][cc]{$\Im[q_3^{1/3}]$}
\psfrag{Re[q3^(1/3)]}[cc][bc]{$\Re[q_3^{1/3}]$}

\begin{document}

\begin{titlepage}
\begin{flushright}
\begin{tabular}{l}
LPT--Orsay--02--30\\
TPJU-08/2002 \\
UB--ECM--PF--02/07\\
hep-th/0204124
\end{tabular}
\end{flushright}

\vskip3cm
\begin{center}
  {\large \bf
  Noncompact Heisenberg spin magnets from high-energy QCD  \\[2mm] II.~Quantization conditions and energy spectrum}

\def\thefootnote{\fnsymbol{footnote}}%
\vspace{1cm}
{\sc S.\'{E}. Derkachov}${}^1$, {\sc G.P.~Korchemsky}${}^2$, {\sc
J.~Kota{\'n}ski}${}^3$ and {\sc A.N.~Manashov}${}^4$\footnote{ Permanent
address:\ Department of Theoretical Physics,  Sankt-Petersburg State University,
St.-Petersburg, Russia}
\\[0.5cm]

\vspace*{0.1cm} ${}^1$ {\it
Department of Mathematics, St.-Petersburg Technology Institute,\\
St.-Petersburg, Russia
                       } \\[0.2cm]
\vspace*{0.1cm} ${}^2$ {\it
Laboratoire de Physique Th\'eorique%
\footnote{Unite Mixte de Recherche du CNRS (UMR 8627)},
Universit\'e de Paris XI, \\
91405 Orsay C\'edex, France
                       } \\[0.2cm]
\vspace*{0.1cm} ${}^3$ {\it
Institute of Physics, Jagellonian University,\\ Reymonta 4, PL-30-059 Cracow,
Poland
                       } \\[0.2cm]
\vspace*{0.1cm} ${}^4$
 {\it
Department d'ECM, Universitat de Barcelona,\\
08028 Barcelona, Spain}

\vskip2cm
{\bf Abstract:\\[10pt]} \parbox[t]{\textwidth}{
We present a complete description of the spectrum of compound states of reggeized
gluons in QCD in multi-colour limit. The analysis is based on the identification
of these states as ground states of noncompact Heisenberg $SL(2,\mathbb{C})$ spin
magnet. A unique feature of the magnet, leading to many unusual properties of its
spectrum, is that the quantum space is infinite-dimensional and conventional
methods, like the Algebraic Bethe Ansatz, are not applicable. Our solution relies
on the method of the Baxter $\mathbb{Q}-$operator. Solving the Baxter equations,
we obtained the explicit expressions for the eigenvalues of the
$\mathbb{Q}-$operator. They allowed us to establish the quantization conditions
for the integrals of motion and, finally, reconstruct the spectrum of the model.
We found that intercept of the states built from even (odd) number of reggeized
gluons, $N$, is bigger (smaller) than one and it decreases (increases) with $N$
approaching the same unit value for infinitely large $N$.}
\vskip1cm

\end{center}

\end{titlepage}

\newpage

\tableofcontents

\newpage

\setcounter{footnote} 0

\section{Introduction}

It has been recently realized that QCD possesses a hidden symmetry at high-energy
\cite{L1,FK}. This symmetry is not seen at the level of classical QCD Lagrangian
and it manifests itself through remarkable integrability properties of the
Schr\"odinger equation for the partial waves of the scattering amplitudes in
perturbative QCD in the so-called generalized leading logarithmic approximation
(GLLA) \cite{B,CDLO}. It turns out that in the multi-colour limit this equation
coincides with the Schr\"odinger equation for two-dimensional quantum-mechanical
completely integrable model, which was dubbed in \cite{FK,DKM-I} as a noncompact
Heisenberg spin magnet.

The asymptotics of the scattering amplitudes ${\cal A}(\textrm{s},\textrm{t})$ at
high energy, $\textrm{s}\gg -\textrm{t}$, is governed by the contribution of an
infinite number of soft gluons exchanged between the scattered particles. In the
GLLA approximation, the scattering amplitude is given by \cite{B,CDLO,K1}
\be
{\cal A}(\textrm{s},\textrm{t}) \sim -i\textrm{s}\sum _{N=2}^\infty (i \asbar)^N
\frac{\textrm{s}^{-\asbar E_N/4}}{(\asbar\sigma_N
\ln \textrm{s})^{1/2}}\,\xi_{a,N}(\textrm{t})
\xi_{b,N}(\textrm{t})\,,
\label{amp}
\ee
where $\asbar=\as N_c/\pi$ is the QCD coupling constant and the sum goes over an
arbitrary number of reggeized gluons exchanged in the $\textrm{t}-$channel,
$N=2,3,...$. In the GLLA approximation, the reggeized gluons interact with each
other elastically and form colour-singlet compound states~\cite{L2}. These states
can be defined as solutions to the Bartels-Kwiecinski-Praszalowicz (BKP)
equation~\cite{B,KP}
\be
{\cal H}_N \Psi(\vec z_1,\vec z_2,...,\vec z_N) = E_N \Psi(\vec z_1,\vec
z_2,...,\vec z_N)\,.
\label{Sch}
\ee
The effective QCD Hamiltonian ${\cal H}_N$ acts on the colour $SU(N_c)$ charges
of $N$ reggeized gluons and their two-dimensional transverse coordinates, $\vec
z_k$ ($k=1,...,N$), which belong to the hyperplane orthogonal to the momenta of
two scattered particles. The contribution to the scattering amplitude \re{amp} of
the compound states built from $N$ reggeized gluons, $\Psi(\vec z_1,\vec
z_2,...,\vec z_N)$, has the standard Regge form $\sim \textrm{s}^{-\asbar
E_N/4}$. It is dominated at large $\textrm{s}$ by the contribution of the ground
state $E_N$ for Eq.~\re{Sch}. Obviously, it increases (or decreases) with
$\textrm{s}$ if the energy of the ground state is negative (or positive). As we
will show below, the spectrum of \re{Sch} is gapless (see Eq.~\re{accum} below)
and, as a consequence, one has to retain in \re{amp} the contribution of the
excited states next to the ground state. This amounts to appearance of the
additional factor $(\asbar\sigma_N\ln \textrm{s})^{-1/2}$ in the r.h.s.\ of
\re{amp}. The residue factors $\xi_{a(b),N}(\textrm{t})$ measure the overlap of
$\Psi(\vec z_1,\vec z_2,...,\vec z_N)$ with the wave functions of the scattered
particles and they depend, in general, on the momentum transferred, $\textrm{t}$,
and the colour factor, $1/N_c^2$.

Calculation of the spectrum of the compound states \re{Sch} for arbitrary number
of reggeized gluons $N$ and eventual resummation of their contribution to the
scattering amplitude \re{amp} is a longstanding problem in high-energy
QCD~\cite{Ven,MS,J,CM}. At $N=2$, the solution to \re{Sch} has been found a long
time ago -- the well-known Balitsky-Fadin-Kuraev-Lipatov (BFKL) Pomeron
\cite{L2}. At $N=3$ the solution to \re{Sch} -- the Odderon state in QCD
\cite{LN}, was formulated only a few years ago by Janik and Wosiek \cite{JW} by
making use of the remarkable integrability properties of the effective QCD
Hamiltonian \cite{L1,FK}. Their solution has been later verified in
Refs.~\cite{check,DL}. The methods employed at $N=3$ in \cite{JW} can not be
generalized, however, to higher $(N\ge 4)$ reggeized gluon compound states and
very little is known about the solutions to \re{Sch} for $N\ge 4$. Recently, a
significant progress has been made in solving the Schr\"odinger equation \re{Sch}
for higher $N$ in the multi-colour limit, $N_c\to \infty$ and $\bar \as={\rm
fixed}$~\cite{DKM-I,DL}. The first results of the calculation of the ground state
energy $E_N$ for higher reggeized gluon compound states in multi-colour QCD were
reported in a letter format \cite{KKM}. In this paper, we shall provide a
detailed account on the approach used in \cite{KKM} and present new results for
the spectrum of the Schr\"odinger equation \re{Sch}. The reader interested in
learning more about the latter could skip the first part of the paper and go
directly to Section 5.

Our approach to solving the BKP equation \re{Sch} is based on the identification
of the effective Hamiltonian  in the multi-colour QCD, ${\cal H}_N$, as the
Hamiltonian of a completely integrable two-dimensional noncompact Heisenberg spin
magnet~\cite{L1,FK}. The latter model describes the nearest neighbour interaction
between spins of $N$ particles ``living'' on a two-dimensional plane of
transverse coordinates $\vec z=(x,y)$. The corresponding spin operators
$\overrightarrow{S_k}$ and $\overrightarrow{\bar S_k}$ (with $k=1,...,N$) are the
generators of the unitary principal series representation of the
$SL(2,\mathbb{C})$ group specified by a pair of complex spins $(s,\bar s)$.
They act on the $\vec z-$plane as the differential operators%
\footnote{That is, the isotopic ``spin'' space coincides with the coordinate space
of $N$ particles.}
\be
S_k^0=z_k\partial_{z_k} +s\,,\quad S_k^-=-\partial_{z_k}\,,\quad
S_k^+=z_k^2\partial_{z_k}+2sz_k\,,
\label{S-op}
\ee
so that $\overrightarrow{S_k}^2=(S_k^0)^2+(S_k^+S_k^-+S_k^-S_k^+)/2=s(s-1)$. The
operators $\bar S_k^{\pm,0}$ are given by similar expressions with $z_k$ and $s$
replaced by $\bar z_k$ and $\bar s$, respectively. Here, the notation was
introduced for the (anti)holomorphic coordinates on a two-dimensional $\vec
z-$plane, $z_k=x_k+iy_k$ and $\bar z_k=z_k^*$, so that $d^2z_k =dz_kd\bar z_k/2$.
By the definition,
%$\overrightarrow{S_k}^2=(S_k^0)^2+(S_k^+S_k^-+S_k^-S_k^+)/2=s(s-1)$,
%$\overrightarrow{\bar S_k}^2=\bar s(\bar s-1)$ and
$[S_k^a,\bar S_n^b]=0$ for
$a,b=\pm,0$. For the principal series of the $SL(2,\mathbb{C})$, the possible
values of the complex spins $(s,\bar s)$ take the form~\cite{group}
\be
s=\frac{1+n_s}2+i\nu_s\,,\qquad
\bar s=1-s^*=\frac{1-n_s}2+i\nu_s
\label{spins}
\ee
with $n_s$ integer and $\nu_s$ real. For the reggeized gluon compound states,
Eq.~\re{Sch}, the $SL(2,\mathbb{C})$ spins take the values $s=0$ and $\bar s=1$,
or equivalently $n_s=-1$ and $\nu_s=0$.

The Hamiltonian of the noncompact $SL(2,\mathbb{C})$ Heisenberg spin magnet is
given by~\cite{FK,DKM-I}
\be
{\cal H}_N = \sum_{k=1}^N \left[ H(J_{k,k+1}) + H(\bar J_{k,k+1}) \right],
\label{Ham}
\ee
where $H(J)=\psi(1-J)+\psi(J)-2\psi(1)$ with $\psi(x)=d\ln\Gamma(x)/dx$,
$J_{k,k+1}$ is the sum of two $SL(2,\mathbb{C})$ spins,
$J_{k,k+1}(J_{k,k+1}-1)=(\vec S_k+\vec S_{k+1})^2 $ with $J_{N,N+1}\equiv
J_{N,1}$, and similar for $\bar J_{k,k+1}$. The model possesses the set of
mutually commuting conserved charges $q_k$ and $\bar q_k$ ($k=2,...,N$). Their
number is large enough for the Schr\"odinger equation \re{Sch} to be completely
integrable. The charges $q_k$ are given by the $k-$th order differential
operators acting on the holomorphic coordinates of particles. They have
particularly simple form for the $SL(2,\mathbb{C})$ spins $s=0$ and $\bar
s=1$~\cite{L1,FK}
\be
q_k\bigg|_{s=0,\bar s=1}=i^k\sum_{1\le j_1 < j_2 < ... < j_k\le N}
z_{j_1j_2}...z_{j_{k-1},j_k}z_{j_k,j_1}\partial_{z_{j_1}}...
\partial_{z_{j_{k-1}}}\partial_{z_{j_k}}
\label{q}
\ee
with $z_{jk}\equiv z_j-z_k$. The charges $\bar q_k$ are given by similar
expressions in the $\bar z-$sector
\be
\bar q_k\bigg|_{s=0,\bar s=1}=i^k\sum_{1\le j_1 < j_2 < ... < j_k\le N}
\partial_{\bar z_{j_1}}...
\partial_{\bar z_{j_{k-1}}}\partial_{\bar z_{j_k}}
\bar z_{j_1j_2}...\bar z_{j_{k-1},j_k}\bar z_{j_k,j_1}\,,
\label{q-bar}
\ee
so that $\bar q_k=q_k^\dagger$ with respect to the $SL(2,\mathbb{C})$ scalar
product
\be
\|\Psi\|^2=\int d^2 z_1 ... d^2 z_N \, |\Psi(\vec z_1,\vec z_2,...,\vec
z_N)|^2\,.
\label{SL2-norm}
\ee
The eigenstates $\Psi(\vec z_1,\vec z_2,...,\vec z_N)$ have to diagonalize these
operators and be normalizable with respect to \re{SL2-norm}. The corresponding
eigenvalues $q\equiv\{q_k,\bar q_k=q_k^*\}$, with $k=2,...,N$, form the complete
set of quantum numbers parameterizing the spectrum of the Schr\"odinger equation
\re{Sch}, $E_N=E_N(q,\bar q)$. The eigenproblem for the operators \re{q} and
\re{q-bar} leads to a complicated system of $(N-1)-$differential equations on
$\Psi(\vec z_1,\vec z_2,...,\vec z_N)$, which was previously solved at $N=2$
\cite{L2} and $N=3$ \cite{JW}. For higher $N$, instead of dealing with this
system, we apply the method developed in \cite{DKM-I}. It represents an
application of the Quantum Inverse Scattering Method~\cite{QISM} to noncompact
Heisenberg spin magnet model.

The noncompact Heisenberg spin magnet, Eq.~\re{Ham}, can be considered as a
generalization of the well-known spin$-1/2$ Heisenberg spin chain model (as well
as its analogs for higher $SU(2)$ spins \cite{XXX}) to arbitrary complex spins
belonging to noncompact, unitary representations of the $SL(2,\mathbb{C})$ group.
As we will demonstrate below, the noncompact and compact Heisenberg magnets have
completely different properties, yet another manifestation of the fact that the
quantum space of the model is infinite-dimensional in the former case. In
particular,  in distinction with the compact spins, the principal series of the
$SL(2,\mathbb{C})$ group does not have the highest weight and, as a consequence,
the conventional Algebraic Bethe Ansatz \cite{ABA} is not applicable to
diagonalization of the Hamiltonian \re{Ham}. To solve the Schr\"odinger equation
\re{Sch} for arbitrary number of particles $N$ we will rely instead on the method
of the Baxter $\mathbb{Q}-$operator~\cite{Bax}.

In this method, the Hamiltonian \re{Ham}, the integrals of motion \re{Sch} and,
in general, all transfer matrices of the noncompact Heisenberg spin magnet are
expressed in terms of a single operator $\mathbb{Q}(u,\bar u)$, which acts on the
quantum space of the model and depends on a pair of spectral parameters, $u$ and
$\bar u$. The explicit form of this operator was found in \cite{DKM-I}. As a
result, the Schr\"odinger equation \re{Sch} turns out to be equivalent to the
eigenproblem for the Baxter $\mathbb{Q}-$operator. It is this problem that we
address in the present paper. Namely, we calculate the eigenvalues of the Baxter
$\mathbb{Q}-$operator, establish the quantization conditions for the integrals of
motion \re{q} and, finally, obtain a complete description of the spectrum of the
Schr\"odinger equation \re{Sch}.

As we will show below, the system \re{Sch} has many features in common with
two-dimensional conformal field theories (CFT)~\cite{BPZ}. Since the Hamiltonian
\re{Ham} is given by the sum of two mutually commuting operators acting in the
$z-$ and $\bar z-$sectors, the dynamics in the two sectors is independent on each
other. As a consequence, the solutions to the Schr\"odinger equation \re{Sch}
have the chiral structure similar to that of correlation functions in the CFT.
Namely, the eigenstates $\Psi(\vec z_1,\vec z_2,...,\vec z_N)$ can be factorized
into the product of ``conformal blocks'' depending on the (anti)holomorphic
coordinates and the conserved charges $q$. Similar factorization holds for the
eigenvalues of the Baxter operator $\mathbb{Q}(u,\bar u)$. For $\Psi(\vec
z_1,\vec z_2,...,\vec z_N)$ to be a single-valued function on the two-dimensional
$\vec z-$plane, the conserved charges $q$ have to satisfy the quantization
conditions. The charges $q$ play the r\^ole analogous to that of the conformal
weights of primary fields in the CFT. As we will show, the spectrum of their
quantized values turns out to be very similar to the Kac spectrum of the
conformal weights in the minimal CFT~\cite{BPZ}.

The paper is organized as follows. In Section 2, we summarize the main properties
of the Baxter $\mathbb{Q}-$operator for noncompact Heisenberg spin magnets. In
Section 3, we show that the problem of finding the eigenvalues of the operator
$\mathbb{Q}(u,\bar u)$ can be reduced to solving the $N$th order Fuchsian
differential equation. Its solution leads to the set of consistency conditions
which can be satisfied only if the integrals of motion $q$ take quantized values.
In Section 4, we calculate the eigenvalues of the operator $\mathbb{Q}(u,\bar u)$
and demonstrate that they are factorized into a product of ``conformal blocks'',
which depend separately on the (anti)holomorphic spectral parameters, $u$ and
$\bar u$. Using these expressions, it becomes straightforward to determine the
exact spectrum of the model for arbitrary number of particles, $N$, and complex
$SL(2,\mathbb{C})$ spins, $s$ and $\bar s$. In Section 5, we present the results
of our calculations for the special values of the $SL(2,\mathbb{C})$ spins, $s=0$
and $\bar s=1$. The obtained expressions define the spectrum of the compound
states of reggeized gluons in multi-colour QCD. Section 6 contains the concluding
remarks. The details of the calculations are summarized in the Appendices.

\section{Baxter $\mathbb{Q}-$operator}

In this Section we shall describe, following \cite{DKM-I}, the general properties
of the Baxter $\mathbb{Q}-$operator for the noncompact Heisenberg spin magnet. We
assume that the number of particles $N$ is arbitrary and the complex
$SL(2,\mathbb{C})$ spins $s$ and $\bar s$ are given by \re{spins}. The operator
$\mathbb{Q}(u,\bar u)$ depends on two complex spectral parameters $u$ and $\bar
u$. It acts on the quantum space of the model $V_N=V\otimes ...\otimes V$, with
$V\equiv V^{(s,\bar s)}$ being the representation space of the principal series
of the $SL(2,\mathbb{C})$ group. For $\mathbb{Q}(u,\bar u)$ to be a well-defined
operator on $V_N$, the spectral parameters have to satisfy the additional
condition
\be
i(u-\bar u) = n
\label{u-bar u}
\ee
with $n$ being an arbitrary integer. The Baxter $\mathbb{Q}-$operator commutes
with the Hamiltonian of the model \re{Ham} and shares the common set of the
eigenstates
\be
\mathbb{Q}(u,\bar u)\,\Psi_{\vec p,\{q,\bar q\}}(\vec z_1,\vec z_2,...,\vec z_N)
= Q_{q,\bar q}(u,\bar u) \,\Psi_{\vec p,\{q,\bar q\}}(\vec z_1,\vec z_2,...,\vec
z_N)\,,
\label{Q-values}
\ee
with $\vec p$ being the total two-dimensional momentum of the state and $\{q,\bar
q\}$ denoting the total set of the quantum numbers, $q_k$ and $\bar q_k$ with
$k=2,...,N$.

The Baxter operator plays the central role in our analysis as the energy spectrum
of the model can be expressed in terms of its eigenvalues, $Q_{q,\bar q}(u,\bar
u)$. Indeed, there exist the following two (equivalent) relations between
$Q_{q,\bar q}(u,\bar u)$ and the energy $E_N=E_N(q,\bar q)$
\be
E_N(q,\bar q)=\varepsilon_N + i\frac{d}{du} \ln\left[ Q_{q,\bar
q}^{}(u+is,u+i\bar s)\, \left(Q_{q,\bar q}^{}(u-is,u-i\bar
s)\right)^*\right]\bigg|_{u=0},
\label{Energy-I}
\ee
where $\varepsilon_N=2N\Re \left[\psi(2s)+\psi(2-2s)-2\psi(1)\right]$ and
\ba
E_N(q,\bar q) &=&- \Im\frac{d}{du}\ln \bigg[u^{2N}Q_{q,\bar
q}(u+i(1-s),u+i(1-\bar s))\,
\label{Energy-II}
\\[-1mm]
&&\hspace*{38mm}
\times Q_{-q,-\bar q}(u+i(1-s),u+i(1-\bar s))\bigg]\bigg|_{u=0}.
\nonumber
\ea
The Hamiltonian \re{Ham} is invariant under cyclic permutations of particles
$[\mathbb{P}\,\Psi](\vec z_1,...,\vec z_{N-1},\vec z_N) = \Psi(\vec z_2,...,\vec
z_N,\vec z_1)$ and, as a consequence, its eigenstates possess a definite value of
the quasimomentum defined as
\be
[\mathbb{P}\,\Psi_{\vec{p}\{q,\bar q\}}](z_1,...,z_N) = \e^{i\theta_N(q,\bar q)}
\,\Psi_{\vec{p}\{q,\bar q\}}(z_1,...,z_N)\,,\quad\theta_N(q,\bar
q)=2\pi{k}/{N}\,,
\label{cyclic}
\ee
with $k$ integer in virtue of $\mathbb{P}^N=1$. The quasimomentum can be
expressed in terms of the eigenvalues of the Baxter operator as
\be
\theta_N(q,\bar q)= i\ln\frac{Q_{q,\bar q}(is,i\bar s)}{Q_{q,\bar q}(-is,-i\bar
s)}\,.
\label{quasi-old}
\ee
Thus, the Schr\"odinger equation \re{Sch} turns out to be equivalent to the
eigenproblem for the Baxter $\mathbb{Q}-$operator, Eq.~\re{Q-values}.

The eigenvalues of the Baxter operator,  $Q_{q,\bar q}(u,\bar u)$, have to fulfil
the following three conditions:

\medskip$(i)$ {\it Baxter equations:}

\medskip\noindent
The function $Q_{q,\bar q}(u,\bar u)$ has to satisfy the holomorphic Baxter
equation
\begin{equation}
t_N(u)\,Q_{q,\bar q}(u,\bar u)\!=\! (u+is)^N\,Q_{q,\bar q}(u+i,\bar u) +
(u-is)^N\,Q_{q,\bar q}(u-i,\bar u)\,,
\label{Bax-eq}
\end{equation}
where
\be
t_N(u)=2u^N+q_2 u^{N-2} + ... +q_N
\label{t_N}
\ee
is the eigenvalue of the auxiliary transfer matrix with $q\equiv(q_2,\ldots,q_N)$
denoting the eigenvalues of the holomorphic integrals of motion. The ``lowest''
integral of motion, $q_2$, is related to the total $SL(2,\mathbb{C})$ spin, $h$,
of the system of $N$ particles
\be
q_2=-h(h-1)+Ns(s-1)\,,\quad h={(1+n_h)}/2+i\nu_h
\label{q2}
\ee
with $n_h$ integer and $\nu_h$ real. In addition, $Q_{q,\bar q}(u,\bar u)$ obeys
the equation similar to \re{Bax-eq} in the $\bar u-$sector with $s$, $h$, and
$q_k$ replaced, respectively, by
\be
\bar s=1-s^*\,,\qquad \bar h=1-h^*\,,\qquad \bar q_k=q_k^*
\label{hbar}
\ee
with $k=2,...,N$. The function $Q_{q,\bar q}(u,\bar u)$ does not depend on the
total momentum of the state, $\vec p=(p,\bar p)$ with $p=-i\sum_k S_k^-$ and
$\bar p=-i\sum_k\bar S_k^-$, and it is invariant under $h\to 1-h$ and $\bar h\to
1-\bar h$. Indeed, the Baxter operator $\mathbb{Q}(u,\bar u)$ commutes with the
generators of the $SL(2,\mathbb{C})$ group and, as a consequence, its eigenvalue
$Q_{q,\bar q}(u,\bar u)$ depends only on the $SL(2,\mathbb{C})$ Casimir
operators, like $q_2$, which are symmetric under the above transformation of the
total spin.

\medskip $(ii)$ {\it Analytical properties:}

\medskip\noindent
The spectral parameters satisfy \re{u-bar u} and their possible values can be
parameterized as $u=\lambda-in/2$ and $\bar u=\lambda+in/2$, with $n$ arbitrary
integer and $\lambda$ complex. Then, $Q_{q,\bar q}(u,\bar u)$ should be a
meromorphic function of $\lambda$ with an infinite set of poles of the order not
higher than $N$ situated at the points
\be
\{u_{m}^{\pm}=\pm i(s-m)\,,\ \bar u_{\bar m}^{\pm}=\pm
i(\bar s-\bar m)\}\,,\qquad m,\bar m=1,2,...
\label{poles}
\ee
The behaviour of $Q_{q,\bar q}(u,\bar u)$ in the vicinity of the pole at $m=\bar
m=1$ can be parameterized as
\be
Q_{q,\bar q}(u_{1}^{\pm}+\epsilon,{\bar u}_{1}^{\pm}+\epsilon)=R^\pm(q,\bar
q)\left[\frac1{\epsilon^N} +\frac{i\,E^\pm(q,\bar q)}{\epsilon^{N-1}}+ ...
\,\right]\!.
\label{Q-R,E}
\ee
\\[-3mm]
The functions $R^\pm(q,\bar q)$ fix an overall normalization of the Baxter
operator, while the residue functions $E^\pm(q,\bar q)$ define the energy of the
system (see Eqs.~\re{RR} and \re{energy} below).

\medskip$(iii)$ {\it Asymptotic behaviour:}

\medskip\noindent
In the above parameterization of the spectral parameters, $Q_{q,\bar q}(u,\bar
u)$ should have the following asymptotic behaviour for $|\Im
\lambda|<1/2$ and $\Re\lambda\to\infty$
\be
Q_{q,\bar q}(\lambda-in/2,\lambda+in/2)\sim \e^{i\Theta_h(q,\bar
q)}\lambda^{h+\bar h-N(s+\bar s)}+\e^{-i\Theta_h(q,\bar q)}\lambda^{1-h+1-\bar
h-N(s+\bar s)}\,,
\label{Q-asym}
\ee
with the phase $\Theta_h(q,\bar q)$ depending on the quantum numbers of the state
and the total $SL(2,\mathbb{C})$ spins $h$ and $\bar h$ defined in \re{q2} and
\re{hbar}.

As we will show in Section 3, the Baxter equation \re{Bax-eq} supplemented with
the additional conditions on the pole structure of its solutions, Eq.~\re{poles},
and asymptotic behaviour at infinity, Eq.~\re{Q-asym}, fixes uniquely the
eigenvalues of the Baxter operator, $Q_{q,\bar q}(u,\bar u)$, and, therefore,
allows us to determine the spectrum of the model.

Additional properties of the function $Q_{q,\bar q}(u,\bar u)$ can be deduced
from the symmetry of the model under permutations of particles. Apart from the
cycle symmetry, Eq.~\re{cyclic}, the Hamiltonian ${\cal H}_N$ is invariant under
mirror permutations $\mathbb{M}\,\Psi(\vec z_1,...,\vec z_{N-1},\vec
z_N)=\Psi(\vec z_N,...,\vec z_2,\vec z_1)$ \cite{DKM-I}. This transformation acts
on the integrals of motion as $q_k\to (-1)^k q_k$ and maps the eigenstate of the
Hamiltonian into another one with the same energy but different set of the
quantum numbers
\be
\left[\mathbb{M}\,\Psi_{q,\bar q}\right](\vec z_1,...,\vec z_{N-1},\vec z_N)
=\Psi_{-q,-\bar q}(\vec z_1,...,\vec z_{N-1},\vec z_N)
\label{mirror}
\ee
with $-q\equiv(q_2,-q_3,...,(-1)^n q_n)$ and similar for $\bar q$. For
$q_{2k+1}=0$, or equivalently $q=-q$, Eq.~\re{mirror} is replaced by
$\mathbb{M}\,\Psi_{q,\bar q}=(-1)^{Nn_s+n_h}\Psi_{q,\bar q}$. This property leads
to the following parity relations for the residue functions $R^+(q,\bar q)$
defined in \re{Q-R,E}
\be
R^+(q,\bar q)/R^+(-q,-\bar q)=\e^{2i\theta_N(q,\bar q)}
\label{RR}
\ee
and for the eigenvalues of the Baxter operator
\be
Q_{q,\bar q}(-u,-\bar u) = \e^{i\theta_N(q,\bar q)} Q_{-q,-\bar q}(u,\bar u)\,.
\label{Q-symmetry}
\ee
We recall that the spectral parameters $u$ and $\bar u$ have to satisfy \re{u-bar
u}. Examining the behaviour of the both sides of \re{Q-symmetry} around the pole
at $u=u_{1}^{\pm}$ and $\bar u={\bar u}_{1}^{\pm}$ and making use of
Eq.~\re{Q-R,E} one gets
\be
R^\pm(q,\bar q)=(-1)^N\e^{i\theta_N(q,\bar q)}R^\mp(-q,-\bar q)\,,\qquad
E^\pm(q,\bar q)=-E^\mp(-q,-\bar q)\,.
\label{sym}
\ee

To obtain the expression for the energy $E_N(q,\bar q)$, we apply \re{Energy-II}.
Calculating the logarithmic derivative of $Q_{q,\bar q}$ in the r.h.s. of
\re{Energy-II}, we replace the function $Q_{q,\bar q}(u\pm i(1-s),u\pm i(1-\bar
s))$ by its pole expansion \re{Q-R,E}. Then, applying the second relation in
\re{sym}, one finds
\be
E_N(q,\bar q)= E^+(-q,-\bar q)+ (E^+(q,\bar q))^*=\Re\left[E^+(-q,-\bar q)+
E^+(q,\bar q)\right]\,,
\label{energy}
\ee
where the last relation follows from hermiticity of the Hamiltonian \re{Ham}. We
conclude from Eqs.~\re{energy} and \re{Q-R,E}, that in order to find the energy
$E_N(q,\bar q)$, one has to calculate the residue of $Q_{q,\bar q}(u,\bar u)$ at
the $(N-1)$th order pole at $u=i(s-1)$ and $\bar u=i(\bar s-1)$.

\section{Quantization conditions}

Let us construct the solution to the Baxter equation \re{Bax-eq} satisfying the
additional conditions \re{Q-R,E} and \re{Q-asym}. It proves convenient to use the
following integral representation for $Q_{q,\bar q}(u,\bar u)$
\be
Q_{q,\bar q}(u,\bar u)= \int\frac{d^2 z}{z\bar z}\, z^{-i u} {\bar z}^{-i\bar
u}\, Q(z,\bar z)\,,
\label{Q-R}
\ee
where integration goes over the two-dimensional $\vec z-$plane with $\bar z=z^*$.
This ansatz is advantageous in many respects. Firstly, the condition \re{u-bar u}
is automatically satisfied since it is only for these values of the spectral
parameters that the $z-$integral in the r.h.s.\ of \re{Q-R} is well-defined.
Secondly, the functional Baxter equation on $Q_{q,\bar q}(u,\bar u)$ is
translated into the $N$th order differential equation for the function  $Q(z,\bar
z)$. Its derivation is based on the identity
\be
P(u)\,Q_{q,\bar q}(u,\bar u)=\int\frac{d^2 z}{z\bar z}\, z^{-i u} {\bar
z}^{-i\bar u}\, P(-iz\partial_z)\,Q(z,\bar z)\,,
\label{identity}
\ee
with $P(u)$ being a polynomial in $u$. Substituting \re{Q-R} into the Baxter
equation \re{Bax-eq} and applying \re{identity}, one arrives at
\be
\left[z^s\lr{z\partial_z}^{N}z^{1-s}+z^{-s}\lr{z\partial_z}^{N}z^{s-1}
-2\lr{z\partial_z}^{N}-\sum_{k=2}^N i^{k}q_k\lr{z\partial_z}^{N-k}
\right]Q(z,\bar z)=0\,.
\label{Eq-1}
\ee
The $\bar z-$dependence of $Q(z,\bar z)$ is constrained by a similar equation in
the antiholomorphic sector with $s$ and $q_k$ replaced by $\bar s=1-s^*$ and
$\bar q_k=q_k^*$, respectively. Finally, as we will show below, the remaining
conditions on the analytical properties and asymptotic behaviour of $Q_{q,\bar
q}(u,\bar u)$, Eqs.~\re{Q-R,E} and \re{Q-asym}, become equivalent to a
requirement for $Q(z,\bar z=z^*)$ to be a single-valued function on the complex
$z-$plane.

Going through a standard analysis \cite{WW}, one finds that the differential
equation \re{Eq-1} is of Fuchsian type with three regular singular points located
at $z=0$, $z=1$ and $z=\infty$. Defining $N$ linear independent solutions to
Eq.~\re{Eq-1}, $Q_a(z)$, and their antiholomorphic counterparts, $\widebar
Q_b(\bar z)$, we construct the general expression for the function $Q(z,\bar z)$
as
\be
Q(z,\bar z) = \sum_{a,b=1}^N Q_a(z)\, C_{ab}\, \widebar Q_b(\bar z)\,,
\label{general-sol}
\ee
with $C_{ab}$ being an arbitrary mixing matrix. The functions $Q_a(z)$ and
$\widebar Q_b(\bar z)$ %are defined on the whole complex plane and
acquire a nontrivial monodromy around three singular points, $z,\,\bar z=0$, $1$
and $\infty$. For $Q(z,\bar z=z^*)$ to be well-defined on the whole plane, the
monodromy should cancel in the r.h.s.\ of \re{general-sol}. This requirement
leads to the set of nontrivial conditions on the matrix $C_{ab}$. Solving them,
we will be able not only to obtain the values of the mixing coefficients,
$C_{ab}$, but also determine the quantized values of the integrals of motion
$q_k$.

We would like to point out that our approach to defining the function $Q(z,\bar
z)$ in Eq.~\re{general-sol} is similar in many respects to the well-known
approach to constructing the correlation functions in the minimal CFT \cite{DF}.
There, $Q(z,\bar z)$ plays the r\^ole of four-point correlation functions
depending on the anharmonic ratios of the coordinates, $z$ and $\bar z$. The
latter satisfy the differential equations (``the null vector condition'') similar
to \re{Eq-1}, in which the integrals of the motion $q_k$ are replaced by some
combinations of the conformal weights of the primary fields. In the minimal CFT,
the conformal blocks $Q_a(z)$ and $\widebar Q_b(\bar z)$ are given by multiple
contour integrals and their monodromy around the singular points $z=0$, $1$ and
$\infty$ can be found in a closed form. Going over to Eq.~\re{Eq-1}, one finds
(see Section 3.4 below) that similar representation exists only at $N=2$ and it
remains unclear whether it can be generalized for arbitrary number of particles
$N$. Our subsequent analysis does not rely on such representation.

To determine the function $Q(z,\bar z)$, we shall construct the r.h.s.\ of
\re{general-sol} in the vicinity of three singular points, $z=0$, $z=1$ and
$z=\infty$, and analytically continue the obtained expressions onto the whole
$z-$plane. Additional simplification occurs due to the symmetry of the
differential equation \re{Eq-1} under the transformation $z\to 1/z$ and $q_k\to
(-1)^k q_k$. This symmetry is a manifestation of a general property of the
eigenvalues of the Baxter operator, Eq.~\re{Q-symmetry}, which leads to
\\[1mm]
\be
Q_{q,\bar q}(z,\bar z)=\e^{i\theta_N(q,\bar q)}Q_{-q,-\bar q}(1/z,1/\bar z)\,.
\label{Qz-symmetry}
\ee
\\[0mm]
Here, we indicated explicitly the dependence of the function $Q(z,\bar z)$ on the
integrals of motion. Applying \re{Qz-symmetry}, one can define $Q(z,\bar z)$
around $z=\infty$ from the solution at $z=0$.

\subsection{Solution around $z=0$}

Looking for the solution to \re{Eq-1} around $z=0$ in the form $Q(z) \sim z^a$,
we find that the exponent $a$ satisfies the indicial equation
\be
(a-1+s)^N=0\,.
\label{indicial-0}
\ee
Since its solution, $a=1-s$, is $N-$times degenerate, the small$-z$ asymptotics
of $Q(z)$ contains terms $\sim (\ln z)^k$ with $k\le N-1$. Let us define the
fundamental set of linear independent solutions to \re{Eq-1} around $z=0$ as
\footnote{The monodromy matrix, defined as $Q_n^{(0)}(z\e^{2\pi i})
=M_{nk}Q_k^{(0)}(z)$, has a Jordan structure and, therefore, it can not be
brought to a diagonal form upon redefinition of the fundamental basis. It is
interesting to note that similar situation occurs in the operator algebra of
primary fields in the so-called Logarithmic CFT~\cite{LCFT}.}
\ba
&&Q_1^{(0)}(z) = z^{1-s} u_1(z)
\nonumber\\
&&Q_m^{(0)}(z)=z^{1-s}\left[u_1(z) (\ln z)^{m-1}+\sum_{k=1}^{m-1}c_{m-1}^k
u_{k+1}(z) (\ln z)^{m-k-1}\right],
\label{Q-0-h}
\ea
with $2\le m\le N$ and the binomial coefficients $c_{m-1}^k=(m-1)!/(k!(m-k-1)!)$
inserted for later convenience. The functions $u_m(z)$ are given by power series
\be
u_m(z) = 1+\sum_{n=1}^\infty z^n\,u^{(m)}_{n}(q)\,,
\label{power-series-0}
\ee
which converge uniformly inside the region $|z|<1$. Inserting \re{Q-0-h} and
\re{power-series-0} into \re{Eq-1}, one finds that the expansion coefficients
$u^{(m)}_{n}(q)$ satisfy three-term (nonhomogeneous) recurrence relations with
respect to $n$. To save space, we do not present here their explicit form.

The fundamental set of solutions to the antiholomorphic differential equation,
$\widebar Q_m^{(0)}(\bar z)$, can be obtained from \re{Q-0-h} by replacing $s$
and $q_k$ by $\bar s=1-s^*$ and $\bar q_k=q_k^*$, respectively. Then, the general
solution for $Q(z,\bar z)$ around $z=0$ is given by
\be
Q(z,\bar z) \stackrel{|z|\to 0}{=} \sum_{m,\bar m=1}^N Q^{(0)}_m(z)\,
C^{(0)}_{m\bar m}\,\widebar{Q}^{(0)}_{\bar m}(\bar z)\,.
\label{Q-0}
\ee
The mixing matrix $C^{(0)}_{m\bar m}$ has to be chosen in such a way that
$Q(z,\bar z)$ should be single-valued at $z=0$, or equivalently, the monodromy of
$Q^{(0)}_m(z)$ and $\widebar{Q}^{(0)}_{\bar m}(\bar z=z^*)$ around $z=0$ should
cancel each other in the r.h.s.\ of \re{Q-0}. According to \re{Q-0-h}, the
monodromy of $Q^{(0)}_m(z)$ is due to $z^{1-s}-$factor and $\ln z-$terms. Taking
into account that $s-\bar s=n_s$ is an integer, we find that the factor
$z^{1-s}\bar z^{1-\bar s}$ does not affect the monodromy of the r.h.s.\ of
\re{Q-0}. Then, for $Q(z,\bar z)$ to be single valued, it should depend only on
$\ln(z\bar z)$ rather than on $\ln z$ and $\ln\bar z$ separately. It is
straightforward to verify that this condition is satisfied provided that the
matrix elements $C^{(0)}_{nm}$ vanish below the main anti-diagonal, that is for
$n+m> N+1$, and have the following form for $n+m\le N+1$
\be
C^{(0)}_{nm}=\frac{\sigma}{(n-1)!(m-1)!}
\sum _{k=0}^{N-n-m+1}{\frac {(-2)^{k}}{k!}\,\alpha_{k+n+m-1}}
\label{C0}
\ee
with $\sigma, \alpha_1,...,\alpha_{N-1}$ being arbitrary complex parameters and $\alpha_N=1$.%
\footnote{To cancel the monodromy, it is enough to require that the sum over $k$ should depend
only on the sum $n+m$. We have chosen the sum in this particular form for later
convenience (see Eq.~\re{Q-small}).}

The mixing matrix $C^{(0)}_{m\bar m}$ depends on $N$ arbitrary complex parameters
$\sigma$ and $\alpha_k$. We can fix the normalization of the function $Q(z,\bar
z)$ by choosing the value of $\sigma$. Since the resulting expression for the
eigenvalue of the Baxter operator has to satisfy simultaneously two parity
relations, Eqs.~\re{RR} and \re{Qz-symmetry}, its value can not be arbitrary.
%$\sigma$ can not be an arbitrary function of $q$ and $\bar q$.
As we will see in a moment, both relations are satisfied for
$\sigma=\exp(i\theta_N(q,\bar q))$, with $\theta_N(q,\bar q)$ being the
quasimomentum. Later, we will use \re{Qz-symmetry} to calculate the eigenvalues
of $\theta_N(q,\bar q)$ (see Eq.~\re{parity-qc}).

Substituting \re{C0} into \re{Q-0} and taking into account \re{Q-0-h}, one gets
small$-z$ expansion of the function $Q(z,\bar z)$. The leading asymptotic
behaviour for $z\to 0$ can be obtained by neglecting ${\cal O}(z)$ corrections in
\re{power-series-0}. In this way, one finds
\be
Q_{q,\bar q}(z,\bar z)=z^{1-s}\bar z^{1-\bar s}\e^{i\theta_N(q,\bar
q)}\left[\frac{\ln^{N-1} (z\bar z)}{(N-1)!} +
\frac{\ln^{N-2} (z\bar z)}{(N-2)!}\, \alpha_{N-1}+... +
\frac{\ln (z\bar z)}{1!}\, \alpha_{2}+\alpha_1\right]\lr{1+{\cal O}(z,\bar z)}\,.
\label{Q-small}
\ee
Using this expression, we can calculate the contribution of the small$-z$ region
to the eigenvalue of the Baxter operator, Eq.~\re{Q-R}. Introducing a cut-off,
$\rho\ll 1$, and replacing $Q(z,\bar z)$ in Eq.~\re{Q-R} by its expansion
\re{Q-small}, one integrates term-by-term over the region $|z| < \rho$ by making
use of the identity
$$
\int_{|z|<\rho}\frac{d^2 z}{z\bar z} z^{-iu}\bar z^{-i\bar u}
\ln^n(z\bar z) z^{m-s} \bar z^{\bar m-\bar s}=\pi\delta_{m-s-iu,\bar m-\bar s-i\bar u}\left[
\frac{(-1)^n\,n!}{(m-s-iu)^{n+1}}+{\cal O}((m-s-iu)^0)\right],
$$
with $m$ and $\bar m$ positive integer. We find that, in agreement with the
general properties of the Baxter operator, Eqs.~\re{poles} and \re{Q-R,E}, the
function $Q_{q,\bar q}(u,\bar u)$ has poles of the order $N$ at the points
$u=i(s-m)$ and $\bar u=i(\bar s-\bar m)$. At $m=\bar m=1$ one finds from
\re{Q-small}
\be
Q_{q,\bar q}(u_{1}^{+}+\epsilon,{\bar u}_{1}^{+}+\epsilon)=-\frac{\pi
\e^{i\theta_N(q,\bar
q)}}{(i\epsilon)^N}
\left[1+i\epsilon\,\alpha_{N-1}+...+(i\epsilon)^{N-2}
\,\alpha_2+(i\epsilon)^{N-1}\,\alpha_1+{\cal O}(\epsilon^N)
\right]
\label{Q-pole-0}
\ee
with $u_{1}^{+}$ and ${\bar u}_{1}^{+}$ defined in \re{poles}. It is easy to see,
using \re{Qz-symmetry} and \re{Q-small}, that the remaining poles of $Q_{q,\bar
q}(u,\bar u)$ are located at $u=-i(s-m)$ and $\bar u=-i(\bar s-\bar m)$ and they
originate from integration in \re{Q-R} over the region of large $|z|> 1/\rho$.

Matching \re{Q-pole-0} into \re{Q-R,E} one gets
\be
R^+(q,\bar q)=-\frac{\pi}{i^N}\e^{i\theta_N(q,\bar q)}\,,\qquad E^+(q,\bar
q)=\alpha_{N-1}(q,\bar q)\,.
\label{E=alpha}
\ee
We verify, using $\theta_N(q,\bar q)=-\theta_N(-q,-\bar q)$, that the obtained
expression for $R^+(q,\bar q)$ satisfies the parity relation \re{RR}. According
to their definition, Eq.~\re{C0}, the $\alpha-$parameters are arbitrary complex.
We indicated in \re{E=alpha} the dependence of the $\alpha-$parameters on the
integrals of motion since we anticipate that their values will be fixed by the
quantization conditions to be discussed below. Insertion of the second relation
in \re{E=alpha} into \re{energy} leads to the following remarkable expression for
the energy
\be
E_N(q,\bar q)=\Re\left[\alpha_{N-1}(-q,-\bar q)+\alpha_{N-1}(q,\bar q)\right],
\label{E-fin}
\ee
where $q=\{q_k\}$, $-q=\{(-1)^k q_k\}$ and similar for $\bar q$.

We conclude that the small$-z$ asymptotics \re{Q-small} leads to the correct
analytical properties of the eigenvalues of the Baxter operator, Eq.~\re{poles}.
Moreover, the energy of the system, $E_N(q,\bar q)$, is related to the matrix
elements of the mixing matrix \re{C0} in the fundamental basis \re{Q-0-h}.

\subsection{Solution around $z=1$}

Substituting $Q(z)\sim(z-1)^b$ into \re{Eq-1} one obtains after some calculation
the following indicial equation
\be
(b+1+h-Ns)(b+2-h-Ns)\prod_{k=0}^{N-3}(b-k)=0\,,
\label{b-exponents}
\ee
with the total $SL(2,\mathbb{C})$ spin $h$ defined in \re{q2}. Since the
solutions $b=k$ with $k=0,...,N-3$ differ from each other by an integer, one
expects to encounter logarithmically enhanced terms $\sim\ln(1-z)$. However, a
close examination of \re{Eq-1} reveals that the solutions to \re{Eq-1} do not
contain such terms provided that $h\neq (1+n_h)/2$, or equivalently $\Im h\neq 0$
(see Eq.~\re{q2}). At $h=(1+n_h)/2$ and $\Im s\neq 0$, the additional degeneracy
occurs between the solutions to \re{b-exponents}, $b=Ns-h-1$ and $b=Ns+h-2$. It
leads to the appearance of the terms $\sim\ln(1-z)$ in the asymptotics of $Q(z)$
for $z\to 1$. Obviously, similar relations hold in the $\bar z-$sector.

The fundamental set of solutions to Eq.~\re{Eq-1} around $z=1$ is defined
similarly to \re{Q-0-h}. For $\Im h\neq 0$ one gets
\ba
&&Q_1^{(1)}(z) = z^{1-s} (1-z)^{Ns-h-1}v_1(z)\,,
\nonumber
\\[2mm]
&&Q_2^{(1)}(z) = z^{1-s} (1-z)^{Ns+h-2}v_2(z)\,,
\nonumber
\\[2mm]
&&Q_m^{(1)}(z) = z^{1-s} (1-z)^{m-3} v_m(z)\,,
\label{set-1}
\ea
with $m=3,...,N$. The functions $v_{i}(z)$ $(i=1,2)$ and $v_m(z)$ are given by
the power series
\be
v_i(z)=1+\sum_{n=1}^\infty (1-z)^n \,v^{(i)}_{n}(q)\,,
\qquad
v_m(z)=1+\sum_{n=N-m+1}^\infty (1-z)^n\, v^{(m)}_{n}(q)\,,
\label{v-series}
\ee
which converge uniformly inside the region $|1-z|<1$. We notice that
$Q_2^{(1)}(z)$ can be obtained from the function $Q_1^{(1)}(z)$ by replacing
$h\to 1-h$. Substituting \re{set-1} into \re{Eq-1}, one finds that the expansion
coefficients $v^{(i)}_{n}$ and $v^{(m)}_{n}$ satisfy the $N-$term homogenous
recurrence relations
with respect to the index $n$.%
\footnote{The factor $z^{1-s}$ was included in the r.h.s.\ of \re{set-1} and \re{Q-deg}
to simplify the form of the recurrence relations. Without this factor, the
recursion will involve $N+1$ terms.} As was already mentioned, at $h=(1+n_h)/2$
the solutions $Q_{1,2}^{_{(1)}}(z)$ become degenerate and one of them,
$Q_1^{_{(1)}}(z)$ for $n_h\ge 0$, has to be redefined to include the additional
$\ln(1-z)-$term
\be
Q_1^{(1)}(z)\bigg|_{h=(1+n_h)/2} = z^{1-s} (1-z)^{Ns-(n_h+3)/2}\left[(1-z)^{n_h}
\ln (1-z)\, v_2(z)+ \widetilde v_1(z)\right]\,.
\label{Q-deg}
\ee
Here the function $v_2(z)$ is the same as before, while $\widetilde
v_1(z)=\sum_{k=0}^\infty\tilde v_k z^k$ and the coefficients $\tilde v_k$ satisfy
the $N-$term recurrence relations with the boundary condition $\tilde v_{n_h}=1$.

The fundamental set in the antiholomorphic sector, $\widebar Q_n^{_{(1)}}(\bar
z)$, is obtained from the functions $Q_n^{_{(1)}}(z)$ by replacing $s$ and $h$ by
$\bar s=1-s^*$ and $\bar h=1-h^*$, respectively. Among all functions in the
fundamental set \re{set-1} only two, $Q_1^{(1)}(z)$ and $Q_2^{(1)}(z)$, are not
analytical at $z=1$. As a consequence, a general solution for $Q(z,\bar z)$
possessing a trivial monodromy around $z=1$ can be constructed as
\be
Q(z,\bar z)\stackrel{z\to 1}{=}\beta_h Q_1^{(1)}(z)\widebar Q_1^{(1)}(\bar
z)+\beta_{1-h} Q_2^{(1)}(z)\widebar Q_2^{(1)}(\bar z) +\sum_{m,\bar m=3}^N
Q_m^{(1)}(z)\,\gamma_{m\bar m}\,\widebar Q_{\bar m}^{(1)}(\bar z)\,.
\label{Q-1}
\ee
The $\beta-$coefficients depend, in general, on the total spin $h$ (and $\bar
h=1-h^*$). They are chosen in \re{Q-1} in such a way that the symmetry of the
eigenvalues of the Baxter operator under $h\to 1-h$ is manifest. It is convenient
to rewrite \re{Q-1} in a matrix form as
\be
Q(z,\bar z)=\overrightarrow{Q^{(1)}}%(z)
\cdot C^{(1)}
\cdot\overrightarrow{\widebar Q^{(1)}}\,,\qquad
C^{(1)}= \left(\begin{array}{cc}
{\begin{array}{cc} \beta_h & 0   \\ 0 &\beta_{1-h}\end{array}} & \mbox{\bf\large
0}
\\[4mm] %& \\
\mbox{\bf\large 0} & \mbox{\Large\boldmath$\gamma$}
\end{array}\right)\,,
\label{C1-exp}
\ee
with $\mbox{\boldmath$\gamma$}\equiv\gamma_{m\bar m}$. The expansion \re{Q-1} is
valid only for $\Im h\neq 0$. For $h=(1+n_h)/2$ the first two terms in the
r.h.s.\ of \re{Q-1} look differently in virtue of \re{Q-deg}
\be
Q(z,\bar z)\bigg|_{h=(1+n_h)/2}=\beta_1\! \left[Q_1^{(1)}(z)\widebar
Q_2^{(1)}(\bar z)+Q_2^{(1)}(z)\widebar Q_1^{(1)}(\bar z)\right]+\beta_2\,
Q_2^{(1)}(z)\widebar Q_2^{(1)}(\bar z) + ...\, ,
\ee
where ellipses denote the remaining terms. Substituting \re{Q-1} into \re{Q-R}
and performing integration over the region of $|1-z|\ll 1$, one can find the
asymptotic behaviour of $Q(u,\bar u)$ at large $u$ and $\bar u$. As we will show
in Section 4, it turns out to be in agreement with the general properties of the
Baxter operator, Eq.~\re{Q-asym}.

The mixing matrix $C^{(1)}$ defined in \re{C1-exp} has a block-diagonal
structure. It depends on $2+(N-2)^2$ complex parameters $\beta_h$, $\beta_{1-h}$
and $\gamma_{m\bar m}$ which, in general, are some functions of the integrals of
motion $(q,\bar q)$ to be fixed by the quantization conditions. Let us take into
account that the function $Q(z,\bar z)$ has to satisfy the duality relation
\re{Qz-symmetry}. For $|1-z|\to 0$, one can apply \re{Q-1} to evaluate the both
sides of \re{Qz-symmetry} in terms of the mixing matrices $C^{(1)}(q,\bar q)$ and
$C^{(1)}(-q,-\bar q)$. This leads to the set of relations on the functions
$\beta_i(q,\bar q)$ and $\gamma_{m\bar m}(q,\bar q)$.

To obtain these relations one uses
%Their derivation is based on
the following property of the fundamental basis \re{set-1}
\be
Q^{(1)}_a(1/z;-q)= \sum_{b=1}^N S_{ab} \,Q^{(1)}_b(z;q)\,,
\label{S-def}
\ee
where $\Im(1/z)>0$. Here, we indicated explicitly the dependence on the integrals
of motion. Since the $Q-$functions in the both sides of \re{S-def} satisfy the
same differential equation \re{Eq-1}, the $S-$matrix does not depend on $z$. As a
consequence, one can evaluate the matrix elements $S_{ab}$ by examining the
leading asymptotic behaviour of the both sides of \re{S-def} for $z\to 1$. In
this way, applying \re{set-1} and \re{v-series}, one finds that the only
nonvanishing matrix elements are given by
\be
S_{11}=\e^{-i\pi(Ns-h-1)}\,,\qquad S_{22}=\e^{-i\pi(Ns+h-2)}
\,,\qquad
S_{k,k+m}=(-1)^{k-3}\frac{(k-2s-1)_m}{m!}
\label{S-matrix}
\ee
with $(x)_m\equiv\Gamma(x+m)/\Gamma(x)$, $3\le k \le N$ and $0\le m \le N-k$.
Similar relations hold in the antiholomorphic sector, $\widebar
S_{11}=\e^{i\pi(N\bar s-\bar h-1)}$, $\widebar S_{22}=\e^{i\pi(N\bar s+\bar
h-2)}$ and $\widebar S_{k,k+m}=(-1)^{k-3}{(k-2\bar s-1)_m}/{m!}$.

Finally, we substitute \re{Q-1} and \re{S-def} into \re{Qz-symmetry} and find
\\[-1mm]
\ba
\beta_h(q,\bar q)&=& \e^{i\theta_N(q,\bar q)} (-1)^{Nn_s+n_h}\beta_h(-q,-\bar q)\,,\qquad
\nonumber
\\[1mm]
\gamma_{m\bar m}(q,\bar q) &=& \e^{i\theta_N(q,\bar q)}
\sum_{n,\bar n\ge 3}^N S_{nm} \gamma_{n\bar n}(-q,-\bar q)\,\widebar S_{\bar n\bar
m}\,.
\label{parity-qc}
\ea
These relations imply that, similar to the energy, Eq.~\re{E-fin}, the
eigenvalues of the quasimomentum, $\theta_N(q,\bar q)$, can be obtained from the
mixing matrix at $z=1$. In particular, it follows from the first relation in
\re{parity-qc} that the quasimomentum of the eigenstates with $q_{2k+1}=\bar
q_{2k+1}=0$ $(k=1,2...)$, or equivalently $q=-q$, is equal to
\be
\e^{i\theta_N(q,\bar q)}=(-1)^{Nn_s+n_h}\,,
\label{quasi-0}
\ee
since $\beta_h(q,\bar q)=\beta_h(-q,-\bar q)$. At $N=2$ one finds from
\re{parity-qc} that $\e^{i\theta_2}=(-1)^{n_h}$.

\subsection{Transition matrices}

Eqs.~\re{Q-0} and \re{Q-1} define the solution for $Q(z,\bar z)$ in the vicinity
of $z=0$ and $z=1$, respectively. To obtain the eigenvalues of the Baxter
$\mathbb{Q}-$operator, Eq.~\re{Q-R}, one has to sew \re{Q-0} and \re{Q-1} inside
the region $|1-z|<1,\,|z| < 1$ and, then, analytically continue the resulting
expression for $Q(z,\bar z)$ into the whole complex $z-$plane by making use of
the duality relation \re{Qz-symmetry}. As we will see in a moment, this can be
done only for the special values of integrals of motion $(q,\bar q)$ satisfying
the quantization conditions (see Eq.~\re{C1-C0} below).

The sewing procedure is based on the relation between two fundamental sets of
solutions, Eqs.~\re{Q-0-h} and \re{set-1}. Choosing $z$ to be inside the region
of convergence of the both series, Eqs.~\re{power-series-0} and \re{v-series}, we
define the transition matrices $\Omega(q)$ and $\widebar \Omega(\bar q)$
\be
Q_n^{(0)}(z)=\sum_{m=1}^N \Omega_{nm}(q)\, Q_m^{(1)}(z)\,,\qquad
\widebar Q_{n}^{(0)}(\bar z)=\sum_{m=1}^N\widebar \Omega_{nm}(\bar q)
 \,\widebar Q_{m}^{(1)}(\bar z)\,.
\label{Omega-def}
\ee
Since the functions $Q_n^{_{(0)}}(z)$ and $Q_m^{_{(1)}}(z)$ satisfy the same
differential equation \re{Eq-1}, the transition matrices are $z-$independent. For
the fundamental set of solutions, Eqs.~\re{Q-0-h} and \re{set-1}, these matrices
are uniquely fixed and they can be calculated as \cite{JW}
\be
\Omega(q)=W^{(0)}[W^{(1)}]^{-1}\,,\qquad W^{(j)}_{nk}=\partial_{z_0}^k Q_n^{(j)}(z_0)
\label{Omega}
\ee
with $j=0\,,1$ and $z_0$ being some reference point, say $z_0=1/2$, and similar
for $\widebar\Omega(\bar q)$. The resulting expressions for the matrices
$\Omega(q)$ and $\widebar\Omega(\bar q)$ take the form of infinite series in $q$
and $\bar q$, respectively. The transition matrices in two sectors are related to
each other as $\Omega(q)=\Omega_{s}(h,q)$ and $\widebar\Omega(q)=\Omega_{\bar
s}(\bar h,\bar q)$.

The transition matrices allow us to analytically continue the solutions \re{Q-0}
valid for $|z|< 1$ to the region $|1-z|< 1$. Substituting \re{Omega-def} into
\re{Q-0} and matching the result into \re{Q-1}, we find that the two expressions
for the function $Q(z,\bar z)$, Eqs.~\re{Q-0} and \re{Q-1}, can be sewed together
provided that the mixing matrices $C^{(0)}$ and $C^{(1)}$ satisfy the following
relation
\be
C^{(1)}(q,\bar q)=\left[\Omega(q)\right]^T C^{(0)}(q,\bar q)\ \widebar
\Omega(\bar{q})\,.
\label{C1-C0}
\ee
This matrix equation provides the quantization conditions for the integrals of
motion of the model, $q_k$ and $\bar q_k$ with $k=3,...,N$. In addition, it
allows us to determine the matrices $C^{(0)}$ and $C^{(1)}$ and, as a
consequence, evaluate the eigenvalues of the Baxter $\mathbb{Q}-$operator,
Eq.~\re{Q-R}. Indeed, replacing in \re{C1-C0} the mixing matrices by their
expressions, Eqs.~\re{Q-0} and \re{Q-1}, we obtain the system of $N^2$ equations
involving $(N-1)$ $\alpha-$parameters  inside the matrix $C^{(0)}$, $2+(N-2)^2$
parameters $\beta_{1,2}$ and $\gamma_{m\bar m}$ inside the matrix $C^{(1)}$, as
well as $(N-2)$ integrals of motion $q_3,...,q_N$ (we recall that $\bar
q_k=q_k^*$). Thus, the system \re{C1-C0} is overdetermined. It allows us to
determine all parameters including the quantized $q$ and, in addition, it
provides $(2N-3)$ nontrivial consistency conditions on the obtained solutions.
Additional consistency conditions follow from \re{parity-qc}.

The solutions to the quantization conditions \re{C1-C0} for different number of
particles $N$ and the emerging properties of the spectrum of the model will be
described in details in Section~5.

\subsection{Special case: $N=2$}

As was already mentioned, the solution to the differential equation \re{Eq-1} for
$N=2$ admit representation in the form of contour integrals and, as a result, the
quantization conditions \re{C1-C0} can be solved exactly. At $N=2$, after the
change of variables $z=(x-1)/x$ and $Q(z)=[x(1-x)]^{1-s}y(x)$, Eq.~\re{Eq-1}
takes the form of the Legendre's differential equation~\cite{WW}
\be
\left[\frac{d}{dx} x(1-x)\frac{d}{dx}+h(h-1)\right] y(x)=0\,.
\label{Legendre}
\ee
Its general solution is well-known as $y(x) = \int_{C_w} {dw} \, w^{h-1}
(w-1)^{h-1} (w-x)^{-h}$, where the integration contour $C_w$ has to be chosen in
such a way that the integrand resumes its original value after encircling $C_w$.
Then, two linear independent solutions to \re{Legendre} are given by the
Legendre's functions of the first and second kind, $\mathbf{P}_{-h}(2x-1)$ and
$\mathbf{Q}_{-h}(2x-1)$, respectively. Using the relation between these functions
\be
-\pi\cot(\pi h)\,\mathbf{P}_{-h}(2x-1)=%-\frac{\tan(\pi h)}{\pi}
\mathbf{Q}_{-h}(2x-1)-\mathbf{Q}_{h-1}(2x-1)
\label{P-Q}
\ee
and going back to the $z-$representation, we choose the fundamental set of
solutions to Eq.~\re{Eq-1} as $Q_s(z;h)$ and $Q_s(z;1-h)$,
where the notation was introduced for%
\footnote{Obviously, this definition is ambiguous.
Instead of using the $\mathbf{Q}-$functions, one may define the fundamental set
entirely in terms of the $\mathbf{P}-$functions (see Eq.~\re{Q0-def} below).}
\be
Q_s(z;h)\equiv\left[\frac{z}{(1-z)^2}\right]^{1-s}\mathbf{Q}_{-h}\lr{\frac{1+z}{1-z}}\,.
\label{Q-function}
\ee
The properties of the function $Q_s(z;h)$, including its relation to the
fundamental set \re{Q-0-h} and \re{set-1}, can be found in
Appendix~\ref{App:N=2}.

Following \re{general-sol}, we construct $Q(z,\bar z)$ as a bilinear combination
of the functions $Q_s(z;h)$ and $Q_s(z;1-h)$ and their antiholomorphic
counterparts, $Q_{\bar s}(\bar z;\bar h)$ and $Q_{\bar s}(\bar z;1-\bar h)$.
Requiring $Q(z,\bar z)$ to have a trivial monodromy around $z=1$ and taking into
account that $Q_s(z;h)\sim (1-z)^{2s-h-1}$ for $z\to 1$ (see Eq.~\re{asym-1}),
one gets
\be
Q(z,\bar z) = c_{h}\,Q_s(z;h)Q_{\bar s}(\bar z;\bar h)+
c_{1-h}\,Q_s(z;1-h)Q_{\bar s}(\bar z;1-\bar h)\,.
\label{N=2-Q}
\ee
To fix the coefficients $c_{h}$ and $c_{1-h}$, one examines the small$-z$
asymptotics of \re{N=2-Q} with a help of Eq.~\re{asym-0} and requires that the
terms $\sim\ln z\ln \bar z$ should cancel and the coefficients in front of $\ln
z$ and $\ln \bar z$ should be the same. Applying the identity
$\psi(1-h)-\psi(h)=\pi\cot(\pi h)$ one finds that the both conditions are
fulfilled provided that $c_{h}=-c_{1-h}$ and $\cot(\pi h)=\cot(\pi \bar h)$. The
second relation is automatically satisfied thanks to the property of the
$SL(2,\mathbb{C})$ spins of the principal series, $h-\bar h=n_h$ with $n_h$
integer. Choosing
\be
c_{h}=\frac2{\pi}\tan(\pi h)(-1)^{n_h}\,,
\label{N=2-c}
\ee
we obtain from \re{N=2-Q} and \re{asym-0} the small$-z$ behaviour of the function
$Q(z,\bar z)$ as
\be
Q(z,\bar z)= z^{1-s}\bar z^{1-\bar s} (-1)^{n_h}\bigg\{\ln(z\bar
z)+2\Re\left[\psi(h)+\psi(1-h)-2\psi(1)\right]+{\cal O}(z,\bar z)\bigg\}\,.
\label{N=2-0}
\ee
Taking into account that the quasimomentum of the $N=2$ states is equal to
$\e^{i\theta_2}=(-1)^{n_h}$, Eq.~\re{quasi-0}, we find that this relation is in
agreement with \re{Q-small}. One determines the $\alpha-$parameter by matching
\re{N=2-0} into \re{Q-small}
\be
\alpha_1(h)=2\Re\left[\psi(h)+\psi(1-h)-2\psi(1)\right]\,.
\label{quasi-N=2}
\ee
Finally, applying \re{E-fin} we calculate the energy at $N=2$ as
\be
E_2(h,\bar h)=2\alpha_1(h)=8\Re\left[\psi\lr{\frac{1+|n_h|}2+i\nu_h}-\psi(1)
\right]\,.
\label{E-2-exact}
\ee
The ground state corresponds to $h=\bar h=1/2$, or equivalently $n_h=\nu_h=0$,
\be
\min E_2(h,\bar h)=-16\ln 2\,.
\ee
and it defines the intercept of the BFKL Pomeron \cite{L2}.

The exact solution at $N=2$ is based on the properties of the Legendre functions,
Eq.~\re{Q-function}. Going over to the systems with the number of particles $N\ge
3$, one encounters the following difficulties. Firstly, representation for the
solution to \re{Eq-1} in the form of contour integrals does not exist or, at
least, it is not warranted. Secondly, for $N\ge 3$ the function $Q(z,\bar z)$
depends on the integrals of motion, $q$ and $\bar q$, whose values should be
determined from the quantization conditions \re{C1-C0}. As we will show in
Section 5, both problems can be solved by using the power series representation
for the fundamental set of solutions, Eqs.~\re{Q-0-h} and \re{set-1}.

In this Section, we have demonstrated that in order for the eigenvalues of the
Baxter $\mathbb{Q}-$operator to possess the prescribed properties,
Eqs.~\re{Bax-eq} -- \re{Q-asym}, the integrals of motion, $q$, have to satisfy
the quantization conditions \re{C1-C0}. In this case, one can construct the
function $Q(z,\bar z)$ in the vicinity of the singular points, $z=0$, $z=1$ and
$z=\infty$, and analytically continue it onto the whole $z-$plane with a help of
the transition matrices \re{Omega-def}. The spectrum of the model -- the energy
and the quasimomentum, can be obtained from the mixing matrices $C^{(0)}$ and
$C^{(1)}$, which define the asymptotic behaviour of $Q(z,\bar z)$ around $z=0$
and $z=1$, respectively.

\section{Eigenvalues of the Baxter $\mathbb{Q}-$operator}

In the previous Section, we established the quantization conditions for the
integrals of motion $q$ and obtained the expression for the energy $E_N$. Let us
now construct the corresponding eigenstates $\Psi_{\vec p,\{q,\bar q\}}(\vec
z_1,\vec z_2,...,\vec z_N)$, Eqs.~\re{Sch} and \re{Q-values}.

The analysis is based on the method of Separated Variables (SoV) developed by
Sklyanin \cite{SoV}. It allows us to find the integral representation for the
eigenstates of the model by going over to the representation of the separated
coordinates $\vec{\mybf{x}}=(\vec x_1,...,\vec x_{N-1})$~\cite{DKM-I}
\be
\Psi_{\vec p,\{q,\bar q\}}(\vec{\mybf{z}}) = \int d^{N-1}\mybf{x}\,\mu(\vec{\mybf{x}})
\, U_{\vec p,\vec{\mybf{x}}}(\vec{\mybf{z}})\,
\lr{\Phi_{\{q,\bar q\}}(\vec{\mybf{x}})}^*\,,
\label{SoV-gen}
\ee
with $\vec{\mybf{z}}=(\vec z_1,...,\vec z_N)$. Here, $U_{\vec
p,\vec{\mybf{x}}}(\vec{\mybf{z}})$ is a kernel of the unitary operator
corresponding to this transformation and $\Phi_{\{q,\bar q\}}(\vec{\mybf{x}})$ is
the wave function in the separated coordinates. The explicit expression for
$U_{\vec p,\vec{\mybf{x}}}(\vec{\mybf{z}})$ for arbitrary $N$ was found in
\cite{DKM-I}. Independently, similar expressions at $N=2$ and $N=3$ were obtained
in \cite{DL}.

Remarkable property of the SoV representation is that $\Phi_{\{q,\bar
q\}}(\vec{\mybf{x}})$ is factorized into the product of the eigenvalues of the
Baxter $\mathbb{Q}-$operator depending on different separated coordinates
\be
\lr{\Phi_{\{q,\bar q\}}(\vec{\mybf{x}})}^*= \e^{i\theta_N(q,\bar q)/2}
\prod_{k=1}^{N-1}
\left[\frac{\Gamma(s+ix_k)\Gamma(\bar s-i\bar x_k)}
{\Gamma(1-s+ix_k)\Gamma(1-\bar s-i\bar x_k)}
\right]^N \,Q_{q,\bar q}(x_k,\bar x_k)\,,
\label{Phi-Q-ops}
\ee
where the additional factor  $\e^{i\theta_N(q,\bar q)/2}$ is needed to ensure
\re{mirror}. Contrary to the $\vec z=(z,\bar z)-$co\-ordinates, the possible
values of the separated coordinates $\vec x_k=(x_k,\bar x_k)$ are quantized as
follows~\cite{DKM-I,DL}
\be
x_k=\nu_k-\frac{in_k}2\,,\qquad \bar x_k=\nu_k+\frac{in_k}2\,,
\label{x-quan}
\ee
with $\nu_k$ real and $n_k$ integer. Integration on the space of separated
variables in Eq.~\re{SoV-gen} implies summation over integer $n_k$ and
integration over continuous $\nu_k$
\be
\int d^{N-1} {\mybf{x}}=\prod_{k=1}^{N-1}
\lr{\sum_{n_k=-\infty}^\infty\int_{-\infty}^\infty d\nu_k}\,,
\qquad
\mu(\vec{\mybf{x}})=\frac{2\pi^{-N^2}}{(N-1)!}
\prod_{j,k=1\atop j>k}^{N-1}{|\vec x_k-\vec x_j|^2}\,,
\label{int-dx}
\ee
where $|\vec x_k-\vec x_j|^2=(\nu_k-\nu_j)^2+(n_k-n_j)^2/4$.

Eqs.~\re{SoV-gen} and \re{Phi-Q-ops} allow us to calculate the eigenfunctions of
the model in terms of the eigenvalues of the Baxter $\mathbb{Q}-$operator. By the
construction, the latter have poles specified in \re{poles}. One verifies,
however, that they lie outside the integration contour in \re{int-dx} and the
integral in \re{SoV-gen} is well-defined. Still, one can make use of the pole
structure of $Q_{q,\bar q}(u,\bar u)$ by closing the integration contour over
$\nu_k$ into the upper (or lower) half-plane and calculating the asymptotics of
the wave function $\Psi_{\vec p,\{q,\bar q\}}(\vec{\mybf{z}})$ in the different
regions of the $\vec{\mybf{z}}-$space.

According to \re{Q-R} and \re{general-sol}, the eigenvalue of the Baxter
$\mathbb{Q}-$operator is given by the following two-dimensional integral
\be
Q_{q,\bar q}(u,\bar u)=\int\frac{d^2 z}{z\bar z}\, z^{-i u} {\bar z}^{-i\bar
u}\,\sum_{a,b=1}^N Q_a(z)\, C_{ab}\, \widebar Q_b(\bar z)\,.
\label{Q-2dim}
\ee
Similar integrals have already appeared in the calculation of correlation
functions in two-dimen\-sional CFT. Applying the results of \cite{M}, one can
convert $Q_{q,\bar q}(u,\bar u)$ into a sum of products of holomorphic and
antiholomorphic contour integrals, $\int_{C}dz\,z^{-1-i\bar u} Q_a(z)$ and
$\int_{\bar C}d\bar z\,\bar z^{-1-i\bar u} \widebar Q_b(\bar z)$, respectively,
with the integration contours $C$ and $\bar C$ starting and ending at one of the
singular points $z,\bar z=0,\,1$ and $\infty$. These contour integrals define the
set of $2N-$functions of the (anti)holomorphic spectral parameters $u$ and $\bar
u$. In analogy with the CFT, we shall refer to them as the $Q-$blocks. The
resulting expression for $Q_{q,\bar q}(u,\bar u)$, Eq.~\re{Q-2dim}, is given by a
bilinear combination of $N$ blocks belonging to two sectors.

We would like to stress that, contrary to $Q_{q,\bar q}(u,\bar u)$, the
definition of the $Q-$blocks is ambiguous. The eigenvalues of the Baxter
operator, $Q_{q,\bar q}(u,\bar u)$, stay invariant if one replaces the $Q-$blocks
by their linear combinations  and redefines appropriately the expansion
coefficients $C_{ab}$ in the r.h.s. of \re{Q-2dim}. Making use of this ambiguity,
one may look for the definition of the blocks, for which the expression for
$Q_{q,\bar q}(u,\bar u)$ looks particularly simple. In this Section, we shall
present such a definition. We will demonstrate that the eigenvalues of the Baxter
operator can be expressed in terms of only two $Q-$blocks -- one in each sector,
defined below in Eq.~\re{blocks-def}. We will also show that the quantization
conditions for the integrals of motion and the expressions for the energy and the
quasimomentum, established in Section~3, can be reformulated in terms of the
$Q-$blocks.

\subsection{Decomposition over the Baxter blocks}

To proceed with calculation of \re{Q-2dim}, one has to specify the mixing matrix
$C_{ab}$, as well as the functions $Q_a(z)$ and $\widebar Q_b(\bar z)$. These
functions have to be defined uniformly on the whole complex $z-$plane with the
cuts and their bilinear combination, Eq.~\re{general-sol}, should match \re{Q-0}
and \re{Q-1} for $z\to 0$ and $z\to 1$, respectively.

Let us choose $C_{ab}$ to be the mixing matrix introduced in \re{Q-1} and define
the functions $Q_a(z)$ in the vicinity of $z=1$ as%
\be
C_{ab}=C_{ab}^{(1)}(q,\bar q)\,,\qquad Q_a(z;q)\stackrel{z\to
1}{=}Q_a^{(1)}(z;q)\,.
\label{C-choice}
\ee
Analytical continuation of $Q_a(z;q)$ to the region $z\to 0$ and $z\to\infty$ can
be obtained from \re{Omega-def} and \re{S-def} as
\ba
&& Q_a(z;q)\stackrel{z\to
0}{=}\sum_{b=1}^N\,[\Omega^{-1}(q)]_{ab}\,Q_b^{(0)}(z;q)\,,
\nonumber
\\[2mm]
&&
Q_a(z;q)\stackrel{z\to
\infty}{=}\sum_{b=1}^N\,[\Omega(-q)
\,S]_{ab}^{-1}\,Q_b^{(0)}(1/z;-q)\,,
\label{global-Q}
\ea
with the matrix $S$ given by \re{S-matrix}. The functions $\widebar Q_b(\bar
z;\bar q)$ are defined similarly.

The functions $Q_a(z)$ defined in this way possess a nontrivial monodromy at
$z=1$. Encircling the point $z=1$ on the complex $z-$plane in anticlockwise
direction, one calculates from \re{set-1} the corresponding monodromy matrix as
\be
Q_a(z) \mapsto M_{ab}Q_b(z)\,,\qquad M={\rm diag}\bigg(\e^{2\pi i(Ns-h)},\e^{2\pi
i(Ns+h)},1,...,1\bigg)\,.
\label{M-matrix}
\ee
Unity entries in this matrix correspond to $(N-2)$ functions in the fundamental
set \re{set-1} analytical at $z=1$. Following \cite{M}, the two-dimensional
integral in Eq.~\re{Q-2dim} can be evaluated as (see Appendix~\ref{App-contour}
for details)
\be
Q_{q,\bar q}(u,\bar u)=\frac1{2i}\sum_{a,b=1}^N
\left[(1-M^T)\,C^{(1)}\right]_{ab}
\int_1^\infty \frac{dz}{z}z^{-iu}\,Q_a(z;q)
\int_0^1 \frac{d\bar z}{\bar z}\bar z^{-i\bar u}
\,\widebar Q_b(\bar z;\bar q)\,,
\label{Q-1dim}
\ee
with $M^T=M$ according to \re{M-matrix}. Here we tacitly assumed that the
bilinear combination Eq.~\re{general-sol} is a single-valued function on the
complex plane. This implies, in particular, that the integrals of motion $q$ have
to satisfy the quantization conditions \re{C1-C0}.

As follows from their definition, Eq.~\re{global-Q}, the functions $Q_a(z)$
satisfy the relation \re{S-def}
\be
Q_a(1/z;-q)= \sum_{b=1}^N S_{ab} \,Q_b(z;q)\,,
\label{Q-duality}
\ee
which holds for arbitrary $z$ such that $\Im(1/z)>0$. Changing the integration
variable in \re{Q-1dim}, $z\to 1/z$, and applying \re{Q-duality} one gets
\be
Q_{q,\bar q}(u,\bar u)=\frac1{2i}\sum_{a,b=1}^N
\left[S^T(1-M)\,C^{(1)}\right]_{ab}
\int_0^1 \frac{dz}{z}z^{iu}\,Q_a(z;-q)
\int_0^1 \frac{d\bar z}{\bar z}\bar z^{-i\bar u}
\,\widebar Q_b(\bar z;\bar q)\,.
\label{Q-1dim-new}
\ee
We recall that the matrices $S$, $M$ and $C^{(1)}$ were defined before in
Eqs.~\re{S-matrix}, \re{M-matrix} and \re{Q-1}, respectively. Notice also that
$M^{-1}=S^2$. Substituting the monodromy matrix \re{M-matrix} into
\re{Q-1dim-new} we find that among $N^2-$terms in the r.h.s.\ of \re{Q-1dim-new}
only two (with $a=b=1$ and $a=b=2$) provide a nonvanishing contribution to
$Q_{q,\bar q}(u,\bar u)$. These two terms correspond to the $Q-$functions
nonanalytical at $z=1$, $Q_1(z) \sim (1-z)^{Ns-h-1}$ and $Q_2(z)
\sim (1-z)^{Ns+h-2}$.

Let us introduce notation for the holomorphic and antiholomorphic Baxter blocks
\ba
Q(u;h,q)&=&\frac1{\Gamma(Ns-h)}\int_0^1 \frac{dz}{z}z^{iu}\,Q_1(z;-q)\,,%\qquad
\nonumber
\\
\widebar Q(\bar u;\bar h,\bar q)&=&\frac1{\Gamma(N\bar s-\bar h)}\int_0^1 \frac{d\bar z}{\bar z}\bar z^{-i\bar u}
\,\widebar Q_1(\bar z;\bar q)\,,
\label{blocks-def}
\ea
with the normalization factors chosen for later convenience (see
Eq.~\re{block-asym} below). Taking into account Eqs.~\re{S-matrix} and \re{Q-1},
one gets from \re{Q-1dim-new}
\ba
Q_{q,\bar q}(u,\bar u)&=&\pi\left[~\frac{\Gamma(N\bar s-\bar
h)}{\Gamma(1-Ns+h)}%\sin(\pi(Ns+1-h))
\beta_h(q,\bar
q)\,Q(u;h,q)\,\widebar Q(\bar u;\bar h,\bar q)\right.
\nonumber\\[2mm]
&&
\left.~~+
%\sin(\pi(Ns+h))
\frac{\Gamma(N\bar s-1+\bar h)}{\Gamma(2-Ns-h)}
\beta_{1-h}(q,\bar q)\,Q(u;1-h,q)\,\widebar Q(\bar u;1-\bar h,\bar
q)\right].
\label{Q-block1}
\ea
Here, we used the fact that the functions $Q_1(z;q)$ and $Q_2(z;q)$ (as well as
$\bar Q_1(\bar z;\bar q)$ and $\bar Q_2(\bar z;\bar q)$) can be obtained one from
another through substitution $h\leftrightarrow 1-h$.

We notice that in Eq.~\re{Q-1dim} the integration contours over $z$ and $\bar z$
are different, whereas in the original two-dimensional integral \re{Q-2dim} the
both variables appear on equal footing. Repeating the calculation of \re{Q-2dim}
and making use of the monodromy of the functions $\widebar Q_b(\bar z;\bar q)$ in
the antiholomorphic sector
\be
\widebar Q_a(\bar z) \mapsto \widebar M_{ab}\widebar Q_b(\bar z)\,,\qquad
\widebar M={\rm diag}\bigg(\e^{-2\pi
i(N\bar s-\bar h)},\e^{-2\pi i(N\bar s+\bar h)},1,...,1\bigg)\,,
\ee
one arrives at another (through equivalent) expression for the eigenvalues of the
Baxter operator
\ba
Q_{q,\bar q}(u,\bar u)&=&\pi\left[~\frac{\Gamma(Ns-h)}{\Gamma(1-N\bar s+\bar h)}
\beta_h(q,\bar q)\,Q(-u;h,-q)\,\widebar Q(-\bar u;\bar h,-\bar q)\right.
\nonumber\\[2mm]
&&
\left.
~~~+\frac{\Gamma(Ns-1+h)}{\Gamma(2-N\bar s-\bar h)}
\beta_{1-h}(q,\bar q)\,Q(-u;1-h,-q)\,\widebar Q(-\bar u;1-\bar h,-\bar
q)\right].
\label{Q-block2}
\ea
One can verify the equivalence of Eqs.~\re{Q-block2} and \re{Q-block1}, by
substituting \re{Q-block2} into the l.h.s.\ of \re{Q-symmetry} and taking into
account the second relation in \re{parity-qc}.

Thus, for a given set of the integrals of motion, $(q,\bar q)$, satisfying the
quantization conditions \re{C1-C0}, the eigenvalue of the Baxter
$\mathbb{Q}-$operator, $Q_{q,\bar q}(u,\bar u)$, is unique. It is expressed in
terms of two chiral blocks introduced in \re{blocks-def} and is given by
Eq.~\re{Q-block1}.

\subsection{Properties of the blocks}

Let us show that the blocks $Q(u;h,q)$ and $\widebar Q(\bar u;\bar h,\bar q)$,
defined in \re{blocks-def}, have the following properties:
\begin{itemize}
\item[({\it i})] $Q(u;h,q)$ satisfies the chiral Baxter equation \re{Bax-eq}.
$\widebar Q(\bar u;\bar h,\bar q)$ obeys similar equation in the antiholomorphic
sector.
\item[({\it ii})] $Q(u;h,q)$ and $\widebar Q(\bar u;\bar h,\bar q)$ are
meromorphic functions~\cite{MW,DKM-I,DL} on the complex $u-$ and $\bar u-$planes,
respectively, with the only poles (which are all of order not higher than $N$)
located at the points $u_m^-=-i(s-m)$ and $\bar u_{\bar m}^+=i(\bar s-\bar m)$,
with $m$ and $\bar m$ positive integer. The same property can be expressed in a
concise form as
%\ba
%&& Q(u;h,q)=\Gamma^N(1-s+iu) f(u)
%\,,
%\nonumber
%\\[3mm]
%&&
%\widebar Q(\bar u;\bar h,\bar q)=\Gamma^N(1-\bar s-i\bar u)\bar f(\bar u)
%\label{poles-block}
%\ea
\be
Q(u;h,q)=\Gamma^N(1-s+iu) f(u)\,,\qquad
\widebar Q(\bar u;\bar h,\bar
q)=\Gamma^N(1-\bar s-i\bar u)\bar f(\bar u)
\label{poles-block}
\ee
with $f(u)$ and $\bar f(\bar u)$ being entire functions.
\item[({\it iii})] At large $u$ and $\bar u$, away from the poles
\re{poles-block}, that is for $\Re(1-s+iu)>0$ and $\Re(1-\bar s-i\bar u)>0$, the
functions $Q(u;h,q)$ and $\widebar Q(\bar u;\bar h,\bar q)$ have the asymptotic
behaviour
%\ba
%&&
%Q(u;h,q)\sim (iu)^{-Ns+h}\left[1+{\cal O}(1/u)\right]\,,
%\nonumber
%\\[3mm]
%&&
%\widebar Q(\bar u;\bar h,\bar q)\sim(-i\bar u)^{-N\bar s+\bar h}\left[1+{\cal
%O}(1/\bar u)\right]\,.
%\label{block-asym}
%\ea
\be
Q(u;h,q)\sim (iu)^{-Ns+h}\left[1+{\cal O}(1/u)\right]\,,\qquad
\widebar Q(\bar u;\bar h,\bar q)\sim(-i\bar u)^{-N\bar s+\bar h}\left[1+{\cal
O}(1/\bar u)\right]\,.
\label{block-asym}
\ee
\end{itemize}
By the definition, Eq.~\re{blocks-def}, the blocks $Q(u;h,q)$ and $\widebar
Q(\bar u;\bar h,\bar q)$ are related to the same universal function calculated
for different values of the parameters
\be
Q(u;h,q)=Q_s(u;h,q)\,,\qquad
\widebar Q(\bar u;\bar h,\bar q)=Q_{\bar s}(-\bar u;\bar h,-\bar q)\,.
\label{Q-univ}
\ee

To verify these properties one uses the integral representation for the blocks,
Eq.~\re{blocks-def}. Then, the first property follows from the fact that the
functions $Q_1(z)$ and $\widebar Q_1(\bar z)$ satisfy the differential equation
\re{Eq-1}. As to the second property, the poles of $Q(u;h,q)$ and $\widebar
Q(\bar u;\bar h,\bar q)$ come from integration in \re{blocks-def} over the region
of small $z$ and $\bar z$. The leading asymptotic behaviour of the functions
$Q_1(z)$ and $\widebar Q_1(\bar z)$ in this region can be obtained from the first
relation in \re{global-Q} as $Q_a(z)\sim z^{1-s}\ln^{N-1}z$ and $\bar Q_b(\bar
z)\sim \bar z^{1-\bar s}\ln^{N-1}\bar z$. Finally, to obtain the asymptotics at
infinity, Eq.~\re{block-asym}, one integrates in \re{blocks-def} over the region
$z\to 1$ and $\bar z\to 1$ and makes use of Eqs.~\re{global-Q} and \re{set-1} to
replace $Q_1(z)\sim (1-z)^{Ns-h-1}$ and $\widebar Q_1(\bar z)\sim (1-\bar
z)^{N\bar s-\bar h-1}$.

The above three properties uniquely specify the blocks for $\Im h\neq 0$. As we
have seen in Section 3.2 (see Eq.~\re{Q-deg}), at $\Im h=0$ one of the
fundamental solutions to the differential equation \re{Eq-1} has to be redefined
in order to avoid a degeneracy. One encounters the same problem trying to define
the block $Q(u;h,q)$ at $h=(1+n_h)/2$. In this case, the linear combination
$Q(u;h,q)+c\,Q(u;1-h,q)$ satisfies the three conditions on the $Q-$block for
arbitrary $c$ and, as a consequence, the blocks $Q(u;h,q)$ and $Q(u;1-h,q)$
become degenerate. The expression for the blocks at $h=(1+n_h)/2$ can be found in
the Appendix~\ref{App-6}.

Using the definition of the blocks $Q(u;h,q)$ and $\widebar Q(\bar u;\bar h,\bar
q)$, one can establish different useful relations between them. In particular, as
shown in the Appendix~\ref{App-4}, the blocks in each sector satisfy nontrivial
Wronskian relations (Eqs.~\re{Wronskian} and \re{Wronskian-anti}). Moreover, the
blocks in two sectors are related to each other as (see Eq.~\re{Q-relation})
\be
Q(u;h,q)=\left[\frac{\Gamma(1-s+iu)}{\Gamma(s+iu)}\right]^N\lr{\widebar
Q(u^*;1-\bar h,\bar q)}^*\,.
\label{Qbar-Q}
\ee

At $N=2$ the eigenvalues of the Baxter $\mathbb{Q}-$operator and the $Q-$blocks
can be expressed in terms of the ${}_3F_2-$hypergeometric series of a unit
argument. As was shown in Section 3.4, at $N=2$ the function $Q_1(z)$ entering
\re{blocks-def} is equal (up to an overall normalization) to the Legendre
function of the second kind. Substituting \re{Q-function} into \re{blocks-def}
and using integral representation of the Legendre functions \cite{WW}, one
obtains after some algebra
\be
Q_s(u;h)=\frac12\Gamma\left[{{1-s+iu,1-s+iu,1-h} \atop
{1+s-h+iu,2-s+iu-h}}\right]{}_3F_2\left({{s+iu,1-s+iu,1-h}\atop
{1+s-h+iu,2-s-h+iu}}\bigg|\,1 \right)\,,
\label{N=2-block}
\ee
where $\Gamma[...]$ denotes the ratio of the products of the $\Gamma-$functions
with the arguments listed in the upper and lower rows, respectively. Together
with \re{Q-block1} and \re{beta-h} this leads to the following expression for the
eigenvalues of the Baxter operator at $N=2$
\ba
&&Q(u,\bar u)=2(-1)^{n_h}\tan(\pi h)
\nonumber
\\[2mm]
&&\qquad\times\bigg[\frac{\Gamma(2\bar s-\bar
h)}{\Gamma(1-2s+h)}\,Q(u;h)\,\widebar Q(\bar u;\bar h)-\frac{\Gamma(2\bar
s-1+\bar h)} {\Gamma(2-2s-h)}\,Q(u;1-h)\,\widebar Q(\bar u;1-\bar h)\bigg],
\label{Q-2-3}
\ea
where $Q(u;h)=Q_s(u;h)$ and $\bar Q(\bar u;\bar h)=Q_{\bar s}(-\bar u;\bar h)$
are (non-normalized) $Q-$blocks at $N=2$.%
\footnote{Due to the additional factor ${\Gamma^2(1-h)}/(2\Gamma(2-2h))$
in the r.h.s.\ of \re{Q-Q-rel2}, this expression has asymptotics at infinity that
differs from \re{block-asym} by the same factor.}

 For $N\ge 3$ one can calculate the blocks by replacing the functions
$Q_1(z;q)$ in \re{blocks-def} by their expressions in terms of the fundamental
solutions around $z=0$ and $z=1$, Eqs.~\re{global-Q} and \re{C-choice},
respectively. As shown in the Appendix~\ref{App-4}, this leads to two different
series representations for the block $Q(u;h,q)$, Eqs.~\re{Q-series-1} and
\re{Q-series-2}, which are valid in the different regions on the complex
$u-$plane.

We are now in position to demonstrate that the eigenvalues of the Baxter operator
constructed in this Section have correct asymptotic behaviour at infinity,
Eq.~\re{Q-asym}. Substituting \re{block-asym} into \re{Q-block1}, we verify that
$Q_{q,\bar q}(u,\bar u)$ satisfies Eq.~\re{Q-asym} with the phase
$\Theta_h(q,\bar q)$ given by
\be
\e^{2i\Theta_h(q,\bar q)}=(-1)^{n_h} \frac{\beta_h(q,\bar q)}{\beta_{1-h}(q,\bar q)}
\frac{\Gamma(2-Ns-h)\Gamma(N\bar s-\bar h)}{\Gamma(1-Ns+h)\Gamma(N\bar s-1+\bar
h)}\,.
\label{Theta}
\ee
In particular, for $h=\bar h=1/2+i\nu_h$ and $\nu_h\to 0$ one has $\beta_h(q,\bar
q)\sim 1/\nu_h$ leading to $\e^{2i\Theta_{1/2}(q,\bar q)}=-1$.

\subsection{Quantization conditions from the $Q-$blocks}

As we have seen in the previous Section, the eigenvalues of the Baxter
$\mathbb{Q}-$operator can be expressed in terms of the $Q-$blocks satisfying the
conditions $(i)$--$(iii)$. In distinction with $Q_{q,\bar q}(u,\bar u)$, the
$Q-$blocks can be constructed for arbitrary values of the integrals of motion $q,
\bar q$. In this Section, we will show that the requirement for $Q_{q,\bar q}(u,\bar u)$ to have
correct analytical properties leads to the quantization conditions for the
integrals of motion which are equivalent to \re{C1-C0}.

The general expression for the eigenvalue of the Baxter operator $Q_{q,\bar
q}(u,\bar u)$ in terms of the blocks $Q(u;h,q)$ and $Q(u;1-h,q)$ and their
antiholomorphic counterparts looks like
\be
Q_{q,\bar q}(u,\bar u) = c_h\, Q(u;h,q)\,\widebar Q(\bar u;\bar h,\bar q) +
c_{1-h}\, Q(u;1-h,q)\,\widebar Q(\bar u;1-\bar h,\bar q)
\label{block-ansatz}
\ee
with $c_h$ (and $c_{1-h}$) being arbitrary function of $h$ and the integrals of
motion. $Q_{q,\bar q}(u,\bar u)$ defined in this way is symmetric under $h\to
1-h$ and $\bar h\to 1-\bar h$, it satisfies the Baxter equations in the
holomorphic and antiholomorphic sectors, and its asymptotic behaviour at infinity
is in agreement with Eqs.~\re{Q-asym}. Therefore, it remains to show that
$Q_{q,\bar q}(u,\bar u)$ has the correct structure of the poles, Eq.~\re{poles}.
To this end, one applies \re{Qbar-Q} (see also \re{Q-relation}) and rewrites
\re{block-ansatz} in two equivalent forms
\ba
\lefteqn{\hspace*{-7mm}Q_{q,\bar q}(u,\bar u)}&&
\nonumber
\\
&&\hspace*{-9mm}=\left[\frac{\Gamma(1-\bar s-i\bar u)}{\Gamma(\bar s-i\bar
u)}\right]^N\!
\bigg\{
c_h\, Q(u;h,q)\lr{Q(\bar u^*; 1-h,q)}^*+c_{1-h}\,Q(u;1-h,q)\lr{Q(\bar u^*;
h,q)}^*
\bigg\}
\nonumber
\\
&&\hspace*{-9mm}=\left[\frac{\Gamma(1-s+iu)}{\Gamma(s+iu)}\right]^N\!
\bigg\{
c_h\widebar Q(\bar u;\bar h,\bar q)\lr{\widebar Q(u^*;1-\bar h,\bar q)}^*
+c_{1-h}\widebar Q(\bar u;1-\bar h,\bar q)\lr{\widebar Q(u^*;\bar h,\bar q)}^*
\bigg\}.
\label{Q-poles-manifest}
\ea
The analytical properties of $Q_{q,\bar q}(u,\bar u)$ are now manifest -- the
poles of $Q_{q,\bar q}(u,\bar u)$ in $\bar u$ and $u$ are generated by the
$\Gamma-$functions in the first and the second relation, respectively. One
deduces from \re{Q-poles-manifest} that for arbitrary $c_h$ and $c_{1-h}$ the
function $Q_{q,\bar q}(u,\bar u)$ has the $N-$th order poles at $\bar u=i(\bar
s-\bar m)$ and, separately, at $u=-i(s-m)$ with $m,\bar m=\mathbb{Z}_+$.

Let us now require that the analytical properties of \re{Q-poles-manifest} should
match similar properties of the Baxter $\mathbb{Q}-$operator, Eq.~\re{poles}. We
remind that the operator $\mathbb{Q}(u,\bar u)$ is well-defined only if the
spectral parameters $u$ and $\bar u$ satisfy \re{u-bar u}. Imposing this
condition, we find that \re{Q-poles-manifest} has the prescribed poles,
\re{poles}, plus additional ``spurious'' $N-$th order poles located at
\ba
&&\{u=i(s+m-1)\,,\ \bar u=i(\bar s-\bar m)\}\,,
\nonumber\\
&&\{u=-i(s-m)\,,\ \bar u=-i(\bar s+\bar m-1)\}\,,
\label{spur-poles-new}
\ea
with $m,\bar m=1,2,...$. Thus, the coefficients $c_h$ (and $c_{1-h}$) and the
integrals of motions $(q,\bar q)$ have to be chosen in such a way that
\re{Q-poles-manifest} has to have vanishing residues at the poles
\re{spur-poles-new}. Introducing the functions
\be
\Phi(\epsilon)=\frac{Q(i(s+\epsilon);h,q)}{Q(i(s+\epsilon);1-h,q)}\,,\qquad
\widebar\Phi(\epsilon)=\frac{\widebar Q(-i(\bar s+\epsilon);\bar h,\bar q)}
{\widebar Q(-i(\bar s+\epsilon);1-\bar h,\bar q)}
\label{phi-1}
\ee
and examining the first and the second relation in \re{Q-poles-manifest} for
$\{u=i(s+m-1+\epsilon),\bar u=i(\bar s-\bar m+\epsilon)\}$ and
$\{u=-i(s-m+\epsilon),\bar u=i(\bar s+\bar m-1+\epsilon)\}$, respectively, as
$\epsilon\to 0$, one finds that this requirement leads to
\ba
&&c_h\,\Phi(m-1+\epsilon)+c_{1-h}\,\lr{\Phi(\bar m-1+\epsilon)}^*={\cal
O}(\epsilon^N)
\nonumber
\\[2mm]
&&c_h\,\widebar\Phi(\bar m-1+\epsilon)+c_{1-h}\,\lr{\widebar\Phi(
m-1+\epsilon)}^*={\cal O}(\epsilon^N)\,,
\label{const-1}
\ea
with $m$ and $\bar m$ being positive integer. It follows from the Wronskian
relations (see Appendix B, Eq.~\re{phi-phi}), that the infinite system of
equations \re{const-1} becomes equivalent to a single condition at $m=\bar m=1$
\be
\frac{c_{1-h}}{c_h}\frac{\lr{\Phi(\epsilon)}^*}{\Phi(\epsilon)}=-1+{\cal
O}(\epsilon^N)\,,\qquad
\frac{c_{1-h}}{c_h}\frac{\lr{\widebar\Phi(\epsilon)}^*}{\widebar\Phi(\epsilon)}=-1+{\cal
O}(\epsilon^N)\,,
\label{const-2}
\ee
with the functions $\Phi(\epsilon)$ and $\widebar\Phi(\epsilon)$ defined in
\re{phi-1}.

One finds from \re{const-2} that $c_h/c_{1-h}$ is a pure phase. Its value can be
obtained by matching \re{block-ansatz} into \re{Q-asym} at large $u$ and $\bar u$
with a help of \re{block-asym},
\be
\frac{c_{1-h}}{c_h}=(-1)^{n_h}\e^{-2i\Theta_h(q,\bar q)}\,,
\label{c-ratio}
\ee
%with $\Theta_h(q,\bar q)$ given by \re{Theta}.
Then, recalling the definition of the function $\Phi(\epsilon)$, Eq.~\re{phi-1},
we obtain from \re{const-2}
\be
{\rm arg}\left[\frac{Q(i(s+\epsilon);h,q)}{Q(i(s+\epsilon);1-h,q)}\right]
=\pi\lr{\frac{n_h+1}2+\ell}-\Theta_h(q,\bar q)+ {\cal O}(\epsilon^N)
\label{ell}
\ee
and similar relation for the antiholomorphic block
\be
{\rm arg}\left[\frac{\widebar Q(-i(\bar s+\epsilon);\bar h,\bar q)}{\widebar
Q(-i(\bar s+\epsilon);1-\bar h,\bar q)}\right]=\pi\lr{\frac{n_h+1}2+\bar
\ell}-\Theta_h(q,\bar q)+ {\cal O}(\epsilon^N)\,,
\label{ell-bar}
\ee
with $\ell$ and $\bar\ell$ integer, such that, in general, $\bar\ell\neq\ell$.
Applying \re{Qbar-Q}, one can express \re{ell-bar} entirely in terms of the
holomorphic blocks. The blocks entering the relations \re{ell} and \re{ell-bar}
are given by \re{Q-series-1}.

Expanding the both sides of \re{ell} and \re{ell-bar} in powers of $\epsilon$,
one obtains the (overdetermined) system of $2N$ real quantization conditions on
$N-2$ complex charges $q_3,..., q_N$ and real phase $\Theta_h(q,\bar q)$. We will
verify in Section 5 that their solutions are consistent with the quantization
conditions \re{C1-C0}.

\subsection{Energy spectrum from the $Q-$blocks}

Let us show that the energy, $E_N(q,\bar q)$, and quasimomentum, $\theta_N(q,\bar
q)$, admit a simple representation in terms of the $Q-$blocks. To this end, we
introduce new blocks
\ba
Q_0(u;q)&=&a_h Q(u;h,q)+ a_{1-h} Q(u;1-h,q)\,,
\nonumber
\\[2mm]
\widebar Q_0(\bar u;\bar q)&=&\bar{a}_{h} \widebar Q(\bar u;\bar h,\bar q)
+ \bar a_{1-h} \widebar Q(\bar u;1-\bar h,\bar q)\,,
\label{Q0-new}
\ea
which are symmetric under $h\to 1-h$ and $\bar h\to 1-\bar h$.

We require that $Q_0(u;q)$ and $\bar Q_0(\bar u;\bar q)$ should have the same
poles as the $Q-$blocks, Eq.~\re{poles-block}, but of the order not be higher
than $N-1$. Applying \re{Qbar-Q}, it is straightforward to verify that the linear
combination of the $Q-$blocks in the r.h.s.\ of \re{Q0-new} has a vanishing
residues at the $N$th pole at $u^-_1=-i(s-1)$ and $\bar u^+_1=i(\bar s-1)$
provided that (up to an overall normalization)
\be
a_h=\tan(\pi h) \lr{\widebar Q(-i\bar s;\bar h,\bar q)}^*\,,\qquad
\bar a_{\bar h}=\tan(\pi \bar h) \lr{Q(is;h,q)}^*\,.
\label{a-def}
\ee
Then, the residues at the remaining $N$th order poles, $\{u^-_m,\bar u^+_{\bar
m}\}$, vanish automatically, since otherwise $Q_0(u;q)$ and $\widebar Q_0(\bar
u;\bar q)$ will not satisfy the Baxter equations \re{Bax-eq}.

Let us now consider the following expressions
\ba
Q_1(u;q)&=&\left[\frac{\Gamma(1-s+iu)}{\Gamma(s+iu)}\right]^N\lr{\widebar
Q_0(u^*;\bar q)}^*=
\lr{\bar a_{1-\bar h}}^* Q(u;h,q)+\lr{\bar a_{\bar h}}^*Q(u;1-h,q)
,
\nonumber
\\
\widebar Q_1(\bar u;\bar q)&=&\left[\frac{\Gamma(1-\bar s-i\bar u)}
{\Gamma(\bar s-i\bar u)}\right]^N\lr{Q_0(\bar u^*;q)}^*=
\lr{a_{1-h}}^* \widebar Q(\bar u;\bar h,\bar q)+\lr{a_{h}}^*\widebar
Q(\bar u;1-\bar h,\bar q),
\label{Q1-def}
\ea
where in the r.h.s.\ we used \re{Q-relation}. In distinction with \re{Qbar-Q},
this transformation maps $Q_0(u;q)$ into another block $\widebar Q_1(\bar u;\bar
q)$ and similar for $\widebar Q_0(\bar u;\bar q)$. The ratio of the
$\Gamma-$functions in the r.h.s.\ of \re{Q1-def} ensures that $Q_1(u;q)$ and
$\widebar Q_1(\bar u;\bar q)$ have the same pole structure as the $Q-$block,
Eq.~\re{poles-block} and, in addition,
\be
Q_1(i(s+n)+\epsilon;q)\sim\epsilon\,,\qquad
\widebar Q_1(-i(\bar
s+n)+\epsilon;\bar q)\sim\epsilon
\label{Q1-zero}
\ee
for $\epsilon\to 0$ and $n$ nonnegative integer. This property plays a crucial
r\^ole in our analysis.

Combining together \re{Q0-new} and \re{Q1-def}, one can express the $Q-$blocks in
terms of new blocks $Q_0(u;q)$ and $Q_1(u;q)$. Substitution of the resulting
expressions into \re{block-ansatz} yields
\be
Q_{q,\bar q}(u,\bar u)= A_{q,\bar q}\, Q_0(u;q) \widebar Q_0(\bar u;\bar q) +
B_{q,\bar q} \, Q_1(u;q) \widebar Q_1(\bar u;\bar q)\,.
\label{diag-Q}
\ee
Notice that the cross-terms, like $Q_0(u;q)\widebar Q_1(\bar u;\bar q)$, do not
appear in this expression. They have nonvanishing residues at the $N$th order
spurious poles \re{spur-poles-new} and their appearance in the r.h.s.\ of
\re{diag-Q} is protected by the quantization conditions \re{ell} and
\re{ell-bar}. The coefficients $A_{q,\bar q}$ and $B_{q,\bar q}$ can be expressed
in terms of $c_h$, $a_h$ and $\bar a_{\bar h}$, defined in Eqs.~\re{c-ratio} and
\re{a-def}, but we will not need these expressions for our purposes.

To find the quasimomentum \re{quasi-old}, one has to evaluate \re{diag-Q} at
$u=\pm is$ and $\bar u=\pm i \bar s$. Since the second term in \re{diag-Q}
vanishes due to \re{Q1-zero}, one gets
\be
\theta_N(q,\bar q)=i\ln\frac{Q_0(is;q)}{Q_0(-is;q)}
+i\ln\frac{\widebar Q_0(i\bar s;\bar q)}{\widebar Q_0(-i\bar s;\bar q)}\,.
\ee
The calculation of the energy \re{Energy-I} is more cumbersome. One finds from
\re{diag-Q} and \re{Q1-zero} that
\be
\frac{d}{du}\ln Q_{q,\bar q}(u-is,u-i\bar s)\bigg|_{u=0}=\lr{\ln Q_0(-is;q)}'
+\lr{\ln\widebar Q_0(-i\bar s;\bar q)}' +\Delta Q\,,
\label{Delta Q}
\ee
where prime denotes the derivative with respect to the spectral parameter and
\be
\Delta Q=\frac{B_{q,\bar q}}{A_{q,\bar q}}
\frac{Q_1(-is){\widebar Q}_1'(-i\bar s)}{Q_0(-is)\widebar Q_0(-i\bar s)}
=\lim_{\epsilon\to 0}\frac1{i\epsilon}
\frac{Q_1(-is) Q_0(-i(s-1+\epsilon))}{Q_0(-is)Q_1(-i(s-1+\epsilon))}\,.
\label{Delta}
\ee
Here, the second relation follows from the requirement for \re{diag-Q} to have
vanishing residues at the ``spurious'' poles \re{spur-poles-new}, or equivalently
$Q_{q,\bar q}(-i(s-1+\epsilon),-i(\bar s+\epsilon))$ to be finite as $\epsilon\to
0$. The residue of $Q_1(-i(s-1+\epsilon))$ at the $N$th order pole at
$\epsilon=0$ can be obtained from the first relation in \re{Q1-def}. To calculate
the residue of $Q_0(-i(s-1+\epsilon))$ at the $(N-1)$th order pole at
$\epsilon=0$ one applies the Wronskian relation
\be
Q_0(u+i;q)Q_1(u;q)-Q_0(u;q)Q_1(u+i;q)= {\rm const}\times
\left[\frac{\Gamma(iu-s)}{\Gamma(iu+s)}\right]^N,
\label{W01}
\ee
which follows from \re{Wronskian} and the definition of the $Q_0-$ and
$Q_1-$blocks, Eqs.~\re{Q0-new} and \re{Q1-def}, respectively. The normalization
constant in the r.h.s.\ of \re{W01} can be obtained for $u=-i(s+\epsilon)$ from
the comparison of the residues of the both sides of \re{W01} at the $N$th order
pole in $\epsilon$. In this way, one finds from \re{Delta Q}
\be
\frac{d}{du}\ln Q_{q,\bar q}(u-is,u-i\bar s)\bigg|_{u=0}=\frac{iN}{1-2s}
+\lr{\ln\widebar Q_0(-i\bar s;\bar q)}'-\lr{\ln[\widebar Q_0(-i\bar s;\bar
q)]^*}'\,.
\label{new-1}
\ee
Repeating similar calculation of the logarithmic derivative of \re{diag-Q} at
$u=is$ and $\bar u=i\bar s$ we find
\be
\frac{d}{du}\ln Q_{q,\bar q}(u+is,u+i\bar s)\bigg|_{u=0}=-\frac{iN}{1-2\bar s}
+\lr{\ln Q_0(is;q)}'-\lr{\ln[Q_0(is;q)]^*}'\,.
\label{new-2}
\ee
Finally, we substitute the last two relations into \re{Energy-I} and obtain the
following expression for the energy
\be
E_N(q,\bar q)=-2\Im\lr{\ln Q_0(is;q)}'+2\Im\lr{\ln\widebar Q_0(-i\bar s;\bar
q)}'+\varepsilon_N\,,
\label{E-1}
\ee
where $\varepsilon_N=2N\Re\left[\psi(2s)+\psi(2-2s)-2\psi(1)\right]$. One can
further simplify this expression by using the relation $(\ln Q_{q,\bar
q}(-is,-i\bar s))'=-(\ln Q_{-q,-\bar q}(is,i\bar s))'$ that follows from
\re{Q-symmetry}. Together with \re{new-1} and \re{new-2}, it leads to
\be
E_N(q,\bar q)=-2\Im\lr{\ln Q_0(is;q)}'-2\Im\lr{\ln Q_0(is;-q)}'+E^{(0)}_N\,,
\label{E-2}
\ee
where $E_N^{(0)}=2N\Re\left[\psi(2s)+\psi(1-2s)-2\psi(1)\right]$. The symmetry of
the energy $E_N(q,\bar q)=E_N(-q,-\bar q)$ becomes manifest in this form.

The results obtained in Sections~3 and 4 provide the basis for calculating the
energy spectrum of the model. We have demonstrated that
\begin{itemize}

\item The eigenvalues of the Baxter operator,
$Q_{q,\bar q}(u,\bar u)$, possessing the correct analytical properties and
asymptotic behaviour at infinity (see Section 2) can be constructed only for the
special values of the integrals of motion $q$ and $\bar q$ satisfying the
quantization conditions, Eqs.~\re{C1-C0}, \re{ell} and \re{ell-bar}.

\item The functions $Q_{q,\bar q}(u,\bar u)$ can be decomposed over the chiral
$Q-$blocks, which depend on the spectral parameters and the integrals of motion
in the (anti)holomorphic sector, Eq.~\re{Q-block1} and \re{diag-Q}. In
distinction with $Q_{q,\bar q}(u,\bar u)$, the definition of the $Q-$blocks is
ambiguous and, therefore,
%they play in our analysis the r\^ole of auxiliary building blocks.
they do not have any physical meaning {\it per se\/}.

\item Once the quantization conditions for the integrals of motion are fulfilled,
the corresponding eigenvalues of the Baxter operator are uniquely fixed (up to an
overall normalization). They allow us to calculate the energy spectrum of the
model by using two different expressions Eqs.~\re{E-fin} and \re{E-2}, which lead
to the same value of $E_N$.

\end{itemize}
We would like to mention that we disagree on these points with the approach of
Ref.~\cite{DL}, in which the Baxter equation for noncompact magnet of spin $s=1$
has been investigated.

\section{Energy spectrum}

In this Section we solve the quantization conditions \re{C1-C0} and describe the
spectrum of the Schr\"odinger equation \re{Sch} for different number of particles
$N$.

For arbitrary $N$, the spectrum of the model -- the energy, $E_N(q,\bar q)$, and
the corresponding eigenstates, $\Psi_{\vec p,\{q,\bar q\}}(\vec {\mybf{z}})$, are
uniquely specified by the total set of quantum numbers in the holomorphic and
antiholomorphic sectors, $q=(q_2,q_3, ..., q_N)$ and $\bar q=(\bar q_2,\bar
q_3,...,\bar q_N)$, respectively. Since $q_k=\bar q_k^*$, the total number of
independent complex valued quantum numbers is equal to $(N-1)$. One of them,
$q_2$, fixes the total $SL(2,\mathbb{C})$ spin of the state $(h,\bar h)$ defined
in \re{q2}. According to Eq.~\re{q2}, the quantized values of $h$, or
equivalently $q_2$, are parameterized by integer $n_h$ and real $\nu_h$
\be
q_2(n_h,\nu_h)=\frac14-\lr{\frac{n_h}2+i\nu_h}^2+Ns(s-1)\,.
\label{q2-branch}
\ee
At $N=2$ this becomes the only quantum number parameterizing the spectrum of the
model \re{E-2-exact}. For $N\ge 3$ one has to deal with a bigger set of the
quantum numbers.

To find the spectrum of the integrals of motion $q_3,...,q_N$ for $N\ge 3$, one
has to solve the overdetermined system of the quantization conditions \re{C1-C0}.
As was explained in Section 3, their solutions give us the expressions for the
$\alpha-$, $\beta-$ and $\gamma-$parameters entering the mixing matrices
$C^{{(0)}}$ and $C^{{(1)}}$, Eqs.~\re{Q-0} and \re{Q-1}. Then, the corresponding
energy, $E_N(q,\bar q)$, and the quasimomentum, $\theta_N(q,\bar q)$, are
calculated by inserting these expressions into \re{E-fin} and \re{parity-qc},
respectively. In distinction with the $N=2$ case, Eqs.~\re{E-2-exact} and
\re{quasi-N=2}, the resulting expressions for the spectrum of the model can not
be expressed for arbitrary $N$ in a closed analytical form. Nevertheless, the
results that we are going to present in this Section exhibit a remarkable
regularity and they suggest that there should exist a WKB-like description of the
spectrum.

To analyze the quantization conditions \re{C1-C0}, one has to specify the values
of the $SL(2,\mathbb{C})$ spins $(s,\bar s)$ defined in \re{spins}. Since our
main motivation for studying the noncompact Heisenberg spin magnets came from
high-energy QCD, we shall fix them to be the same as for the $N-$reggeized gluons
compound state, Eqs.~\re{amp} and \re{Sch},
\be
s=0\,,\qquad \bar s=1\,.
\label{s=0}
\ee
Another specifics of QCD is that, solving the Schr\"odinger equation \re{Sch},
one is mainly interested in finding the ground state. It is this state that
provides the dominant contribution to the asymptotic behaviour of the scattering
amplitudes \re{amp} at high-energy. Additional constraint on the solutions to
\re{Sch} is imposed in QCD by the Bose symmetry of the $N-$reggeized gluon
compound states. As we will argue below, this condition selects among all
eigenstates only those with the quasimomentum $\exp(i\theta_N(q,\bar q))=1$. It
is automatically fulfilled for the ground state for which one expects
$\theta_N^{\rm ground}(q,\bar q)=0$.

To solve the quantization conditions \re{C1-C0}, one needs the expressions for
the transition matrices $\Omega(q)$ and $\widebar\Omega(\bar q)$, defined in
\re{Omega-def}. For $N=2$ they can be determined exactly from the properties of
the Legendre functions (see Appendix A). For $N\ge 3$ the calculation of these
matrices is based on the relation \re{Omega}, which in turn relies on the power
series representation of the fundamental solutions, Eqs.~\re{Q-0-h} and
\re{set-1}. In general, the resulting expressions for the transition matrices
take the form of an infinite series in $q$. Solving the quantization conditions,
we truncated infinite series in Eqs.~\re{power-series-0} and \re{v-series} and
retained a sufficiently large number of terms ($n_{\rm max}\sim 10^3$). This
allowed us to calculate the energy spectrum numerically with a rather high
accuracy (see Table~\ref{tab:Summary} below).

As was already mentioned, the system of $N^2$ quantization conditions \re{C1-C0}
is overdetermined. Using the subset of $(N^2-3N+5)$ conditions, one can calculate
the $\alpha-$, $\beta-$ and $\gamma-$parameters in terms of the charges $q$ and
define the following test function
\be
f(q_2,q_3,...,q_N)=\Tr (T T^\dagger)\,,\qquad T=C^{(1)}(q,\bar
q)-\left[\Omega(q)\right]^T C^{(0)}(q,\bar q)\ \widebar
\Omega(\bar{q})\,.
\label{test-func}
\ee
Obviously, $f(q)$ is a positive definite function on the $(N-1)-$dimensional
complex moduli space $q=(q_2,q_3,...,q_N)$. The solutions to the quantization
conditions \re{C1-C0} correspond to points on this space, in which the test
function vanishes, $f(q)=0$. Since the dimension of the moduli space increases
with $N$, the problem of finding zeros of $f(q)$ becomes very nontrivial at
higher
$N$. To solve it, we applied the algorithm developed in \cite{KoP}.%
\footnote{This method was implemented in the form of the Fortran-90 code, which
is available from the authors upon request.} It allowed us to identify the zeros
of the test function \re{test-func} by the ``steepest descent'' method starting
from some reference point on the moduli space. As yet another test of our
approach, we verified that the obtained expressions for the integrals of motion
satisfy the conditions \re{ell} and \re{ell-bar}.

\subsection{Fine structure of the spectrum}

Before summarizing the results of our calculations for $N\ge 3$, let us describe
a general structure of the spectrum. We find that for given total
$SL(2,\mathbb{C})$ spin of the system, $h=(1+n_h)/2+i\nu_h$, the quantization
conditions \re{C1-C0} provide us with an infinite number of discrete values of
the integrals of motion. They can be parameterized by the set of integers as
\be
q_k=q_k(\nu_h;n_h,\{\ell\})\,,\qquad E_N=E_N(\nu_h;n_h,\{\ell\})\,,\qquad
\{\ell\}=(\ell_1,\ell_2,...,\ell_{2(N-2)})
\label{branches}
\ee
with $k=3,...,N$. The explicit form of this dependence and interpretation of
integers $\{\ell\}$ will be given below. Eqs.~\re{q2-branch} and \re{branches}
imply that, as a function of the total spin $h$, the quantized values of $q_k$
form the family of one-dimensional continuous nonintersecting trajectories in the
$(N-1)-$dimensional space of $q=(q_2,q_3,...,q_N)$. The ``proper time'' along
each trajectory, $\nu_h$, is defined by the imaginary part of the total spin,
$\Im h=\nu_h$, whereas the integers $n_h,\ell_1,,...,\ell_{2(N-2)}$ specify
different members of the family. Each trajectory in the $q-$space induces the
corresponding trajectory for the energy $E_N=E_N(q,\bar q)$. We recall that, in
contrast with the integrals of motion, $q$ and $\bar q$, the energy takes only
real values.

In what follows, we shall separate all eigenstates of the model into two groups
according to the value of the ``highest'' charge $q_N$: $q_N\neq 0$ and $q_N=0$.
The reason for this is that, as we will show in Section~5.5, there exists an
intrinsic relation between the eigenstates with $q_N=0$ and the eigenstates of
the system with the number of particles equal to $N-1$ and $q_{N-1}\neq 0$.
Namely, they share the same spectrum of the energy and the integrals of motion
$q_k$ (with $k=2,...,N-1$) and the corresponding wave functions are related to
each other in a simple way.

For the moment, we shall exclude the eigenstates with $q_N=0$ from our
consideration and return to them in Section 5.5. We find that among all
eigenstates of the $N-$particle system with $q_N\neq 0$, the minimal energy
occurs at $\nu_h=n_h=0$, or equivalently $h=1/2$,
\be
E_N^{\rm ground}\equiv \min E_N(\nu_h;n_h,\{\ell\})=E_N(0;0,\{\ell_{\rm
ground}\})
\ee
It belongs to the special trajectory $\{\ell_{\rm ground}\}$, to which we shall
refer as {\em the ground state trajectory}. The energy along this trajectory is a
continuous function of $\nu_h$ and it approaches its minimal value, $E_N^{\rm
ground}$, at $\nu_h=0$. In the vicinity of $\nu_h=0$, one finds an accumulation
of the energy levels
\be
E_N(\nu_h;0,\{\ell_{\rm ground}\})-E_N^{\rm ground} = \sigma_{N}
\nu_h^2+ {\cal O}(\nu_h^4)\,,
\label{accum}
\ee
with $\sigma_N$ being the diffusion coefficients.

The spectrum of quantized $q_2,...,q_N$ possesses the following symmetry
\be
q_k \to (-1)^k q_k \to q_k^*
\label{map}
\ee
with $k=2,...,N$. Here, the first relation follows from invariance of the
Hamiltonian under mirror permutations of particles, Eq.~\re{mirror}. The second
relation is a consequence of the symmetry of the model at $s=0$ and $\bar s=1$
under interchange of the $z-$ and $\bar z-$sectors, or equivalently
$q_k\leftrightarrow\bar q_k$ and $s(s-1)\leftrightarrow\bar s(\bar s-1)$. The
relation \re{map} implies that if the quantization conditions \re{C1-C0} are
satisfied at some point $\{q_k\}$ on the moduli space, then the same holds true
at the points $\{(-1)^k q_k\}$, $\{q_k^*\}$ and $\{(-1)^k q_k^*\}$.

As we will see in the next Section, the spectum of quantized charges $q$ has a
hidden structure, which can be revealed by examining the distribution of the
quantized values of the ``highest'' charge $q_N^{1/N}$. Since $q_N^{1/N}$ is a
multi-valued function of complex $q_N$, each eigenstate of the model
%, specified by a definite value of $q_N$,
will be represented on the complex
$q_N^{1/N}-$plane by $N$ different points. Together with the symmetry property of
the spectrum, Eq.~\re{map}, this leads to the following transformation on the
moduli space
\be
q_N^{1/N} \to \exp\lr{\pi i k/N} \,q_N^{1/N}\,,\qquad [0< k< 2N]
\label{map1}
\ee
with $k$ integer for odd $N$ and even for even $N$. It maps one of the
eigenstates into itself or into another one with the same energy.

In the rest of this Section we shall present the results of our calculations of
the spectrum of the model for the number of particles $3\le N \le 8$.

\subsection{Quantum numbers of the $N=3$ states}

At $N=3$ the eigenstates depend on the quantum number $q_3$, which is an
eigenvalue of the operator $q_3$ defined in \re{q}. Some of the eigenvalues of
this operator have been already found in \cite{JW,check,DL} using different
methods. The eigenstates found in \cite{JW} have pure imaginary values of
quantized $q_3$ and their quasimomentum is equal to $\theta_3=0$. We will
demonstrate in this Section, that contrary to the statements made in \cite{DL}
the spectrum of the operator $q_3$ is not exhausted by pure imaginary values,
even in the sector with $\theta_3=0$.

\begin{figure}[th]
\vspace*{3mm}
\centerline{{\epsfysize8cm \epsfbox{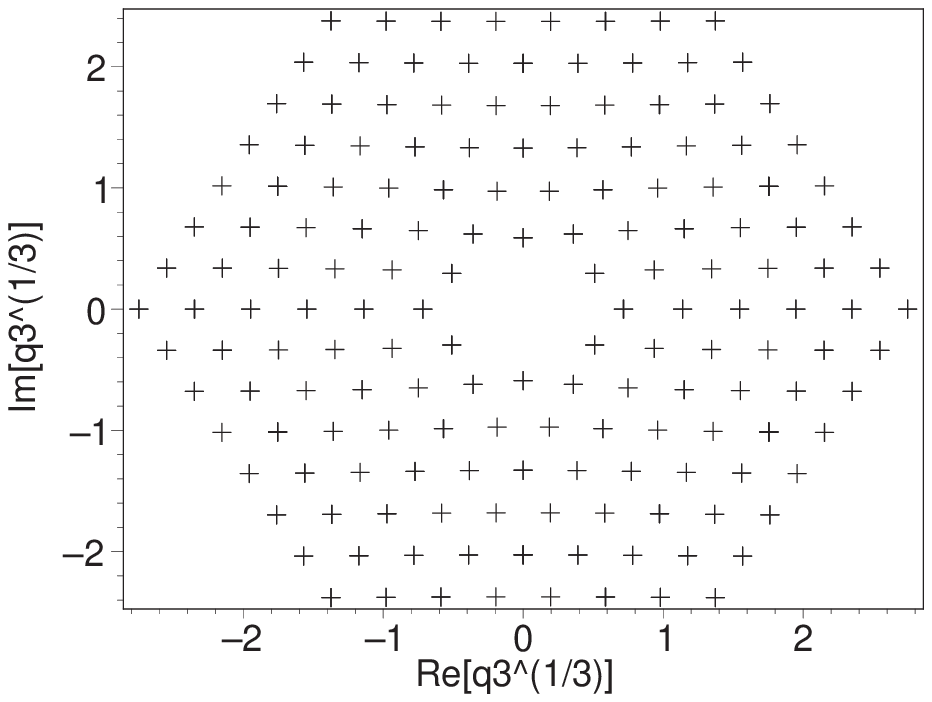}}}
\caption[]{The spectrum of quantized $q_3^{1/3}$ for the system of $N=3$ particles.
The total $SL(2,\mathbb{C})$ spin of the system is equal to $h=1/2$.}
\label{Fig-N=3}
\end{figure}

Solving the quantization conditions \re{C1-C0} at $N=3$, we reconstructed the
full spectrum of quantized $q_3$. We found that apart from pure imaginary $q_3$
calculated in \cite{JW,check}, the spectrum also contains (an infinite number of)
complex values of $q_3$, including pure real ones. Enumerating the quantized
$q_3$ according to their absolute value starting from the smallest one, we notice
that $|q_3|$ grows cubically with its number. This suggests to describe the
spectrum in terms of $q_3^{1/3}$ rather than $q_3$. To illustrate this point, we
present in Figure~\ref{Fig-N=3} the results of our calculations of quantized
$q_3^{1/3}$ for the total $SL(2,\mathbb{C})$ spin of the system $h=1/2$, or
equivalently $n_h=\nu_h=0$. Similar picture emerges at $h=(1+n_h)/2+i\nu_h$ for
$n_h=1,2,...$ and $\nu_h$ real.

%%%%%%%%%%%%%%%%%%%%%%%%%%%%%%%%%%%%%%%%%%%%%%%%%%%%%%%%%%%%%%%%%%%%%%%%%%%%%%%%%%
%%%%%%%%%%%%%%%%%%%%%%%%%%%%%%%%%%%%%%%%%%%%%%%%%%%%%%%%%%%%%%%%%%%%
%%%%%%%%%%%%%%%%%%%%%%%%%%%%%%%%%%%%%%%%%%%%%%%%%%%%%%%%%%%%%%%%%%%%%%%%%%%%%%%%%%
\begin{figure}[ht]
\vspace*{3mm}
\centerline{{\epsfysize8cm \epsfbox{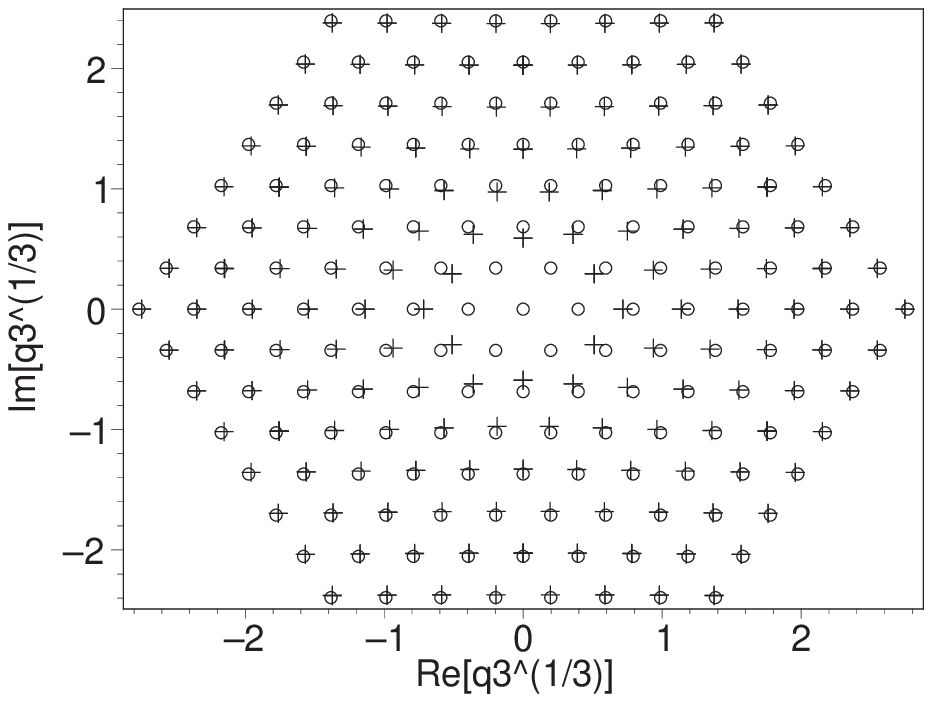}}}
\caption[]{Comparison of the exact spectrum of quantized $q_3^{1/3}$ at $h=1/2$ (crosses) with the WKB
expression \re{q3-WKB} (circles).}
\label{Fig-WKB-N=3}
\end{figure}
%%%%%%%%%%%%%%%%%%%%%%%%%%%%%%%%%%%%%%%%%%%%%%%%%%%%%%%%%%%%%%%%%%%%

The spectrum of quantized $q_3^{1/3}$, shown in Figure~\ref{Fig-N=3}, is in
agreement with the symmetry properties \re{map} and \re{map1}. Defining the
fundamental domain as $0\le\arg (q_3^{_{\scriptstyle 1/3}})< \pi/3$, we find that
the whole spectrum of quantized $q_3^{1/3}$ can be obtained by applying the
transformations \re{map} and \re{map1} to the points belonging to this domain.

It is difficult do not notice a remarkable regularity in the distribution of
quantized charges in Figure~\ref{Fig-N=3}. Apart from a few points close to the
origin, the quantized values of $q_3^{1/3}$ are located at the vertices of the
lattice built from equilateral triangles. As a consequence, they can be
parameterized as
\be
\left[q_{3}^{\rm WKB}(\ell_1,\ell_2)\right]^{1/3}=\Delta_{N=3}\cdot
\left(\frac12\ell_1+i\frac{\sqrt{3}}2\ell_2\right),
%\equiv
%\Delta_{N=3}
%\left[\ell_1\cos\frac{\pi}3+i\ell_2\,\sin\frac{\pi}3\right]
\label{q3-WKB}
\ee
where $\ell_1$ and $\ell_1$ are integers, such that their sum $\ell_1+\ell_1$ is
even. Here, $\Delta_{3}$ denotes the lattice spacing. Its value can be calculated
from the leading-order WKB solution of the Baxter equation~\cite{WKB}
\be
\Delta_{3}
=\left[\frac{3}{4^{1/3}\pi}\int_{-\infty}^1\frac{dx}{\sqrt{1-x^3}}\right]^{-1}
=\frac{\Gamma^3(2/3)}{2\pi}=.395175...
\ee
Quantized $q_3^{1/3}$ occupy the whole complex plane except the interior of the
disk of the radius $\Delta_{3}$
\be
|q_3^{1/3}| > \Delta_{3}\,.
\ee
The comparison of \re{q3-WKB} with the exact expressions for $q_3$ at $h=1/2$ is
shown in Figure~\ref{Fig-WKB-N=3} and Table~\ref{tab:WKB}.%
\footnote{To save space, we truncated in the Table the last few digits in the
obtained numerical values of the charges and the energy.\label{foot}} We find
that the expression \re{q3-WKB} describes the excited eigenstates with a high
accuracy. The agreement becomes less impressive for the eigenstates with smaller
$q_3$. For instance, for the ground state with $iq_3=0.205258...$ and the first
excited state with $q_3=0.368293...$ the accuracy of \re{q3-WKB} is $\sim 16\%$
and $\sim 10\%$, respectively. Eq.~\re{q3-WKB} can be systematically improved by
including subleading WKB corrections. Notice that the same WKB formula
\re{q3-WKB} is valid not only at $h=1/2$ but also for arbitrary spin $h$. In the
latter case, it describes correctly the excited states with $|q_3^{1/3}|\gg
|q_2^{1/2}|$.

\begin{table}[ht]
\begin{center}
\begin{tabular}{|c||c|c|c|c|c|c|}
\hline
 $(\ell_1,\ell_2)$&  $(0,2)$ &  $(2,2)$ &
     $(4,2)$ &  $(6,2)$ &
     $(8,2)$ &  $(10,2)$
\\
\hline
$\left(q_3^{\rm \,exact}\right)^{1/3}$ & $.590\,i$ & $.358+.621\,i$ &
$.749+.649\,i$ & $1.150+.664\,i$ & $1.551+.672\,i$ & $1.951+.676\,i$
\\
\hline
$\left(q_3^{\rm WKB}\right)^{1/3}$ & $.684\,i$ & $.395+.684\,i$ & $.790+.684\,i$
& $1.186+ .684\,i$ &  $1.581+ .684\,i$ & $1.976+ .684\,i$
\\
\hline
$-E_3/4$ & $-.2472$ & $-.6910$ & $-1.7080$ & $-2.5847$ & $-3.3073$ & $-3.9071$
\\
\hline
\end{tabular}
\end{center}
\caption{Comparison of the exact spectrum of $q_3^{1/3}$ at $h=1/2$ with the approximate
WKB expression \re{q3-WKB}. The last line defines the corresponding energy
$E_3(\nu_h=0;\ell_1,\ell_2)$.}
\label{tab:WKB}
\end{table}

According to \re{q3-WKB}, the quantized values of $q_3$ are parameterized by a
pair of integers $\ell_1$ and $\ell_2$ which define the coordinates on the
lattice shown in Figures~\ref{Fig-N=3} and \ref{Fig-WKB-N=3}. Eq.~\re{q3-WKB}
provides the WKB approximation to the function $q_3(\nu_h;n_h,\ell_1,\ell_2)$
introduced in \re{branches}. Calculating the quasimomentum of the eigenstate
specified by the pair of integers, $\ell_1$ and $\ell_2$, one finds that it is
given by
\be
\theta_3(\ell_1,\ell_2)=\frac{2\pi}3\ell_1\quad ({\rm mod}~2\pi)
\,.
\label{quasi-3}
\ee
In particular, the quasimomentum vanishes for the eigenstates with $\ell_1=0\
({\rm mod}~3)$. It is easy to see from \re{q3-WKB} that the corresponding $q_3$
take, in general, complex values. There are, however, special cases, like
$\ell_1=0$ or $\ell_2=0$, when $q_3$ becomes, respectively, pure imaginary or
real. The former values have been previously found in \cite{JW,check}.

We recall that the spectrum of $q_3$, shown in Figure~\ref{Fig-N=3}, corresponds
to the total spin $h=1/2$, or equivalently $n_h=\nu_h=0$ in Eq.~\re{branches}. In
general, quantized $q_3$ depend on integer $n_h$ and continuous $\nu_h$. For
simplicity, we present here our results only at $n_h=0$, or equivalently
$h=1/2+i\nu_h$. For $n_h=1\,,2\,,3\,,...$ the spectrum of $q_3$ exhibits a
similar structure.%
\footnote{For pure imaginary $q_3$, the $n_h-$dependence has been studied in
\cite{KoP}.}

\begin{figure}[ht]
\vspace*{3mm}
\centerline{{\epsfysize6.5cm \epsfbox{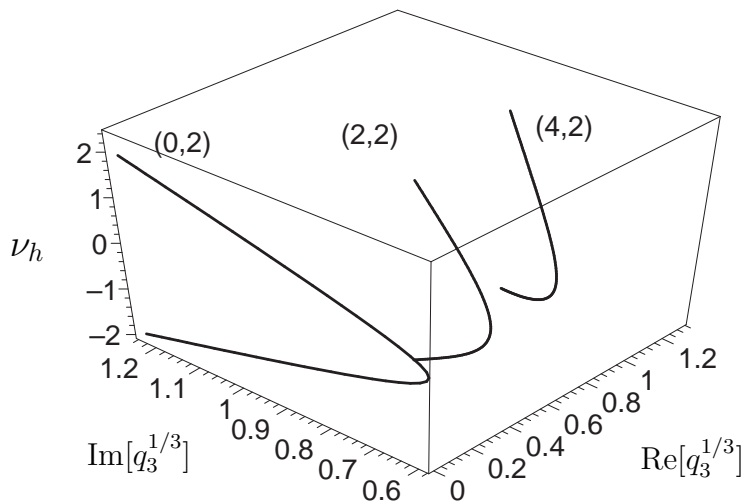}}}
\caption[]{The dependence of quantized $q_3(\nu_h;\ell_1,\ell_2)$ on the total spin $h=1/2+i\nu_h$.
Three curves correspond to the trajectories with $(\ell_1,\ell_2)=(0,2)\,, (2,2)$
and $(4,2)$.}
\label{Fig-3D}
\end{figure}

We find that quantized $q_3=q_3(\nu_h,0,\ell_1,\ell_2)$ are continuous functions
of $\nu_h$. For different integers $\ell_1$ and $\ell_2$, the functions
$q_3(\nu_h,0,\ell_1,\ell_2)$ define an infinite set of trajectories in the
three-dimensional $(\nu_h,\Re(q_3^{1/3}),\Im(q_3^{1/3}))$ moduli space. The
trajectories cross the hyperplane $\nu_h=0$ at the points shown in
Figure~\ref{Fig-N=3} and go to infinity for $\nu_h\to\pm\infty$. To illustrate
the properties of these trajectories, three representatives, corresponding to
$(\ell_1,\ell_2)=(0,2)\,, (2,2)$ and $(4,2)$, are shown in Figure~\ref{Fig-3D}.
The quasimomentum $\theta_3$ takes a constant value along each trajectory,
$\theta_3(\ell_1,\ell_2)=0\,,4\pi/3$ and $2\pi/3$, respectively.

%%%%%%%%%%%%%%%%%%%%%%%%%%%%%%%%%%%%%%%%%%%%%%%%%%%%%%%%%%%%%%%%%%%%%%%%%%%%%%%%%%
%%%%%%%%%%%%%%%%%%%%%%%%%%%%%%%%%%%%%%%%%%%%%%%%%%%%%%%%%%%%%%%%%%%%

\begin{figure}[ht]
\vspace*{3mm}
\centerline{{\epsfysize7cm \epsfbox{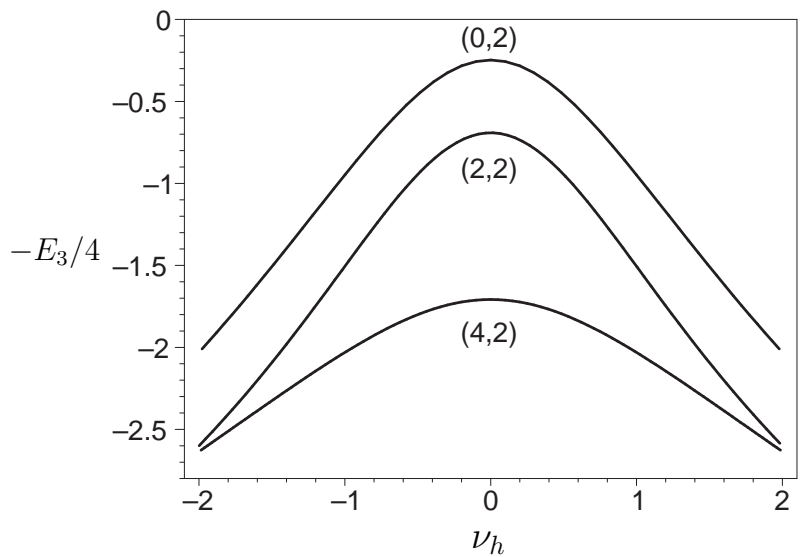}}}
\caption[]{The energy spectrum corresponding to three trajectories shown
in Figure~\ref{Fig-3D}. The ground state is located on the $(0,2)-$trajectory at
$\nu_h=n_h=0$.}
\label{Fig-energy3}
\end{figure}

Let us now consider the energy spectrum at $N=3$. Since the energy is a function
of the total spin $h$ and the charge $q_3$, each trajectory
$q_3=q_3(\nu_h;\ell_1,\ell_2)$ shown in Figure~\ref{Fig-3D} induces the
corresponding trajectory for the energy, $E_3=E_3(\nu_h;\ell_1,\ell_2)$. Solving
the quantization conditions \re{C1-C0} and applying \re{E-fin} we obtain the
energy spectrum shown in Figure~\ref{Fig-energy3}.

%%%%%%%%%%%%%%%%%%%%%%%%%%%%%%%%%%%%%%%%%%%%%%%%%%%%%%%%%%%%%%%%%%%%%%%%%%%%%%%%%%
%%%%%%%%%%%%%%%%%%%%%%%%%%%%%%%%%%%%%%%%%%%%%%%%%%%%%%%%%%%%%%%%%%%%

We find that the energy is a continuous function of $\nu_h$ along each
$(\ell_1,\ell_2)-$trajectory and it approaches its {\it minimal\/} value,
$\min_{\nu_h}E_3(\nu_h;\ell_1,\ell_2)$, at $\nu_h=0$, or equivalently $h=1/2$.
Examining the value of $E_3(0;\ell_1,\ell_2)$ for different sets of integers
$(\ell_1,\ell_2)$, we find that $E_3(0;\ell_1,\ell_2)$ increases as one goes
towards larger $|q_3^{1/3}(0;\ell_1,\ell_2)|$ (see Table~\ref{tab:WKB}). This
implies that the ground state corresponds to the point(s) on the plane of
quantized $q_3^{1/3}$ (see
Figure~\ref{Fig-N=3}) closest to the origin.%
\footnote{As we will show below, this property is rather general and it holds
for the energy $E_N$ as a function on the $(N-2)-$dimensional space of the
integrals of motion $(q_3,...,q_N)$.} It is easy to see from Figure~\ref{Fig-N=3}
that, in total, there are six such points, $(\ell_1,\ell_2)=(0,\pm 2),(\pm 3,\pm
1)$ and $(\mp 3,\pm 1)$. According to \re{quasi-3}, their quasimomentum is equal
to zero. Going over from $q_3^{1/3}$ to $q_3$, we find that these six points
define two nontrivial eigenstates, which have opposite values of $q_3$ and the
same energy
\be
iq_3^{\rm ground}=\pm 0.205258...\,,\qquad E_3^{\rm ground}=.988678...
\label{N3-ground}
\ee
This implies that at $N=3$ and $q_3\neq 0$ the ground state is {\it double
degenerate\/}. As we will demonstrate below, this property is rather general --
the ground state is double degenerate for the systems with {\it odd\/} number of
particles $N$ and $q_N\neq 0$, but it is unique for even $N$. Eq.~\re{N3-ground}
is in agreement with the results of the previous calculations \cite{JW,check}. To
verify that for $q_3\neq 0$ the ground state occurs at $n_h=0$, one has to
compare \re{N3-ground} with the minimal energy in the sectors with higher $n_h$.
At $n_h=1\,,2\,,3$ our results can be found in the first three columns of the
Table~\ref{tab:n_h} (see below). They indicate that for $q_3\neq 0$ the minimal
energy grows as the Lorentz spin of the system, $n_h$, increases.

At $N=3$ the ground state is located on the $(0,2)-$trajectory at $\nu_h=n_h=0$
as shown in Figure~\ref{Fig-energy3}. We notice that close to $\nu_h=0$ there is
an accumulation of the energy levels. At small $\nu_h$ the energy of excited
states is described by a general expression \re{accum} with the dispersion
parameter $\sigma_3$ given below in the Table~\ref{tab:Summary}.

\subsection{Quantum numbers of the $N=4$ states}

At $N=4$ the spectrum of the model depends on two complex quantum numbers $q_3$
and $q_4$. Similar to the $N=3$ case, we shall determine their spectrum by
solving the quantization conditions \re{C1-C0} under the additional condition
$q_4\neq 0$. The only difference with the $N=3$ case is that one has to increase
the dimension of the mixing matrices, $C^{(0)}$ and $C^{(1)}$, and recalculate
the transition matrices, $\Omega$ and $\widebar\Omega$, as functions of $q_3$ and
$q_4$. As before, to understand the structure of the spectrum at $h=1/2$, it
becomes convenient to deal with multi-valued complex variables
$q_4^{_{\scriptstyle 1/4}}$. Then, each eigenstate will be represented on the
complex $q_4^{_{\scriptstyle 1/4}}-$plane by four different points.

%%%%%%%%%%%%%%%%%%%%%%%%%%%%%%%%%%%%%%%%%%%%%%%%%%%%%%%%%%%%%%%%%%%%%%%%%%%%%%%%%%
\medskip
\begin{figure}[ht]
\vspace*{5mm}
\centerline{{\epsfysize6cm \epsfbox{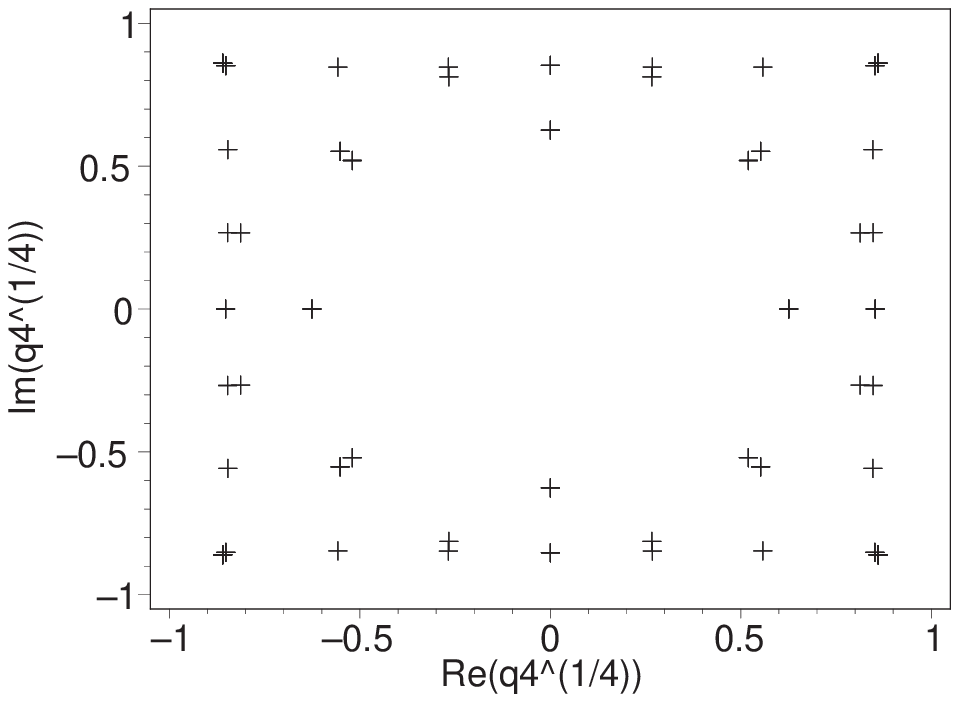}}\qquad{\epsfysize6cm \epsfbox{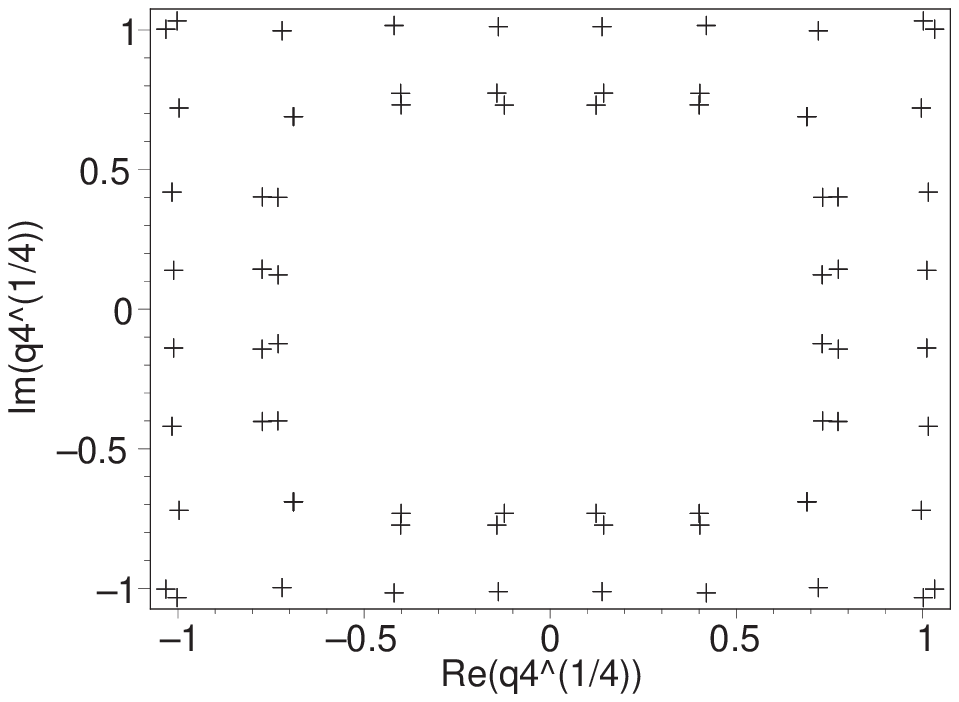}}}
%\medskip
%\centerline{(a) \hspace*{8cm} (b)}
\caption[]{The spectrum of the integrals of motion $q_4$ at $N=4$ and the total spin $h=1/2$.
The left and the right panels correspond to the eigenstates with the
quasimomentum $\e^{i\theta_4}=\pm 1$ and $\pm i$, respectively.}
\label{Fig-Q4}
\end{figure}
%%%%%%%%%%%%%%%%%%%%%%%%%%%%%%%%%%%%%%%%%%%%%%%%%%%%%%%%%%%%%%%%%%%%

At $h=1/2$ the quantized values of $(q_3,q_4)$ are parameterized by four real
numbers and, as a consequence, they do not admit a simple pictorial
representation. The results of our calculations are presented in
Figure~\ref{Fig-Q4}. There, each point on the complex $q_4^{1/4}-$plane has
additional complex coordinate defined by the quantized values of the charge
$q_3$. The latter are not displayed for simplicity. It is convenient to separate
the eigenstates into two sets according to their quasimomentum $\theta_4=(2\pi
k)/4$, $k=0,2$ and $k=1,3$ as shown in Figure~\ref{Fig-Q4}. We notice that some
points have very close values of $q_4$. Nevertheless, the charge $q_3$ and the
energy $E_4$ corresponding to these points are different. Here is an example of a
pair of such states at $h=1/2$ and $\theta_4=0$:
$(q_3=0,q_4=-2.185790,E_4=12.563898)$ and
$(q_3=1.524585\,i,q_4=-2.195368,E_4=12.383790)$. Aside from this spurious
degeneracy, the spectrum of quantized $q_4$ has many features in common with the
spectrum of $q_3$ at $N=3$ shown in Figure~\ref{Fig-N=3}.

\begin{figure}[th]
\vspace*{5mm}
\centerline{{\epsfysize7cm \epsfbox{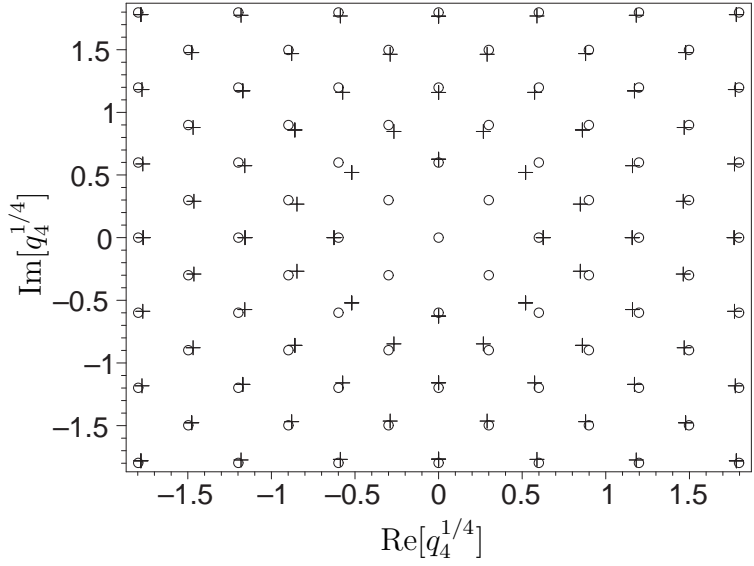}}}
\caption[]{The spectrum of quantized $q_4$ for $h=1/2$ and $q_3=0$. The exact results are shown
by crosses and the WKB predictions based on Eq.~\re{WKB-N4} are denoted by
circles.}
\label{Fig-WKB-N4}
\end{figure}

We notice that quantized $q_4$ are located close to the vertices of a square-like
lattice. To verify this property we selected among all eigenstates of the $N=4$
system only those with $h=1/2$, $q_3=0$ and nonzero values of $q_4$ (see
Figure~\ref{Fig-WKB-N4}). These states have the quasimomentum $\theta_4=0$ and
they play an important r\^ole in our discussion as the ground state of the system
has the same quantum numbers. The WKB analysis of the Baxter equation at $N=4$
leads to the following expression for quantized $q_4$ at $q_3=0$ (see
Eq.~\re{WKB-quan} below)
\be
\left[q_4^{\rm WKB}(\ell_1,\ell_2)\right]^{1/4}
=\Delta_{N=4}\cdot\left(\frac{\ell_1}{\sqrt{2}}+i\frac{\ell_2}{\sqrt{2}}\right),
\label{WKB-N4}
\ee
where the integers $\ell_1$ and $\ell_2$ are such that their sum $\ell_1+\ell_2$
is even and the lattice spacing is
\be
\Delta_{4}
=\left[\frac{4^{3/4}}{\pi}\int_{-1}^1\frac{dx}{\sqrt{1-x^4}}\right]^{-1}
=\frac{\Gamma^2(3/4)}{2\sqrt{\pi}}=0.423606...
\ee
As before, the leading-order WKB formula \re{WKB-N4} is valid only for
$|q_4^{1/4}|\gg |q_2^{1/2}|$. The comparison of \re{WKB-N4} with the exact
results for $q_4$ at $h=1/2$ is shown in Figure~\ref{Fig-WKB-N4} and
Table~\ref{tab:WKB-4} (see footnote~\ref{foot}). We find that quantized
$q_4^{_{\scriptstyle 1/4}}$ occupy the whole complex plane except the interior of
the disk of radius $\Delta_{4}$
\be
|q_4^{1/4}| > \Delta_{4}\,,
\ee
and the WKB formula \re{WKB-N4} describes their spectrum with a good accuracy.

%%%%%%%%%%%%%%%%%%%%%%%%%%%%%%%%%%%%%%%%%%%%%%%%%%%%%%%%%%%%%%%%%%%%
%%%%%%%%%%%%%%%%%%%%%%%%%%%%%%%%%%%%%%%%%%%%%%%%%%%%%%%%%%%%%%%%%%%%

\begin{table}[ht]
\begin{center}
\begin{tabular}{|c||c|c|c|c|c|c|}
\hline
 $(\ell_1,\ell_2)$&  $(2,0)$ &  $(2,2)$ & $(3,1)$ & $(4,0)$ &
     $(3,3)$ &  $(4,2)$
\\
\hline
$\left(q_4^{\rm \,exact}\right)^{1/4}$ & $.626 $ & $.520+.520\,i$ & $.847+.268\,i
$ & $1.158$  & $.860+.860\,i$  & $1.159+.574\,i$
\\
\hline
$\left(q_4^{\rm WKB}\right)^{1/4}$ & $.599 $ & $.599+.599\,i$ & $.899+.299\,i$  &
$1.198$ & $.899+.899\,i$  & $1.198+.599\,i$
\\
\hline
$-E_4/4$ & $0.6742$ & $-1.3783$ & $-1.7919$& $-2.8356$ & $-3.1410$  & $-3.3487$
\\
\hline
\end{tabular}
\end{center}
\caption{Comparison of the exact spectrum of $q_4^{1/4}$ at $q_3=0$ and $h=1/2$ with the approximate
WKB expression \re{WKB-N4}. The last line defines the exact energy $E_4$.}
\label{tab:WKB-4}
\end{table}

We recall that the points shown in Figure~\ref{Fig-WKB-N4} describe the $N=4$
eigenstates with $h=1/2$, $q_3=0$ and the quasimomentum $\theta_4=0$. Similar to
the $N=3$ case, the spectrum of $q_4$ is parameterized by the pair of integers
$(\ell_1,\ell_2)$, Eq.~\re{WKB-N4}. Still, as one can see from
Figure~\ref{Fig-Q4}, there exist the eigenstates which have the same
quasimomentum, close value of $q_4$ but $q_3\neq 0$. Similar phenomenon also
occurs for other values of the quasimomentum. In order to distinguish these
additional eigenstates, one has to introduce the second pair of integers
$(\ell_3,\ell_4)$. The dependence of $q_3$ and $q_4$ on $(\ell_3,\ell_4)$ leads
to the ``fine splitting'' of quantized $q_4$ in Figure~\ref{Fig-Q4}.
%It also allows us to describe a nontrivial spectrum of $q_3$ at $N=4$.
As we will argue below, the WKB approach allows us to describe a nontrivial
spectrum of $q_3$ and $q_4$.

Let us now examine the energy spectrum at $N=4$. Calculating the energy of the
eigenstates shown in Figure~\ref{Fig-Q4}, we find that $E_4>0$ for all points on
the $q_4^{1/4}-$plane except of four points with the coordinates
$(\ell_1,\ell_2)=(\pm 2,0)$ and $(0,\pm 2)$. Due to a residual symmetry
$q_4^{1/4}\to \exp(ik \pi/2)\, q_4^{1/4}$, Eq.~\re{map1}, they describe a single
eigenstate, which can be identified as the ground state of the $N=4$ system with
$h=1/2$ and
\be
q_3^{\rm ground}=0\,,\qquad q_4^{\rm ground}=0.153589...\,,\qquad E_4^{\rm
ground}=-2.696640...
\label{N=4-ground}
\ee
It has the quasimomentum $\theta_4=0$ and, in contrast with the $N=3$ case, it is
unique.

%%%%%%%%%%%%%%%%%%%%%%%%%%%%%%%%%%%%%%%%%%%%%%%%%%%%%%%%%%%%%%%%%%%%%%%%%%%%%%%%%%
\begin{figure}[h]
\vspace*{5mm}
\centerline{{\epsfysize6cm \epsfbox{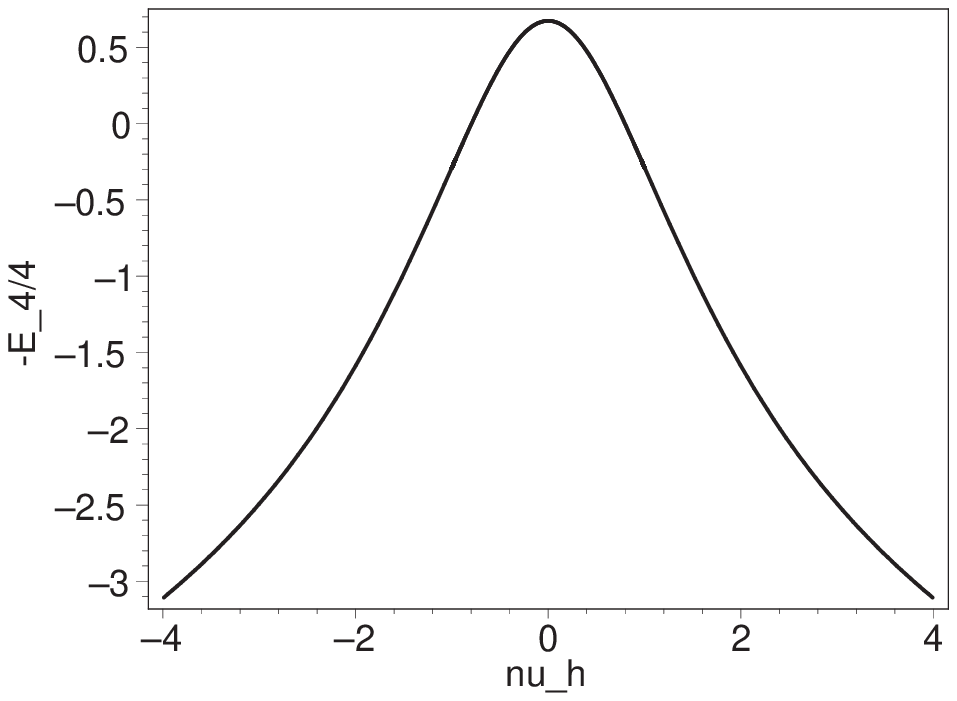}}\qquad{\epsfysize6cm \epsfbox{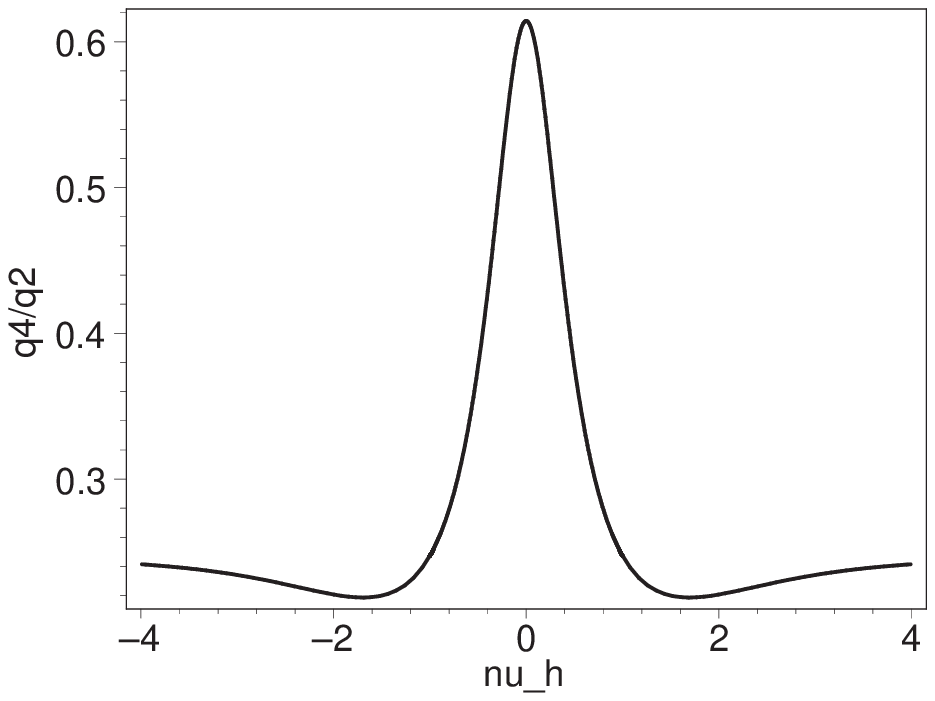}}}
\caption[]{The dependence of the energy, $-E_4/4$, and the quantum number, $q_4/q_2$,
with $q_2=1/4+\nu_h^2$, on the total spin $h=1/2+i\nu_h$ along the ground state
trajectory at $N=4$.}
\label{N4-flow}
\end{figure}
%%%%%%%%%%%%%%%%%%%%%%%%%%%%%%%%%%%%%%%%%%%%%%%%%%%%%%%%%%%%%%%%%%%%

Going over from $\nu_h=0$ to nonzero values of $\nu_h$, one finds that, similar
to the $N=3$ case (see Figure~\ref{Fig-3D}), the eigenvalues of the integrals of
motion flow along the trajectories on the moduli space $(\nu_h,q_3,q_4)$. Namely,
each point shown in Figure~\ref{Fig-Q4} creates its own trajectory labelled by
the set of integers $\{\ell\}$. Among all trajectories the one with
$(\ell_1,\ell_2)=(2,0)$ plays the special r\^ole as it contains the ground state
of the model. We find that $q_3=\Im q_4=0$ for arbitrary $\nu_h$ on the ground
state trajectory, whereas $\Re(q_4)$ and $E_4$ vary with $\nu_h$ as shown in
Figure~\ref{N4-flow}. Accumulation of the energy levels next to the ground state
at $\nu_h=0$ is described by Eq.~\re{accum} with the dispersion parameter
$\sigma_4$ given below in the Table~\ref{tab:Summary}.

The ground state \re{N=4-ground} has the Lorentz spin $n_h=0$. In analogy with
the $N=3$ case, we calculated the minimal energy in the sectors with
$n_h=1\,,2\,,3$ (see the first four columns in the Table~\ref{tab:N4} below) and
verified that it increases with $n_h$.

As we have seen before, the spectrum of quantized charges at $N=3$ and $N=4$ is
described by the simple formulae, Eqs.~\re{q3-WKB} and \re{WKB-N4}. Their
derivation is based on the WKB approach to the eigenproblem for the Baxter
operator~\cite{WKB}. In this approach, $Q_{q,\bar q}(x,\bar x)$ is constructed as
a quasiclassical wave function in the separated coordinates $\vec x=(x,\bar x)$.
The underlying classical dynamics is described by the spectral curve (``equal
energy level'' equation)
\be
\Gamma_N: \qquad
y^2=t_N^2(x)-4x^{2N}=(4x^N + q_2 x^{N-2} + ... + q_N)(q_2 x^{N-2} + ... + q_N)\,,
\label{Gamma_N}
\ee
where $t_N(x)$ was defined in \re{t_N} and $y(x)=2x^N\sinh p_x$ is related to the
holomorphic part of the momentum of a particle in the separated coordinates,
$p_x$. For arbitrary complex $x$, the equation \re{Gamma_N} has two solutions for
$y(x)$. As a consequence, $y(x)$ becomes a single-valued function on the Riemann
surface corresponding to the complex curve $\Gamma_N$. This surface is
constructed by gluing together two copies of the complex $x-$plane along the cuts
$[\sigma_1,\sigma_2]$, $...$, $[\sigma_{2N-3},\sigma_{2N-2}]$ running between the
branching points $\sigma_j$ of the curve \re{Gamma_N}. The latter are defined as
simple roots of the equation $t_N^2(\sigma_j)=4\sigma_j^{2N}$. Their positions on
the complex plane depend on the values of the integrals of motion $q_2$, $q_3$,
.$..$, $q_N$. In general, the Riemann surface defined in this way has a genus
$g=N-2$, which depends on the number of reggeons, $N$. It is a sphere at $N=2$, a
torus at $N=3$ and so on.

Let us define on $\Gamma_N$ the set of oriented closed $\alpha-$ and
$\beta-$cycles. The cycles $\alpha_k$ encircle the cuts
$[\sigma_{2k-1},\sigma_{2k}]$ with $k=1,...,N-2$ and belong to the both sheets of
$\Gamma_N$. The cycles $\beta_k$ run from the cut $[\sigma_{2N-3},\sigma_{2N-2}]$
to $[\sigma_{2k-1},\sigma_{2k}]$ on the upper sheet and, then, back on the lower
sheet. Then, classical trajectories of a particle in the separated coordinates
correspond to wrapping around $\alpha-$ and $\beta-$cycles on the Riemann surface
\re{Gamma_N}. Requiring $Q_{q,\bar q}^{\rm WKB}(x,\bar x=x^*)$ to be
single-valued on the complex $x-$plane, one obtains the following WKB
quantization conditions%
\footnote{We are most grateful to A.~Gorsky for collaboration on this point.}
\be
\Re \oint_{\alpha_k} d S_0 = \pi \ell_{2k-1}\,,\qquad
\Re \oint_{\beta_k} d S_0 = \pi \ell_{2k}\,,\qquad (k=1,...,N-2)
\label{WKB-quan}
\ee
where  $\ell_k$ are integer and $dS_0$ is the ``action'' differential on the
curve \re{Gamma_N}~\cite{WKB}
\be
dS_0=dx\, p_x \cong \frac{N t_N(x)-xt'_N(x)}{\sqrt{t_N^2(x)-4x^{2N}}} dx\,.
\ee
Solving \re{WKB-quan}, one can find the explicit expressions for the integrals of
motion, Eq.~\re{branches}. At $N=3$ and $N=4$, for $|q_N^{_{\scriptstyle
1/N}}|\gg q_2^{1/2}$, one arrives at \re{q3-WKB} and \re{WKB-N4}. The general
analysis of \re{WKB-quan} turns out to be rather involved and it will be
presented in the forthcoming publication.

\subsection{Quantum numbers of the states with higher $N$}

In this Section we will describe the results of our calculations of the spectrum
of the model for higher $N$ and $q_N\neq 0$. As we have seen in the previous
Sections, the structure of the spectrum gets more complicated as one increases
$N$. Therefore, instead of presenting a detailed description of the whole
spectrum, as was done before for $N=3$ and $N=4$, we will restrict our analysis
to the properties of the ground state trajectory only.

Solving the quantization conditions \re{C1-C0} for $N\ge 5$, we found that the
ground state trajectories have different properties for even and odd number of
particles. Namely, for even $N$ the integrals of motion $q_k$ with odd indices
$k$ vanish and for even $k$ they take pure real values
\be
q_{3}=q_5=...=0 \,,\qquad \Im q_4 =\Im q_6=...=0\,,\qquad [N=\rm even]
\label{q-even-N}
\ee
For odd $N$, the integrals of motion $q_k$ take nonvanishing values. They are
pure imaginary for odd $k$ and real otherwise
\be
\Re q_{3}=\Re q_5=...=0 \,,\qquad \Im q_4 =\Im q_6=...=0\,,\qquad [N=\rm odd]
\label{q-odd-N}
\ee
We recall that the total spin of the system on the ground state trajectory is
equal to $h=1/2+i\nu_h$, so that the quantized charges $q_k$ and the energy $E_N$
are functions of $\nu_h$. At $N=3$ and $N=4$ the $\nu_h-$dependence of $q_N$ and
$E_N$ is shown in Figures~\ref{Fig-3D}, \ref{Fig-energy3} and \ref{N4-flow},
respectively.

It follows from Eqs.~\re{q-even-N} and \re{q-odd-N} that for even $N$ the ground
state trajectory is invariant under the symmetry transformations \re{map},
whereas for odd $N$ it is mapped into another trajectory with the different
quantum numbers $(-1)^k q_k$ but the same energy $E_N$. This implies, that the
ground state of the system is double degenerate for odd number of particles $N$
and it is unique for even $N$. The degeneracy is related to the properties of the
system under mirror permutations of particles~\cite{K1,DKM-I}.

\begin{table}[t]
\begin{center}
\begin{tabular}{|c||c|c|c|c|c|c||r|r|}
\hline
  &   $ iq_3$ &  $q_4$ & $iq_5$ &  $q_6$ & $iq_7$ & $q_8$ & $-E_N/4$
  & $\sigma_N/4\,\,\,$
\\
\hline
${N=2}$ &            &           &           &    &&       & 2.772589 &16.8288
\\
${N=3}$ & .205258 &           &           &    &&       & -.247170 &.9082
\\
${N=4}$ &   0        & .153589 &           &     &&      &
\phantom{-}.674160 & 1.3176
\\
${N=5}$ & .267682 & .039452 & .060243 &     &&      & -.127516 & .4928
\\
${N=6}$ &          0 & .281825 & 0         & .070488 &&&
\phantom{-}.394582 & .5644
\\
$N=7$ & .313072 & .070993 & .128455 & .008494 & .019502 & & -.081410 & .3194
\\
$N=8$ & 0 & .391171 & 0 & .179077 & 0 & .030428 & .280987 & .3409
\\
\hline
\end{tabular}
\end{center}
\caption{The exact quantum numbers, $q$, and the energy, $E_N$, of the ground state of $N$
reggeized gluons in multi-colour QCD. The dispersion parameter, $\sigma_N$,
defines the energy of the lowest excited states, Eq.~\re{accum}.}
\label{tab:Summary}
\end{table}

\begin{figure}[h]
\vspace*{3mm}
\centerline{\epsfysize7cm \epsfbox{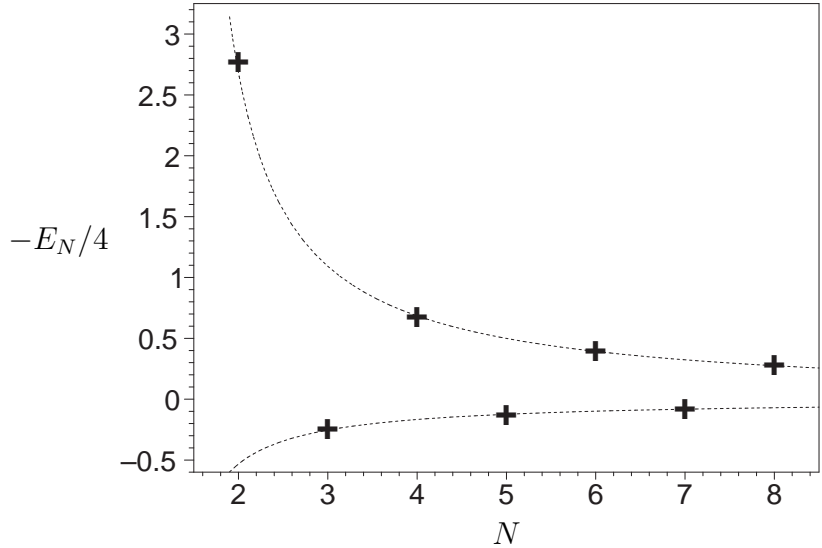}}
\caption[]{The dependence of the ground state energy, $-E_N/4$, on the number of
particles $N$. The exact values of the energy are denoted by crosses. The upper
and the lower dashed curves stand for the functions $1.8402/(N-1.3143)$ and
$-2.0594/(N-1.0877)$, respectively.}
\label{Fig:E_N}
\end{figure}

The results of our calculations of the ground state for the system with the
number of particles $N\le 8$ are summarized in the Table~\ref{tab:Summary} and
Figure~\ref{Fig:E_N}. We recall that for odd $N$ there exists the second
eigenstate with the same energy $E_N$ and the charges $(-1)^k q_{k}$.

The energy of the ground state $E_N$, as a function of the number of particles
$N$, has a number of interesting properties. $E_N$ changes a sign as one
increases the number of particles -- it is negative for even $N$ and positive for
odd $N$. Our results also indicate that the absolute value of the energy
decreases with $N$, $|E_N|\sim 1/N$ for $N\to\infty$ (see Figure~\ref{Fig:E_N}).
For large even and odd number of particles $N$ it approaches the same asymptotic
value $E_{2\infty}=E_{2\infty+1}=0$.

These properties are unique (and quite unexpected) features of noncompact
$SL(2,\mathbb{C})$ Heisenberg spin magnets. It is instructive to compare them
with similar properties of the $SL(2,\mathbb{R})$ Heisenberg magnets studied
earlier in \cite{BDM,AB,BDKM,DKM} in the relation with the Evolution Equations
for high-twist operators in high-energy QCD. There, the number of sites of the
spin magnet is equal to the twist of the operators and the ground state energy
defines the minimal anomalous dimension of these operators. Roughly speaking, the
Hamiltonian of the $SL(2,\mathbb{R})$ magnet is given by the holomorphic part of
the reggeon Hamiltonian \re{Ham} with the only difference that the holomorphic
$z-$coordinates take real values and the eigenstates $\Psi(z_1,...,z_N)$ are
polynomials in $z$. One finds \cite{AB,BDM,BDKM,DKM}, that for the
$SL(2,\mathbb{R})$ spin magnet all charges $q_3,...,q_N$ take real quantized
values, the ground state energy $E_N$ is positive and it monotonically increases
with $N$. Thus, the properties of the $SL(2,\mathbb{C})$ and $SL(2,\mathbb{R})$
Heisenberg spin magnets turn out to be completely different.

%%%%%%%%%%%%%%%%%%%%%%%%%%%%%%%%%%%%%%%%%%%%%%%%%%%%%%%%%%%%%%%%%%%%%%%%%%%%%%%%%%
%%%%%%%%%%%%%%%%%%%%%%%%%%%%%%%%%%%%%%%%%%%%%%%%%%%%%%%%%%%%%%%%%%%%%%%%%%%%%%%%%%

The spectrum of the $SL(2,\mathbb{C})$ magnet is gapless for arbitrary $N$.
Accumulation of the energy levels next to the ground state is described by
\re{accum} with the dispersion parameter $\sigma_N$ given in the Table~3. The
states belonging to the ground state trajectory have the total $SL(2,\mathbb{C})$
spin $h=1/2+i\nu_h$ and their quantum numbers $q_3,...,q_N$ satisfy \re{q-even-N}
and \re{q-odd-N} for even and odd $N$, respectively. The explicit expressions for
$q_k$ can be obtained in the WKB approach from Eq.~\re{WKB-quan}.

%%%%%%%%%%%%%%%%%%%%%%%%%%%%%%%%%%%%%%%%%%%%%%%%%%%%%%%%%%%%%%%%%%%%%%%%%%%%%%%%%%
\begin{figure}[h]
\vspace*{3mm}
\centerline{\epsfysize6.0cm \epsfbox{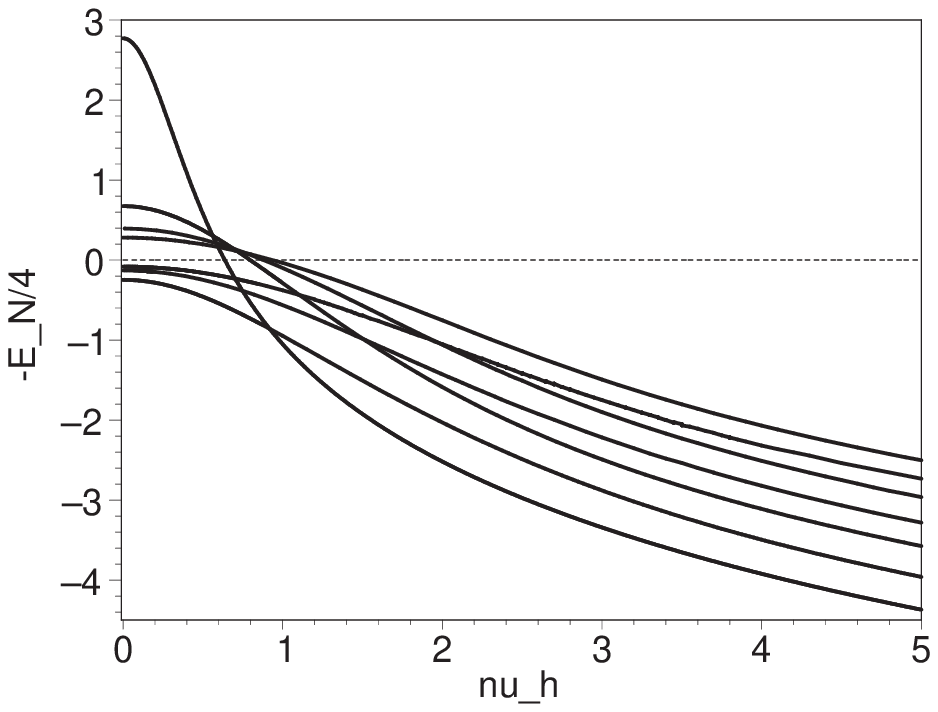}\qquad\epsfysize5.92cm \epsfbox{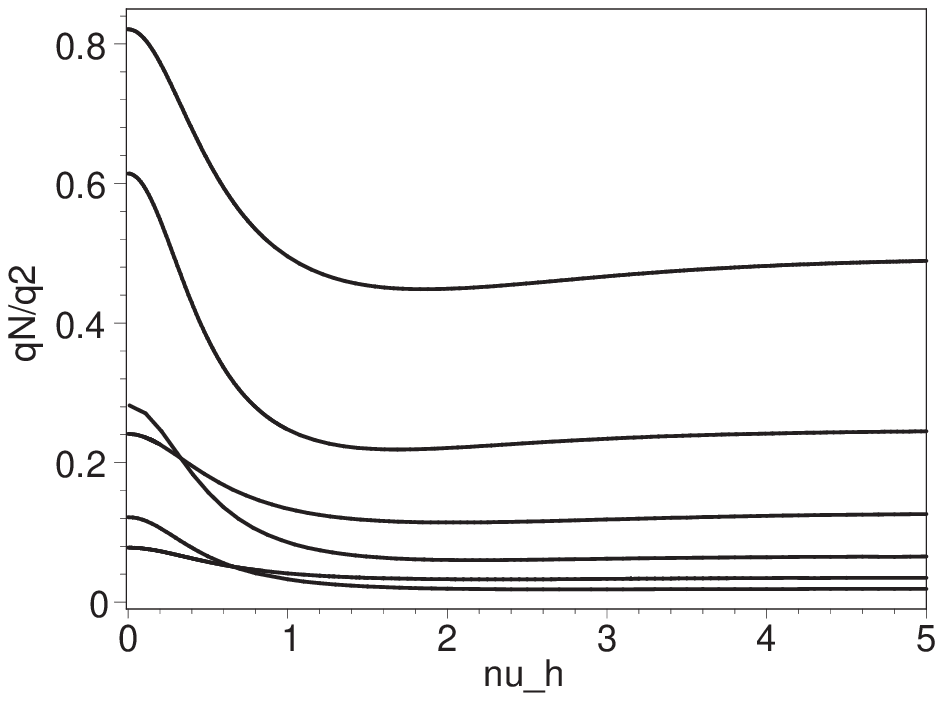}}
\caption[]{The dependence of the energy $-E_N(\nu_h)/4$ and the ``highest'' integral
of motion $|q_N|/q_2$ with $q_2=(1/4+\nu_h^2)$ on the total spin $h=1/2+i\nu_h$
along the ground state trajectory for different number of particles $2\le N\le
8$. At large $\nu_h$, $-E_8>...>-E_3>-E_2$ on the left panel and $|q_8/q_2| <
...< |q_3/q_2|$ on the right panel.}
\label{Fig:E-flow}
\end{figure}
%%%%%%%%%%%%%%%%%%%%%%%%%%%%%%%%%%%%%%%%%%%%%%%%%%%%%%%%%%%%%%%%%%%%%%%%%%%%%%%%%%

Another unusual feature of the $SL(2,\mathbb{C})$ magnet can be revealed by
examining the dependence $E_N=E_N(\nu_h)$ on the total $SL(2,\mathbb{C})$ spin
$h=1/2+i\nu_h$ along the ground state trajectory for different number of
particles $N$. For $2\le N\le 8$ this dependence is shown in
Figure~\ref{Fig:E-flow}. We notice that the flow of the energy $E_N$ with $\nu_h$
is such that the hierarchy of the energy levels at $\nu_h=0$ and large $\nu_h$ is
completely different. At $\nu_h=0$ the values of $E_N(0)$ coincide with those
depicted by crosses in Figure~\ref{Fig:E_N}. For large $\nu_h$ the system
approaches a quasiclassical regime \cite{K2,WKB}, in which the energy
$E_N(\nu_h)$ and the quantum numbers $q_N$ have a universal scaling behaviour
\be
E_N(\nu_h)\sim 4\ln |q_N| \,,\qquad |q_N| \sim  C^N\,\nu_h^2\,,
\ee
with $C\approx 0.52$. As can be seen from the right panel in
Figures~\ref{N4-flow} and \ref{Fig:E-flow}, this regime starts already at $\nu_h
\approx 2$. These results suggest that the ground state of the $SL(2,\mathbb{C})$
magnet has the properties of quantum antiferromagnet, whereas the excited states
are ferromagnetic.

\subsection{Descendent states}

Until now, we have excluded from consideration the eigenstates with $q_N=0$. Let
us now demonstrate that these states can be expressed in terms of the
$(N-1)-$particle eigenstates. That is why we shall refer to them as descendent
states.

To see this, one examines the Baxter equation \re{Bax-eq} for the
$SL(2,\mathbb{C})$ spins $s=0$ and $\bar s=1$. One finds from \re{t_N} that for
$q_N=\bar q_N=0$ the transfer matrix factorizes as $t_N(u)=u t_{N-1}(u)$ and the
holomorphic Baxter equation \re{Bax-eq} at $s=0$ takes the form
\be
t_{N-1}(u)\, Q_N(u,\bar u)=u^{N-1}\left[ Q_N(u+i,\bar u)+Q_N(u-i,\bar u)\right].
\label{Q-N-1}
\ee
Here, we introduced a subscript to indicate that $Q_N(u,\bar u)$ corresponds to
the $N-$particle system. Eq.~\re{Q-N-1} coincides with the Baxter equation for
the eigenvalue of the Baxter operator for the system of $(N-1)-$particles,
$Q_{N-1}(u,\bar u)$. The same happens in the antiholomorphic sector.  One
verifies that for $\bar s=1$ the function $\bar u\,Q_N(u,\bar u)$ satisfies the
same antiholomorphic Baxter equation as $Q_{N-1}(u,\bar u)$. This suggests that
up to a unessential normalization factor
\be
Q_N^{(q_N=0)}(u,\bar u)=Q_{N-1}(u,\bar u)/\bar u\,.
\label{BaxNp1}
\ee
However, in order to identify the r.h.s.\ of this relation as the eigenvalue of
the Baxter operator for $q_N=\bar q_N=0$, one has to verify that $Q_{N}(u,\bar
u)$ has appropriate pole structure, Eqs.~\re{Q-R,E} and \re{poles}. By the
construction, the function $Q_{N-1}(u,\bar u)$ possesses correct analytical
properties. $Q_{N}(u,\bar u)$ inherits $(N-1)$th order poles of the function
$Q_{N-1}(u,\bar u)$, and, in addition, it acquires a spurious pole at $u=\bar
u=0$. Therefore, Eq.~\re{BaxNp1} defines the eigenvalue of the Baxter operator
only if the residue at this spurious pole vanishes, or equivalently
$Q_{N-1}(0,0)=0$. As we will show in a moment, this condition is satisfied
provided that $q_{N-1} \neq 0$ and the quasimomentum of the $(N-1)-$particle
state defined by the function $Q_{N-1}(u,\bar u)$ is equal to
$\e^{i\theta_{N-1}}=(-1)^{N}$.

To evaluate $Q_{N-1}(0,0)$, we examine of the both sides of \re{Q-N-1} at $u=\bar
u=\varepsilon$ as $\varepsilon\to 0$. Taking into account that $Q_{N-1}(u,\bar
u)$ is finite at $u=\bar u=0$ and it has $(N-1)$th order poles at $(u=\pm
i\,,\bar u=0)$, Eq.~\re{Q-R,E},
\be
Q_{N-1}(\pm i+\varepsilon,\varepsilon)=\frac{R_{N-1}^\pm(q,\bar
q)}{\varepsilon^{N-1}}\left[1+{\cal O}(\varepsilon)\right]\,,
\ee
we find from \re{Q-N-1} that
\be
q_{N-1}\,Q_{N-1}(0,0)=R_{N-1}^+(q,\bar q)+R_{N-1}^-(q,\bar q)\,.
\label{rel}
\ee
The value of the residue functions $R_{N-1}^\pm(q,\bar q)$ depends on the
normalization of the Baxter operator, while their ratio is fixed by the
quasimomentum of the state. Indeed, it follows from \re{RR} and \re{sym} that
$R_{N-1}^+(q,\bar q)/R_{N-1}^-(q,\bar q)=(-1)^{N-1} \e^{i\theta_{N-1}(q,\bar
q)}$. Combining this
relation together with \re{rel}, we obtain\\[0.mm]
\be
q_{N-1} \,Q_{N-1}(0,0)= \left[1-(-1)^N \e^{-i\theta_{N-1}}\right]R_{N-1}^+(q,\bar
q)\,.
\label{master}
\ee\\[0.mm]
This relation is rather general and it holds for all eigenstates of the
$(N-1)-$particle system with the single-particle $SL(2,\mathbb{C})$ spin $s=0$
and $\bar s=1$. In particular, replacing $N\to N+1$ and putting $q_N=0$, we
deduce from \re{master} that the descendent states have the quasimomentum
\be
\e^{i\theta_N}\bigg|_{q_N=0}=(-1)^{N+1}\,.
\label{quas-0}
\ee
Going back to \re{master}, one finds that for $q_{N-1}\neq 0$ the condition
$Q_{N-1}(0,0)=0$ is satisfied
%and, therefore, Eq.~\re{BaxNp1} defines the
%eigenvalue of the Baxter operator at $q_{N}=0$,
provided that the quasimomentum of the ``ancestor'' $(N-1)-$particle state equals
$\e^{i\theta_{N-1}}=(-1)^{N}$.

Thus, starting from arbitrary $(N-1)-$particle eigenstate with the quasimomentum
$\e^{i\theta_{N-1}}=(-1)^{N}$ and applying \re{BaxNp1}, one can obtain the
$N-$particle solution to the Baxter equation with $q_N=0$. Remarkably enough, the
$N-$particle state defined in this way has exactly the same energy as its
ancestor $(N-1)-$particle state~\cite{K1}
\be
E_{N}(q_2,....,q_{N-1},q_{N}=0)=E_{N-1}(q_2,....,q_{N-1})\,.
\label{E-deg}
\ee
Indeed, as follows from \re{energy} and \re{Q-R,E}, at $s=0$ and $\bar s=1$ the
energy $E_N$ is related to the behaviour of $u^N Q_N(u+i,u)$ around $u=0$.
According to \re{BaxNp1}, at $q_N=0$ this function coincides with $u^{N-1}
Q_{N-1}(u+i,u)$.

We conclude that the spectrum of the $N-$particle system contains (an infinite)
number of descendent $q_N=0$ states, which have the same energy as
$(N-1)-$particle states. An example of such states at $N=3$ and $N=4$ can be
found in the Tables~\ref{tab:n_h} and \ref{tab:N4}. To establish the
correspondence between two different sets of the states, it is convenient to
define a new quantum number, $\vartheta_N=(-1)^{N+1} e^{-i\theta_N}$, so that
$\vartheta_N^N=1$. It follows from our analysis that there is the one-to-one
correspondence between the $(N-1)-$particle eigenstates with $\vartheta_{N-1}=1$
and $q_{N-1}\neq 0$ and the $N-$particle eigenstates
with $\vartheta_{N}=1$ and $q_{N}=0$.%
\footnote{We would like to stress that the charges $q_k$ (with $k=2,...,N$) are
continuous functions of the parameter $\nu_h$ defining the total spin
$h=(1+n_h)/2+i\nu_h$. The statement $q_N=0$ means that $q_N(\nu_h)$ vanishes for
all $\nu_h$, that is along the whole trajectory on the moduli space. If $q_N$
vanishes only for some $\nu_h$, the corresponding eigenstate is {\it not\/}
descendant. Similarly, $q_{N-1}\neq 0$ means that $q_{N-1}(\nu_h)$ does not
vanish except may be for some distinct $\nu_h$.\label{foot1}} This agrees with
the results found previously in \cite{Vacca}. One can argue that all $N-$particle
eigenfunctions with the highest charge $q_{N}=0$ can be obtained in this way. We
demonstrated that the $(N-1)-$particle state with $\vartheta_{N-1}=1$ and
$q_{N-1}\neq 0$ can be transformed into the $N-$particle state with $q_N=0$. Is
it possible to continue this process and construct the $(N+1)-$particle state
with $q_{N-1}\neq 0$ and $q_N=q_{N+1}=0$? In this case, applying \re{BaxNp1} one
gets $Q_{N+1}(u,\bar u)=Q_{N-1}(u,\bar u)/\bar u^2$. To compensate the spurious
pole at $u=\bar u=0$ one has to require that $Q_{N+1}(0,0)=0$, or equivalently
$Q_{N-1}(\epsilon,\epsilon)\sim\epsilon^2$. Substituting this relation into the
Baxter equation \re{Q-N-1} for $u=\bar u=\epsilon$ and comparing the small
$\epsilon-$behaviour of the both sides, one obtains the relation between the
residue functions $E^\pm(q,\bar q)$, Eq.~\re{Q-R,E}, $E_{N-1}^+(q,\bar
q)=E_{N-1}^-(q,\bar q)$. Together with \re{sym} and \re{energy}, this leads to
$E_{N-1}(q_2,...,q_{N-1})=0$. Thus, a positive answer to the above question would
imply that the energy of the $(N-1)-$particle state should vanish for arbitrary
values of $\nu_h$, that is along the whole trajectory on the moduli space. We do
not find such states in the spectrum of the model.

This implies that if one defines a linear operator $\Delta$ that maps the
subspace $V_{N-1}^{(\vartheta_{N-1}=1)}$ of the eigenstates of the
$(N-1)-$particle system with $q_{N-1}\neq 0$ and $\vartheta_{N-1}=1$ into the
$N-$particle states with $q_N=0$ and $\vartheta_{N}=1$,
\be
\label{delta}
\Delta:\quad \,V_{N-1}^{(\vartheta_{N-1}=1)}\to V_{N}^{(\vartheta_{N}=1)}\,,
\ee
then this operator is nilpotent~\cite{Vacca}. In addition, it has the following
properties
\be
\Delta^2=0\,,\qquad \Delta\cdot \mathcal{H}_{N-1}=\mathcal{H}_{N}\cdot\Delta\,,
\qquad\Delta\cdot \mathbb{P}_{N-1}=- \mathbb{P}_{N}\cdot\Delta=(-1)^N \Delta\,,
\label{cohom}
\ee
where the last relation follows from $\vartheta_N=\mathbb{P}_N (-1)^{N+1}$ with
$\mathbb{P}_N$ being the generator of cyclic permutations of $N$ particles,
Eq.~\re{cyclic}. Then, one can show that the wave function of the $N-$particle
state with $q_N=0$ is related to its ``ancestor'' $(N-1)-$particle state as
\ba
\Psi_{N}^{(q_N=0)}(\vec z_1,\ldots,\vec z_{N})&=&
\Delta\cdot\Psi_{N-1}(\vec z_1,\ldots,\vec z_{N-1})
\nonumber
\\
&=& N\Pi_N\,
\left[\frac{1}{\bar z_{N,1}}~+~\frac{1}{\bar z_{N-1,N}}\right]\,\Pi_{N-1}\,
\Psi_{N-1}(\vec z_1,\ldots,\vec z_{N-1})\,,
\label{delta-z}
\ea
with $\bar z_{kn}=\bar z_k-\bar z_n$ and $\Pi_N=\sum_{n=0}^{N-1}
\left((-1)^{N+1}\,\mathbb{P}_{N}\right)^n/N$ being the projector onto eigenstates with
the quasimomentum \re{quas-0}, $\mathbb{P}_N\Pi_N =(-1)^{N+1} \mathbb{P}_N$. The
explicit expressions at $N=3$ and $N=4$ look as follows
\ba
\Psi_3^{(q_3=0)}%(\vec z_1,\vec z_2,\vec z_3)
&=&\frac{\bar z_{12}}{\bar z_{23}\bar z_{31}}\Psi_2(\vec z_1,\vec z_2)+\frac{\bar
z_{23}}{\bar z_{31}\bar z_{12}}\Psi_2(\vec z_2,\vec z_3)+\frac{\bar z_{31}}{\bar
z_{12}\bar z_{23}}\Psi_2(\vec z_3,\vec z_1)\,,
\nonumber\\
\Psi_4^{(q_4=0)}%(\vec z_1,\vec z_2,\vec z_3,\vec z_4)
&=&\frac{\bar z_{31}}{\bar z_{34}\bar z_{41}}\Psi_3(\vec z_1,\vec z_2,\vec z_3)
-\frac{\bar z_{42}}{\bar z_{12}\bar z_{41}}\Psi_3(\vec z_2,\vec z_3,\vec z_4)
\nonumber
\\
%&&\qqqquad\ \+
&+&\frac{\bar z_{13}}{\bar z_{12}\bar z_{23}}\Psi_3(\vec z_3,\vec
z_4,\vec z_1) -\frac{\bar z_{24}}{\bar z_{23}\bar z_{34}}\Psi_3(\vec z_4,\vec
z_1,\vec z_2)\,,
\label{examples}
\ea
where the states $\Psi_2$ and $\Psi_3$ have the quasimomentum $\e^{i\theta_2}=-1$
and $\e^{i\theta_3}=1$, respectively. According to \re{quasi-0},
$\e^{i\theta_2}=(-1)^{n_h}$ and, therefore, the eigenstate $\Psi_3^{(q_3=0)}$ can
be defined only for odd Lorentz spins $n_h$. For $N\ge 4$ the spin $n_h$ of the
state $\Psi_{N}^{(q_N=0)}$ can be arbitrary integer.

Let us transform the descendent states \re{delta-z} into the momentum space using
the relation
\be {k_1...k_N} \cdot\widehat \Psi_N(\vec k_1,...,\vec k_N)
=\int
\prod_{k=1}^N d^2 z_k \e^{i\vec z_k \cdot \vec k_k}
\Psi_N(\vec z_1,...,\vec z_N)\,,
\label{Fourier}
\ee
where $k=(k_x+ik_y)/2$ is holomorphic component of $\vec k=(k_x,k_y)$ and the
additional factor in the l.h.s.\ was introduced for later convenience. One finds
from \re{examples} after some algebra that up to a normalization factor
\ba
\widehat \Psi_3^{(q_3=0)}&=&
\frac{(\vec k_3+\vec k_1)^2}{\vec k_3^2\,\vec k_1^2}
\widehat\Psi_2(\vec k_3+\vec k_1,\vec k_2)
+ \frac{(\vec k_1+\vec k_2)^2}{\vec k_1^2\,\vec k_2^2}\widehat
\Psi_2(\vec k_1+\vec k_2,\vec k_3)
+ \frac{(\vec k_2+\vec k_3)^2}{\vec k_2^2\,\vec k_3^2}\widehat
\Psi_2(\vec k_2+\vec k_3,\vec k_1)\,,
\nonumber
\\
\widehat \Psi_4^{(q_4=0)}&=&
\frac{(\vec k_4+\vec k_1)^2}{\vec k_4^2\,\vec k_1^2}
\widehat \Psi_3(\vec k_4+\vec k_1,\vec k_2,\vec k_3)
-\frac{(\vec k_3+\vec k_4)^2}{\vec k_3^2\,\vec k_4^2}
\widehat \Psi_3(\vec k_3+\vec k_4,\vec k_1,\vec k_2)
\nonumber
\\
&+&\frac{(\vec k_2+\vec k_3)^2}{\vec k_2^2\,\vec k_3^2}\widehat
\Psi_3(\vec k_2+\vec k_3,\vec k_4,\vec k_1) -
\frac{(\vec k_1+\vec k_2)^2}{\vec k_1^2\,\vec k_2^2}\widehat
\Psi_3(\vec k_1+\vec k_2,\vec k_3,\vec k_4)
\,.
\label{Psi-mom}
\ea
Generalization to arbitrary $N$ is straightforward. The expression for
$\widehat\Psi_N^{(q_N=0)}$ coincides with the $q_N=0$ solution found at $N=3$ in
Ref.~\cite{Bartels} and for arbitrary $N$ in Ref.~\cite{Vacca} by different
methods.

One can verify Eq.~\re{delta-z} in a number of different ways. For instance,
applying the operators \re{q} to the both sides of \re{delta-z} one finds that
the state $\Psi_N^{(q_N=0)}$ is annihilated by the operator $q_N$ and has the
same spectrum $q_k$ (with $k=2,...,N-1$) as the state $\Psi_{N-1}$ entering the
r.h.s.\ of \re{delta-z}. It is more tedious to verify that \re{delta-z} is in
agreement with \re{BaxNp1}. To show this, one examines the action of the Baxter
operator on the state \re{delta-z}, $\mathbb{Q}(u,\bar
u)\Psi_{N}^{_{(q_N=0)}}(\vec z_1,\ldots,\vec z_{N})$, replaces the
$\mathbb{Q}-$operator by its integral representation and uses the Feynman diagram
technique to calculate the emerging two-dimensional integrals (see
Ref.~\cite{DKM-I} for details). Finally, one can check by explicit calculation
that $\mathbb{P}_N\,\Psi_{N}^{(q_N=0)}=(-1)^{N+1}\Psi_{N}^{(q_N=0)}$ and
$\Delta\cdot\Psi_{N}^{(q_N=0)}=0$, in agreement with Eqs.~\re{quas-0} and
\re{cohom}, respectively.

\begin{table}[ht]
\begin{center}
\begin{tabular}{|c||c|c|c||c|c|}
\hline
 $n_h$&  $q_3$ &  $E_3$ &
     $\ell$ &  $E_{3,\,\rm d}$ &
     $\ell_{\rm d}$
\\
\hline
$0$ & $.2052\,i$ & $ 0.9884$ & $0$ & $-$ & $-$
\\
\hline
$1$ &  $.3315$ & $1.5368$ & $1$ & $0$ & $0$
\\
\hline
$2$ & $ -.4017+.4201\,i $ & $4.6077$ & $1$ & $-$ & $-$
\\
\hline
$3$ & $1.1766\,i$ & $6.9592$ & $0$ & $8$ & $0$
\\
\hline
\end{tabular}
\end{center}
\caption{The dependence of the minimal energy on the Lorentz spin
$n_h$ at $N=3$. The total $SL(2,\mathbb{C})$ spin equals $h=(1+n_h)/2$ for all
states except $n_h=2$ when $h=3/2+ .176\,i$. Integer $\ell$ defines the
quasimomentum, $\theta_3=2\pi\ell/3$. The subscript $\rm (d)$ refers to the
descendant states (see Section~5.5). The states with $q_3\neq 0$ are degenerate
with respect to \re{map}.}
\label{tab:n_h}
\end{table}

\begin{table}[ht]
\begin{center}
\begin{tabular}{|c||c|c|c|c||c|c|c|}
\hline
 $n_h$&  $q_3$ & $q_4$ & $E_4$ &
     $\ell$ & $q_{3,\,\rm d}$  &  $E_{4,\,\rm d}$ &
     $\ell_{\rm d}$
\\
\hline
$0$ & $0$ & $.1535$ & $-2.6964$ & $0$ & $.2052\,i$ & $0.9884$ & $2$
\\
\hline
%$1$ & $-.2705-.3185\,i$ & $.2030-.1601\,i$ & $\phantom{-}2.5616$ & $1$ & $0$ &
%$0$ & $2$
$1$ & $0$ & $0$ & $0$ & $2$ & $1.3659$ & $8.1080$ & $2$
\\
\hline
$2$ & $0$ & $-.2869$ & $\phantom{-}2.6268$ & $0$ & $1.0236$ & $6.8888$ & $2$
\\
\hline
$3$ & $-.6951$ & $-.6337$ & $\phantom{-}5.8836$ & $1$ & $1.1766\,i$ & $6.9592$ &
$2$
\\
\hline
\end{tabular}
\end{center}
\caption{The dependence of the minimal energy on the Lorentz spin
$n_h$ at $N=4$. The total $SL(2,\mathbb{C})$ spin equals $h=(1+n_h)/2$.
% for all
%states except the one with $n_h=1$ and $q_3\neq 0$ for which $h=1+.049\,i$.
The last three columns correspond to the descendant states. The quasimomentum is
defined as $\theta_4=(2\pi\ell)/4$. The states are degenerate with respect to
\re{map}.}
\label{tab:N4}
\end{table}

Using the results of the previous Sections, one can apply Eqs.~\re{E-deg} and
\re{delta-z} and reconstruct the spectrum of the descendent states for arbitrary
$N$. As before, we shall identify among these states the one with the minimal
energy, $E_N^{\rm (min)}(q_N=0)$. In virtue of \re{E-deg}, this amounts to
finding the minimal energy in the spectrum of the $(N-1)-$particle system with
$q_{N-1}\neq 0$ and the quasimomentum $\e^{i\theta_{N-1}}=(-1)^N$. Obviously, it
can not be smaller then the energy of the ground state, $E_{N-1}^{\rm ground}$,
whose quasimomentum is  $\theta^{\rm ground}_{N-1}=0$ (see
Table~\ref{tab:Summary}). The results of our calculations at $N=3$ and $N=4$ are
given in the Tables~\ref{tab:n_h} and \ref{tab:N4}, respectively. As was already
mentioned, at $N=3$ the descendant states can be constructed only for odd $n_h$.
For $n_h=1$ and $n_h=3$ they are descendants of the $N=2$ states with the energy
\re{E-2-exact}. At $N=4$ the $q_4=0$ states are descendants of the $N=3$ states
with the quasimomentum $\e^{i\theta_3}=1$. Notice that the $N=4$ state with
$n_h=1$ and $q_3=q_4=0$ is not descendant since $q_4\neq 0$ for $\nu_h\neq 0$
(see footnote~\ref{foot1}).

In general, for {\it even\/} $N$ one has $\e^{i\theta_{N-1}}=1$ and, therefore,
the eigenstate  with the mininal energy, $\Psi_{N}^{(q_N=0)}$, is a descendant of
the ground state, $\Psi^{\rm ground}_{N-1}$. As follows from our results (see
Table~\ref{tab:Summary} and Figure~\ref{Fig:E_N}), its energy is positive,
$E_{N}^{\rm min}(q_N=0)=E_{N-1}^{\rm ground}>0$. For {\it odd\/} $N$ one has
$\e^{i\theta_{N-1}}=-1$ and, therefore, this state can not be a descendant of the
ground state. In this case, the minimal value of the energy is given by
$E_{N}^{\rm min}(q_N=0)=0$ and the corresponding integrals of motion are equal to
$q_2=q_3=...=q_{N-1}=0$. This eigenstate is located on the trajectory on the
moduli space at $h=1+i\nu_h$ and $\nu_h=0$. Going over to $\nu_h\neq 0$ we find
that $q_2,q_4,..., q_{N-1}\neq 0$ while $q_3=q_5=...=q_{N-2}=0$ for arbitrary
$\nu_h$. The energy $E_N(q_N=0)$ is a continuous function of $\nu_h$ along this
trajectory and it takes its minimal value, $E_N^{\rm min}(q_N=0)=0$, at
$\nu_h=0$. We notice that it is smaller than the energy of the ground state on
the subspace with $q_N\neq 0$.

At $q_2=...=q_N=0$ the eigenvalue of the Baxter $\mathbb{Q}-$operator can be
easily found from the integral representation \re{Q-R}. Solving the differential
equations \re{Eq-1}, one can construct a single-valued function $Q(z,\bar z)$ as
\be
Q^{(q=0)}(z,\bar z) = \frac{z}{(1-z)^2}\left[c_0\ln^{N-1}(z\bar
z)+c_1\ln^{N-2}(z\bar z) +...+ c_{N-1}\right]\,,
\label{Q-gen}
\ee
with $c_i$ being arbitrary coefficients. To fix the coefficients, one substitutes
\re{Q-gen} into \re{Q-R} and compares the asymptotics of the resulting expression
for $Q(u,\bar u)$ at large $u$ and $\bar u$ with Eq.~\re{Q-asym} for $h=1$, $s=0$
and $\bar s=1$. One finds that $c_1=...=c_{N-1}=0$ and up to a normalization
factor
\be
Q_N^{(q=0)}(u,\bar u)=\frac{u-\bar u}{\bar u^N}=-\frac{in}{\bar u^N}\,,
\label{Q-deg-exp}
\ee
with integer $n$ defined in \re{u-bar u}. In this expression, the numerator
compensates a spurious pole at $u=\bar u=0$. Substituting \re{Q-deg-exp} into
\re{Energy-II} one calculates the corresponding energy, $E_N(q=0)=0$.

\section{Summary}

In this paper, we have continued the study of noncompact Heisenberg
$SL(2,\mathbb{C})$ spin magnets initiated in \cite{DKM-I}. Having solved this
model, we obtained for the first time a complete description of the spectrum of
the multi-reggeon compound states in QCD at large $N_c$.

{}From point of view of integrable models, the results presented in this paper
provide an exact solution of the spectral problem for completely integrable
quantum mechanical model of $N$ interacting spinning particles in two-dimensional
space. A unique feature of this model, leading to many unusual properties of the
energy spectrum, is that its quantum space is infinite-dimensional for finite $N$
and conventional methods, like the Algebraic Bethe Ansatz, are not applicable. To
overcome this problem, we applied the method of the Baxter $\mathbb{Q}-$operator
developed in application to the $SL(2,\mathbb{C})$ spin magnets in~\cite{DKM-I}.
Solving the Baxter equations, we were able to find the exact expressions for the
eigenvalues of the $\mathbb{Q}-$operator. They allowed us to establish the
quantization conditions for the integrals of motion and, finally, reconstruct the
spectrum of the model.

{}From point of view of high-energy QCD, we calculated the spectrum of the
colour-singlet compound states built from $N$ reggeized gluons for $N\le 8$ in
the multi-colour limit, $N_c\to\infty$. The obtained expressions allowed us to
reveal the general properties of the spectrum for arbitrary $N$. Our analysis was
based on the identification of the $N-$reggeized gluon states in multi-colour QCD
as the ground states for the noncompact Heisenberg magnet of the length $N$ and
the $SL(2,\mathbb{C})$ spins $(s=0,\bar s=1)$. The identification however is not
straightforward. The contribution of the $N-$reggeon states to the scattering
amplitude \re{amp} takes the form ${\cal A}= i\textrm{s}\sum_N(i\asbar)^N{\cal
A}_N$ with
\be
{\cal A}_N(\textrm{s},\textrm{t}) = \int d^2 z_0 \e^{i\vec z_0\cdot\vec p}\,
\vev{\Phi_a(\vec z_0)|\e^{-Y\cdot{\cal H}_N^{_{\rm (QCD)}}}
\lr{\vec \partial_1^2\,...\,\vec \partial_N^2}^{-1}|\Phi_b(0)}\,,
\label{AN-rep}
\ee
where $\bar\partial_k\equiv\partial/\partial\bar z_k$, the rapidity
$Y=\ln\textrm{s}$ plays the role of the ``evolution time''. The wave functions
$\ket{\Phi_{a(b)}(z_0)}\equiv\Phi_{a(b)}(\vec z_1-\vec z_0,...,\vec z_N-\vec
z_0)$ describe the coupling of $N-$gluons to the scattered particles. The $\vec
z_0-$integration fixes the momentum transfer, $t=-\vec p^2$. The operators
$1/\vec \partial_k^2$ stand for two-dimensional transverse propagators of
$N-$gluons and the scalar product is taken with respect to the $SL(2,\mathbb{C})$
scalar product \re{SL2-norm}. The Hamiltonian ${\cal H}_N^{_{\rm (QCD)}}
=H^{_{\rm (BFKL)}}_{12} + ... + H^{_{\rm (BFKL)}}_{N,1}$ describes the
interaction between $N$ reggeized gluons in multi-colour QCD and it is given by
the sum of the BFKL kernels~\cite{L1,L2}. Notice that it is different from the
Hamiltonian of the magnet, Eq.~\re{Ham}.

The Hamiltonians ${\cal H}_N^{_{\rm (QCD)}}$ and ${\cal H}_N$ act on different
Hilbert spaces. For the magnet, it coincides with the representation space of the
principal series of the $SL(2,\mathbb{C})$ endowed with the scalar product
\re{SL2-norm}. For the QCD Hamiltonian, the choice of the scalar product is
dictated by physical requirements that the wave functions $\Phi_{a(b)}$ have to
be normalizable and the Hamiltonian ${\cal H}^{_{\rm (QCD)}}_N$ has to be bounded
from below. These conditions are satisfied if one normalizes the eigenstates of
${\cal H}_N^{_{\rm (QCD)}}$ as~\cite{L1}
\be
\|\Psi_N^{_{\rm (QCD)}}\|^2=\int \prod_{k=1}^N
d^2z_k \left|\,\bar\partial_1\,...\,\bar \partial_N {\Psi_N^{_{\rm (QCD)}}(\vec
z)}\right|^2.
\label{norma}
\ee
Under this choice of the normalization condition, ${\cal H}_N^{_{\rm (QCD)}}$ is
related to the Hamiltonian of the $SL(2,\mathbb{C})$ magnet of the spin
$(s=0,\bar s=1)$ as~\cite{FK}
\be
{\cal H}_N^{_{\rm (QCD)}}= \frac{\asbar}4 \lr{\bar \partial_1\,...\,\bar
\partial_N}^{-1}{\cal H}_N\,\lr{\bar \partial_1\,...\,\bar \partial_N}\,,
\label{rel-H-H}
\ee
where $\bar\partial_k=\partial/\partial \bar z_k$. One can verify using \re{Ham}
and \re{S-op} that this transformation changes the spin in the antiholomorphic
sector from $\bar s=1$ to $\bar s=0$.%
\footnote{This corresponds to switching from strength
tensors to gauge potentials in the description of gluon correlations.} Thus, on
the space of functions normalizable with respect to \re{norma}, two Hamiltonians
have the same energy spectrum and their eigenstates are related as
$\Psi_{N\!,q}(\vec z)=\bar\partial_1...\bar \partial_N
\Psi_N^{_{\rm (QCD)}}(\vec z)$. It remains unclear however whether physical
solutions for ${\cal H}_N^{_{\rm (QCD)}}$ can be constructed on a larger class of
functions.

The transformation \re{rel-H-H} is not well defined on the subspace of zero modes
of the operator $\bar\partial_1\,...\,\bar\partial_N$. However, these modes do
not contribute to \re{AN-rep} due to gauge-invariance of the wave functions
$\Phi_{a(b)}(\vec z_1,...,\vec z_N)$. This property can be expressed as
$\ket{\Phi_{a(b)}}=\partial_1\,...\,\partial_N\ket{\Psi_{a(b)}}$ with the states
$\ket{\Psi_{a(b)}}$ normalizable with respect to the $SL(2,\mathbb{C})$ scalar
product \re{SL2-norm}. Then, substituting \re{rel-H-H} into \re{AN-rep}, one gets
\be
{\cal A}_N(\textrm{s},\textrm{t})=\int d^2 z_0 \e^{i\vec z_0\cdot\vec
p}\,\vev{\Psi_{a}(\vec z_0)|\e^{-\asbar Y{\cal H}_N/4} |\Psi_{b}(0)}=
\sum_{q,\bar q} \textrm{s}^{-\asbar E_{N}(q,\bar q)/4}\vev{\Psi_{a}|\Psi_{\vec p,\{q,\bar q\}}}
\vev{\Psi_{\vec p,\{q,\bar q\}}|\Psi_{b}}\,,
\label{AN-dec}
\ee
where the scalar product is taken with respect to \re{SL2-norm}. In the second
relation we decomposed the Hamiltonian over the complete set of its eigenstates
$\Psi_{\vec p,\{q,\bar q\}}$ which have the total momentum $\vec p$ and the
quantum numbers $q=(q_2,...,q_N)$. The sum is dominated by the ground state
contribution. Taking into account \re{accum} and performing integration over
$\nu_h$ one arrives at \re{amp}.

The wave functions $\Psi_{a(b)}(\vec z_1,...,\vec z_N)$ are Bose symmetric --
they are invariant under interchange of any pair of reggeon coordinates and their
colour indices. Therefore the r.h.s.\ of \re{AN-dec} receives a nonzero
contribution only from the states $\Psi_{\vec p,\{q,\bar q\}}$, which have the
same Bose properties. In the multi-colour limit, the Bose symmetry is reduced to
the invariance of the wave function under the cyclic and mirror permutations of
reggeons. Since the wave function of the $N-$reggeon state is factorized, as
$N_c\to\infty$, into a product of the colour tensor and the function of reggeon
coordinates, the both factors have to possess the same parity under the cyclic
and mirror permutations simultaneously, $\mathbb{P}=1$ and $\mathbb{M}=\pm 1$.
Since the operators of the corresponding transformations, $\mathbb{P}$ and
$\mathbb{M}$, do not commute with each other, this requirement leads to the
selection rules on the eigenstates of the noncompact Heisenberg magnet.

By the construction, the eigenstates $\Psi_{\vec p,\{q,\bar q\}}(\vec
z_1,...,\vec z_N)$ diagonalize the operator of cyclic permutations,
Eq.~\re{cyclic}. Then, making use of \re{mirror}, the eigenstates of the operator
of mirror permutations, $\mathbb{M}\,\Psi_{\vec{p}\{q,\bar
q\}}^{(\pm)}=\pm~\Psi_{\vec{p}\{q,\bar q\}}^{(\pm)}$, can be defined as
\be
\Psi_{\vec{p}\{q,\bar q\}}^{(\pm)}(\vec{\mybf{z}})
=\frac{1\pm \mathbb{M}}2~\Psi_{\vec{p}\{q,\bar q\}}(\vec{\mybf{z}})
=\frac12\left[
\Psi_{\vec{p}\{q,\bar q\}}(\vec{\mybf{z}})
\pm\Psi_{\vec{p}\{-q,-\bar q\}}(\vec{\mybf{z}})\right]\,,
\ee
where $q=\{q_k\}$ and $-q=\{(-1)^kq_k\}$. Although these states do not
diagonalize the integrals of motion, $\{q,\bar q\}$, they are the eigenstates of
the Hamiltonian having the same energy $E_N=E_N(q,\bar q)=E_N(-q,-\bar q)$. Using
$\theta_N(-q,-\bar q)=-\theta_N(q,\bar q)$, we find from \re{cyclic} that the
operator of cyclic permutations acts on them as
\be
\mathbb{P}\,\Psi_{\vec{p}\{q,\bar q\}}^{(\pm)}(\vec{\mybf{z}})
=\cos(\theta_N(q,\bar q))\cdot\Psi_{\vec{p}\{q,\bar q\}}^{(\pm)}(\vec{\mybf{z}})+
i\sin(\theta_N(q,\bar q))\cdot\Psi_{\vec{p}\{q,\bar
q\}}^{(\mp)}(\vec{\mybf{z}})\,.
\ee
Thus, the eigenstates $\Psi_{\vec{p}\{q,\bar q\}}^{(\pm)}(\vec{\mybf{z}})$
diagonalize the operators $\mathbb{P}$ and $\mathbb{M}$ simultaneously only at
$\sin(\theta_N(q,\bar q))=0$. Together with \re{cyclic}, this condition selects
among all eigenstates of the magnet only those with the quasimomentum
$\e^{i\theta_N(q,\bar q)}=\pm 1$ for even $N$ and $\e^{i\theta_N(q,\bar q)}=1$
for odd $N$. Obviously, it is satisfied for the ground state, $\theta_N^{\rm
ground}(q,\bar q)=0$.

We conclude that the ground state of the magnet has a definite parity with
respect to the cyclic and mirror permutations simultaneously and, therefore, its
wave function $\Psi_{\vec{p}\{q,\bar q\}}^{_{(\pm)}}(\vec{\mybf{z}})$ can be
identified as the $\vec z-$dependent part of the full wave function of the
compound states of $N$ reggeized gluons as $N_c\to\infty$. The ground state of
the magnet has different properties for even and odd number of particles. For
even $N$, its quantum numbers satisfy $q_{2k+1}=0$, Eq~\re{q-even-N}, and, as a
consequence, two sets of the quantum numbers, $(q,\bar q)$ and $(-q,-\bar q)$
coincide leading to $\Psi_{\vec{p}\{q,\bar q\}}^{_{(+)}}(\vec{\mybf{z}})=
\Psi_{\vec{p}\{q,\bar q\}}(\vec{\mybf{z}})$ and
$\Psi_{\vec{p}\{q,\bar q\}}^{_{(-)}}(\vec{\mybf{z}})=0$. For odd $N$, the ground
state in the sector with $q_N\neq 0$ is double degenerate. The degeneracy occurs
due to the symmetry of the energy $E_N$ under $q_{2k+1}\!\to\!-q_{2k+1}$,
Eq.~\re{energy}. This allows us to construct two mutually orthogonal ground
states, $\Psi_{\vec{p}\{q,\bar q\}}^{_{(\pm)}}(\vec{\mybf z})$, which are
invariant under the cyclic permutations, $\theta_N=0$, and possess a definite
parity under the mirror permutations, $\mathbb{M}\,\Psi_{\vec{p}\{q,\bar
q\}}^{_{(\pm)}}(\vec{\mybf{z}}) =\pm\Psi_{\vec{p}\{q,\bar
q\}}^{_{(\pm)}}(\vec{\mybf{z}})$. In virtue of the Bose symmetry, the colour part
of the wave function of the $N$ reggeized gluon state should have the same parity
under the charge conjugation, $t^a \to -(t^a)^T$, with $t^a$ being the $SU(N_c)$
generators in the quark representation. This allows us to distinguish the ground
states according to their $C-$parity. For even $N$ the ground states with the
parity $\mathbb{M}=1$ have the same $C-$parity as the Pomeron, $C=1$. For odd
$N$, in the sector with $q_N\neq 0$, the ground states with the parity
$\mathbb{M}=1$ and $\mathbb{M}=-1$ have the $C-$parity of the Odderon \cite{LN},
$C=-1$, and the Pomeron, $C=1$, respectively. For the descendant states, $q_N=0$,
one deduces from \re{quas-0} that the physical states with $\theta_N=0$ can be
constructed only for odd $N$~\cite{Vacca}. Their minimal energy is $E_{N,\rm
min}^{(q_N=0)}=0$ and the corresponding state with $q_2=...=q_N=0$ has the
quantum numbers of the Odderon.

Our results indicate that, in the multi-colour limit, in the Pomeron sector, only
compound states built from even number of reggeized gluons $N$ provide the
contribution to the scattering amplitude ${\cal
A}(\textrm{s},\textrm{t})/\textrm{s}$, Eq.~\re{amp}, rising with the energy
$\textrm{s}$. Their intercept $\alpha_N\equiv 1-\asbar E_N/4$ is bigger than one,
but it decreases at large $N$ as $\alpha_N-1\sim 1/N$.

In the Odderon sector, the situation is different. Depending on the value of the
``highest'' charge, $q_N\neq 0$ and $q_N=0$, one can construct two different
solutions for the Odderon state. For the first solution, $q_N\neq 0$, the
intercept of the compound states is smaller than one, but it increases with $N$
as $\alpha_N-1\sim -1/N$. As a consequence, the contribution to the scattering
amplitude \re{amp} from the $N=3$ state (``bare Odderon'') is subdominant at
high-energy with respect to the contribution of the $N=5$ state and so on. The
high-energy asymptotics of the scattering amplitude \re{amp} in the Odderon
sector with $q_N\neq 0$ is governed, as $N_c\to\infty$, by the contribution of
the states with an arbitrary large odd number of reggeized gluons. It increases
the effective value of the Odderon intercept and leads to $\alpha_{\rm
Odderon}=\alpha_{2\infty+1}=1$. For the second solution, $q_N=0$, the intercept
of the compound states equals $1$ for arbitrary odd $N$. At $N=3$ such state was
first constructed in \cite{Bartels}. To calculate the contribution of the $q_N=0$
states to the scattering amplitude, one has to resum in \re{amp} an infinite
number of terms with $N=3,5,...$. They have the same energy behaviour
$\sim\textrm{s}^1/(\sigma_N\ln \textrm{s})^{1/2}$, with the dispersion parameter
$\sigma_N$, which scales at large $N$ as $\sigma_N\sim 1/N^2$ and, therefore,
enhances the contribution of higher reggeon states.

Thus, the two solutions, $q_N\neq 0$ and $q_N=0$, lead to the same value of the
Odderon intercept, $\alpha_{\rm Odderon}=1$, but the properties of the underlying
Odderon states are quite different. The Odderon state with $q_N\neq 0$ does not
couple to a point-like hadronic impact factors of the form~\cite{CKMS}
$\ket{\Phi_a}\sim
\delta(\vec z_1-\vec z_2) \phi(\vec z_2,...,\vec z_N)+\mbox{[cyclic permutations]}$,
like the one for the $\gamma^*\to\eta_c$ transition, whereas the Odderon state
with $q_N=0$ provides a nontrivial contribution~\cite{Bartels}. Another
difference comes from the analysis of the dependence of the scattering amplitude
on the invariant mass of one of the scattered hadrons, $Q^2$. One can show that
in the limit $x_{\rm Bj}=Q^2/\textrm{s}\ll 1$ and $Q^2\to\infty$, the Odderon
states with $q_N\neq 0$ and $q_N=0$ provide a contribution to the scattering
amplitude $\sim 1/Q^p$ of the twist-4 ($p=4$) and twist-3 ($p=3$), respectively.
Both properties have to do with the fact that the wave function of the
$N-$reggeon state, $\Psi_{\vec p,\{q,\bar q\}}(\vec z_1,...,\vec z_N)$, vanishes
as $|\vec z_k -\vec z_{k+1}|\to 0$ for $q_N\neq 0$ and it stays finite for
$q_N=0$. It remains unclear which of these solutions corresponds to a physical
Odderon state in QCD.

We found that in the Pomeron and the Odderon sectors, the intercept of the
$N-$reggeon states approaches the same value $\alpha_\infty=1$ as $N\to\infty$
and their contribution to the scattering amplitude, ${\cal
A}(\textrm{s},\textrm{t})/\textrm{s}$, ceases to depend on the energy
$\textrm{s}$ as $N\to\infty$. It is interesting to notice that this result has
been anticipated a long time ago within the bootstrap approach \cite{Ven}. It is
also in agreement with the upper bound on the energy of the compound reggeized
gluon states established in~\cite{CM}. We would like to remind, however, that the
calculations were performed in the multi-colour limit and the important question
remains: may the nonplanar corrections change the $N-$dependence of the energy
$E_N$ of the compound reggeized gluon states? One expects that the nonplanar
$1/N_c^2$ corrections to the reggeon Hamiltonian  will break integrability of the
Schr\"odinger equation \re{Sch} and calculations will be more involved. This
problem deserves additional studies.

Analyzing the high-energy asymptotics of the scattering amplitudes, one is trying
to identify the effective theory, which describes the QCD dynamics in the Regge
limit. The main objects of this effective theory are the $N=2,3,...$ reggeon
compound states constructed in this paper. In the generalized leading logarithmic
approximation, these states propagate between the scattered hadrons and do not
interact with each other. In the topological $1/N_c^2-$expansion~\cite{Ven},
these states emerge from the summation of cylinder-like diagrams, whose walls are
built from the reggeized gluons. These diagrams can be interpreted as describing
the propagation a closed string between two scattered hadrons. This suggests that
the effective dynamics of the multi-reggeon compound states in multi-colour QCD
has to admit a stringy representation~\cite{GKK}.

\section*{Acknowledgements}

We would like to thank A.~Gorsky, I.~Kogan, N.~Nekrasov, F.~Smirnov, A.~Turbiner,
G.~Veneziano and J.~Wosiek for illuminating discussions. This work was supported
in part by the Blaise Pascal Research Chair awarded to G.~Veneziano, by the grant
KBN-PB-2-P03B-19-17 (J.K.), by the grant of Spanish Ministry of Science (A.M.),
by the grant 00-01-005-00 of the Russian Foundation for Fundamental Research
(S.D. and A.M.) and by the NATO Fellowship (A.M.).

\section*{Note added}

After this paper has been submitted for publication, the preprint by H.~de Vega
and L.~Lipatov (dVL) appeared, [arXiv:~hep-ph/0204245], in which the spectrum of
the $N=3$ and $N=4$ reggeon compound states was investigated. It contains a
number of statements, which are in contradiction with our results. We would like
to comment on them below.

In the dVL paper, it is claimed that, contrary to our findings (see %{\it e.g.\/},
Figures~\ref{Fig-N=3}, \ref{Fig-Q4} and Tables~\ref{tab:WKB}, \ref{tab:WKB-4}),
the quantized valued of the integrals of motion $q_k$ for arbitrary $N$ can take
only pure imaginary values for odd $k$ and real values for even $k$. In
particular, at $N=3$ this would imply that $\Re q_3=0$ for all eigenstates of the
model. Moreover, for complex values of the charges $q_k$ found in our paper, the
dVL approach leads to complex values of the energy $E_N$. The authors attributed
a disagreement between their and our results to the fact that the quantization
procedure proposed in our paper is erroneous. They did not offer, however, any
further explanations and refer instead to \cite{JW}. In order to check the above
assertions and, at the same time, to test the dVL approach, we decided to verify
our results at $N=3$ using the approach by R.~Janik and J.~Wosiek~\cite{JW}. We
found that {\it all\/} eigenstates at $N=3$ constructed in our paper, including
those with $\Re q_3\neq 0$, satisfy the quantization conditions formulated
in~\cite{JW}.% and have the same, real values of the energy $E_3(q_2,q_3)$.%
\footnote{We are grateful to J.~Wosiek for making the Mathematica code
used in~\cite{JW} available to us.} As an example, we present two such states
with $(\ell_1,\ell_2)=(3,3)$ and $(3,5)$, which have the total spin $h=1/2$ and
the quasimomentum $\theta_3=0$ (see Eqs.~\re{q3-WKB} and \re{quasi-3})
\baa
&& q_3(3,3)=-1.475327 \,,\hspace*{31mm} E_3=8.469248\,,
\\
&& q_3(3,5)=-4.752678-3.048722\,i\,,\qquad E_3=13.850368\,.
\eaa
In the notations of~\cite{JW}, the wave functions of these states are specified
by the parameters, correspondingly,
\baa
&&
\alpha_{_{\rm JW}}=0.4332\,,\quad
\beta_{_{\rm JW}}=-0.6345-0.0361\,i\,,\quad \gamma_{_{\rm JW}}=0.6391\,,
\\
&&
\alpha_{_{\rm JW}}=0.0895+0.0007\,i\,,\quad
\beta_{_{\rm JW}}=0.8582\,,\quad \gamma_{_{\rm JW}}=0.5055+0.0037\,i\,.
\eaa
Moreover, one can argue that the set of eigenstates with $\Re q_3=0$ can not be
complete. If it were complete, the operator $q_3+\bar q_3$ defined in Eqs.~\re{q}
and \re{q-bar} would be identically equal to zero on the
$SL(2,\mathbb{C})$ representation space %of functions normalizable with respect to
\re{SL2-norm}. Applying this operator to an arbitrary test function on this space
$\Psi(\vec z_1,\vec z_2,\vec z_3)$, one verifies that $(q_3+\bar q_3)\Psi\neq 0$.
In a similar manner, one can show that for higher $N$ the eigenvalues of the
integrals of motion $q_3,...,\,q_N$ can not take only real or pure imaginary
values.

Another issue concerns the ground state at $N=4$, Eq.~\re{N=4-ground}. In the
first version of the dVL paper, it was claimed that it is located at $h=1$ and
its energy is smaller than the energy of the state defined in Eq.~\re{N=4-ground}
above. Later, in the second version of the paper, the authors found yet another
state at $h=3/2$, which has even smaller energy. We remind that for the principal
series of the $SL(2,\mathbb{C})$ group the total spin $h$ has the form \re{q2},
so that $h=1/2\,, 1$ and $3/2$ correspond to $\nu_h=0$ and the Lorentz spin
$n_h=0\,,1$ and $2$, respectively. Thus, the results of the dVL paper imply that
the minimal energy of the system of $N=4$ particles decreases as the total
angular momentum of their rotation on two-dimensional plane, $n_h$, increases.
Our results (see Table~\ref{tab:N4}, 4th column) indicate that the dependence is
just opposite. In addition, in our approach we do not find such states with $h=1$
and $h=3/2$ in the spectrum of the $SL(2,\mathbb{C})$ magnet and the ground state
occurs at $h=1/2$, Eq.~\re{N=4-ground}. Analyzing the quantization conditions
\re{C1-C0}, we were able to identify the physical meaning of the solutions found
in the dVL paper. They fulfil the quantization conditions at $N=4$, but have
unusual quantum numbers, $h=\bar h=1$ and $h=\bar h=3/2$, which do not match the
principal series of the $SL(2,\mathbb{C})$ group, Eq.~\re{q2}. One can show that
these states are located on the trajectory with $h=\bar h=1/2+\nu_h$, which is
obtained from the ground state trajectory with $h=\bar h=1/2+i\nu_h$ by {\it
analytical continuation\/} from real to imaginary $\nu_h$. The same happens for
the $q_3=0$ solution at $N=3$ found there -- its total spin is $h=\bar h=1$. As a
consequence, the states found in the dVL paper do not belong to the quantum space
of the $SL(2,\mathbb{C})$ magnet and, therefore, are unphysical.

We conclude that the criticism of our results in the dVL paper is groundless. At
$N=3$ the quantization procedure proposed in that paper does not reproduce
correctly the part of the spectrum corresponding to $\Re q_3\neq 0$, whereas at
$N=4$ it generates spurious states, which do not belong to the quantum space of
the model.

\appendix
\renewcommand{\theequation}{\Alph{section}.\arabic{equation}}
\setcounter{table}{0}
\renewcommand{\thetable}{\Alph{table}}

\section{Appendix: Solution to the Baxter equation at
$N=2$}\label{App:N=2}

In this Appendix we summarize the properties of the eigenvalues of the Baxter
operator $Q(u,\bar u)$ at $N=2$. As was shown in the Section 3, $Q(u,\bar u)$ is
equal to the integral \re{Q-R} of the function $Q(z,\bar z)$ defined in
\re{N=2-Q} over the two-dimensional plane.

The function $Q_s(z;h)$ entering \re{N=2-Q} is expressed in terms of the Legendre
function on the second-kind, Eq.~\re{Q-function}. Using the properties of the
Legendre functions~\cite{WW}, one finds the behaviour of $Q_s(z;h)$ around
singular points $z=0$ and $z=1$ as
\ba
Q_s(z;h)&\stackrel{z\to 0}{\sim}& z^{1-s}\left[-\frac12\ln z
-\psi(1-h)+\psi(1)+{\cal O}(z)\right]\,,
\label{asym-0}
\\
Q_s(z;h)&\stackrel{z\to 1}{\sim}&
(1-z)^{2s-h-1}\left[\frac{\Gamma^2(1-h)}{2\Gamma(2-2h)}+{\cal O}(1-z)\right]\,,
\label{asym-1}
\ea
Comparing \re{asym-0} and \re{asym-1} with the asymptotic behaviour of the
functions $Q_n^{(0)}(z)$ and $Q_m^{(1)}(z)$ defined in \re{Q-0-h} and \re{set-1},
one finds that $Q_s(z;h)$ can be decomposed over the fundamental set of solutions
around $z=0$ as
\ba
&&Q_s(z;h)=-\left[\psi(1-h)-\psi(1)-\frac12\right]Q_1^{(0)}(z) -\frac12
Q_2^{(0)}(z)\,,
\nonumber\\
&&Q_s(z;1-h)=-\left[\psi(h)-\psi(1)-\frac12\right]Q_1^{(0)}(z) -\frac12
Q_2^{(0)}(z)\,,
\label{Q-Q-rel1}
\ea
and over the fundamental set of solutions around $z=1$ as
\be
Q_s(z;h)=\frac{\Gamma^2(1-h)}{2\Gamma(2-2h)}Q_1^{(1)}(z)
\,,\qquad
Q_s(z;1-h)=\frac{\Gamma^2(h)}{2\Gamma(2h)}Q_2^{(1)}(z)\,.
\label{Q-Q-rel2}
\ee
The functions $Q_n^{(0)}(z)$ and $Q_m^{(1)}(z)$ are related to each other by the
transition matrix $Q_n^{(0)}(z)=\sum_m \Omega_{nm}\, Q_m^{(1)}(z)$ defined in
\re{Omega-def}. Comparison of the r.h.s.\ of \re{Q-Q-rel1} and \re{Q-Q-rel1}
yields
\be
\Omega(h)=\left(\begin{array}{cc} \Delta(h) & \Delta(1-h)
\\ \delta(h) & \delta(1-h)
\end{array}\right),
\ee
where $\Delta(h)={\Gamma(2h-1)}/{\Gamma^2(h)}$ and
$\delta(h)=-2\left[\psi(h)-\psi(1)-\frac12\right]\Delta(h)$. Similar matrix in
the antiholomorphic sector equals $\widebar\Omega=\Omega(\bar h)$.

The general expression for the function $Q(z,\bar z)$ is given by \re{N=2-Q} with
the expansion coefficients $c_h$ defined in \re{N=2-c}. Substituting
\re{Q-Q-rel1} and \re{Q-Q-rel2} as well as analogous relations in the
antiholomorphic sector into \re{N=2-Q}, one arrives at Eqs.~\re{Q-0} and
\re{C1-exp} with the mixing matrices given by
\be
C^{(0)}=\left(\begin{array}{cc} \alpha_1(h)-2 & 1 \\ 1 & 0
\end{array}\right)
\,,\qquad
C^{(1)}=\left(\begin{array}{cc} \beta_h & 0 \\
0 & \beta_{1-h} \end{array}\right)\,,
\ee
where $\alpha_1(h)$ was defined in \re{quasi-N=2} and
\be
\beta_h
=(-1)^{n_h+1} \frac{\Gamma^2(1-\bar
h)}{\Gamma^2(h)}\frac{\Gamma(2h-1)}{\Gamma(2-2\bar h)}\,.
\label{beta-h}
\ee
It is straightforward to verify that the matrices $C^{(0)}$ and $C^{(1)}$ satisfy
the quantization conditions \re{C1-C0}.

To obtain the $Q-$block at $N=2$ one inserts \re{Q-function} into \re{blocks-def}
\be
Q(u;h)=\frac1{\Gamma(2s-h)}\int_0^1
dz\,z^{iu-s}(1-z)^{2(s-1)}\,\mathbf{Q}_{-h}\lr{\frac{1+z}{1-z}}\,.
\ee
Integration can be performed by replacing the Legendre function by its integral
representation. In this way, one arrives at \re{N=2-block}. The function
$Q(u,\bar u)$ with required analytical properties is given by the bilinear
combination of the holomorphic and antiholomorphic $Q-$blocks, Eq.~\re{Q-2-3}.

As was explained in Section 4.4, to calculate the energy and quasimomentum it is
convenient to introduce the blocks, $Q_0(u)$ and $Q_1(u)$, Eqs.~\re{Q0-new} and
\re{Q1-def}, respectively. At $N=2$ the block $Q_0(u)$ is fixed (up to an overall
normalization) by the requirement to have simple poles at the points $u=-i(s-m)$
with $m>0$~\cite{FK,K1}
\ba
Q_0(u)&=&\frac1{\Gamma(2s-1+h)\Gamma(2s-h)}\int_0^1
dz\,z^{iu-s}(1-z)^{2(s-1)}\,\mathbf{P}_{-h}\lr{\frac{1+z}{1-z}}
\nonumber
\\[2mm]
&=&\frac{1}{\Gamma^2(2s)}\,\,{}_3F_2\left({{s-iu,2s-h,2s-1+h}\atop
{2s,2s}}\bigg|\,1
\right)\,.
\label{Q0-def}
\ea
The antiholomorphic block $\widebar Q_0(\bar u)$ can be obtained from \re{Q0-def}
by replacing $u\to -\bar u$, $s\to \bar s$ and $h\to \bar h$. The block $Q_1(u)$
is defined according to \re{Q1-def} as
\be
\label{Q-1-2}
Q_1(u)~=~\left[\frac{\Gamma(1-s+iu)}{\Gamma(s+iu)}\right]^2\,
\lr{\widebar Q_0(u^*)}^*~=~
\left[\frac{\Gamma(1-s+iu)}{\Gamma(s+iu)}\right]^2\, Q_0(u;1-s)\,,
\ee
where in the last relation we indicated explicitly the dependence of the block
$Q_0$ on the spin $s$.

The two sets of the blocks are linearly dependent
\ba
\label{Q0vQh}
\frac{1}{\Gamma(2s-1+h)}\, Q(u;h)&-&
\frac{1}{\Gamma(2s-h)}\, Q(u;1-h)~=~-\pi\cot(\pi h)\,Q_0(u)\,,\nonumber\\
\frac{1}{\Gamma(1-2s+h)}\, Q(u;h)&-&\frac{1}{\Gamma(2-2s-h)}\,
Q(u;1-h)~=~-\pi\cot(\pi h)\,Q_1(u)\,.
\ea
The inverse relation reads
\ba
\label{QhvQ0}
&&Q(u,h)=\rho(s,h)\left[\frac{1}{\Gamma(2-2s-h)}\, Q_0(u)-
\frac{1}{\Gamma(2s-h)}\, Q_1(u)\right],
\nonumber
\\
&&Q(u,1-h)=\rho(s,h)\left[\frac{1}{\Gamma(1-2s+h)}\,
Q_0(u)-\frac{1}{\Gamma(2s-1+h)}\, Q_1(u)\right],
\ea
where $\rho(s,h)=\pi^2/(2\sin(2\pi s)\sin(\pi h))$. Substituting \re{QhvQ0} into
\re{Q-2-3} one finds that the resulting expression matches \re{diag-Q}.

Using the properties of the ${}_3F_2-$series \cite{Prudnikov}, one can show that
the block $Q_0(u)$, Eq.~\re{Q0-def}, satisfies the following relation
\be
 Q_0(-u;s)-\frac{\sin(\pi h)}{\sin(2\pi s)} Q_0(u;s)=-
\Gamma\left[{{2s,1-2s,1-s-i u,1-s+i
u}\atop{2s-h,2s-1+h,s-iu,s+iu}}\right]
 Q_0(u;1-s),
\label{prop-Q0}
\ee
where we indicated explicitly the dependence on the spin $s$. Applying
\re{Q-1-2}, \re{QhvQ0} and \re{prop-Q0} one can verify that the eigenvalue of the
Baxter operator at $N=2$, Eq.~\re{Q-2-3}, can be rewritten (up to an overall
normalization) as
\be
Q(u,\bar u)\simeq
\Gamma\left[{{1-\bar s-i\bar u,1-\bar s+i\bar
u}\atop{\bar s-i\bar u,\bar s+i\bar u}}\right]
\bigg\{Q_0(u)\, \left(Q_0(-\bar u^*)\right)^* +(-1)^{n_h}\,
Q_0(-u)\,\left(Q_0(\bar u^*)\right)^*\!\bigg\}.
\label{Q-2-2}
\ee
Finally, two equivalent expressions for the eigenvalues of the Baxter
$\mathbb{Q}-$operator at $N=2$, Eqs.~\re{Q-2-2} and \re{Q-2-3}, admit an elegant
representation in terms of two-dimensional Feynman diagrams (see Figures~10a and
b in Ref.~\cite{DKM-I}).

\section{Appendix: Properties of the $Q-$blocks}\label{App-4}

In this Appendix we establish different useful relations between the blocks
$Q(u;h,q)$ and $\widebar Q(\bar u;\bar h,\bar q)$ defined in Eq.~\re{blocks-def}.

\subsection*{Intertwining relations}

It is well-known that the $SL(2,\mathbb{C})$ representations of the principal
series of the spins $(s,\bar s)$ and $(1-s,1-\bar s)$ are unitary equivalent. At
the level of the eigenvalues of the Baxter $\mathbb{Q}-$operator, this property
leads to the following intertwining relation between the blocks
\be
Q_s(u;h,q)=\left[\frac{\Gamma(1-s+iu)}{\Gamma(s+iu)}\right]^N\,Q_{1-s}(u;h,q)\,,
\label{inter-1}
\ee
where subscript indicates the corresponding value of the holomorphic spin.
Indeed, one verifies that the both sides of this relation satisfy the Baxter
equation \re{Bax-eq}, have the same analytical properties \re{poles-block} and
asymptotic behaviour \re{block-asym} at infinity. In a similar manner, using the
identities $s^*=1-\bar s$, $h^*=1-\bar h$ and $\bar q_k=q_k^*$, one can show that
\be
(Q_{1-s}(\bar u^*;h,q))^*=Q_{\bar s}(-\bar u;1-\bar h, -\bar q)\,.
\label{inter-2}
\ee
Combining together Eqs.~\re{inter-1}, \re{inter-2} and \re{Q-univ}, we obtain the
relations between the holomorphic and antiholomorphic blocks
\ba
Q(u;h,q)&=&\left[\frac{\Gamma(1-s+iu)}{\Gamma(s+iu)}\right]^N\lr{\widebar
Q(u^*;1-\bar h,\bar q)}^*\,,
\nonumber
\\
\widebar Q(\bar u; \bar h,\bar q)&=&\left[\frac{\Gamma(1-\bar s-i\bar u)}{\Gamma(\bar s-i\bar u)}\right]^N
\lr{Q(\bar u^*; 1-h,q)}^*\,.
\label{Q-relation}
\ea
Here, the ratio of $\Gamma-$functions compensates the difference in the
analytical properties of two blocks and their asymptotic behaviour at infinity.

\subsection*{Wronskian relations}

The functions $Q(u;h,q)$ and $Q(u;1-h,q)$ satisfy the same Baxter equation
\re{Bax-eq}. This suggests to define their Wronskian as
\be
W(u)=\left[\frac{\Gamma(iu+s)}{\Gamma(iu-s)}\right]^N\bigg[
Q(u+i;h,q)\,Q(u;1-h,q)-Q(u;h,q)\,Q(u+i;1-h,q)\bigg]\,.
\label{Wronskian}
\ee
It follows from \re{Bax-eq} that $W(u)$ is periodic, $W(u+i)=W(u)$. In addition,
taking into account the properties of the blocks, Eqs.~\re{poles-block} and
\re{block-asym}, one finds that $W(u)$ is analytical in the half-plane
$\Re(iu-s+1)>0$ and behaves there at large $u$ as $W(u)\sim u^0$. This implies
that $W(u)$ takes constant values for arbitrary $u$. Then, substituting
\re{block-asym} into \re{Wronskian}, one finds
$
W(u)=1-2h\,.
$
Similar consideration in the antiholomorphic sector leads to
\be
\widebar Q(\bar u-i;\bar h,\bar q)\,\widebar Q(\bar u;1-\bar h,\bar q)
-\widebar Q(\bar u;\bar h,\bar q)\,\widebar Q(\bar u-i;1-\bar h,\bar q) =(1-2\bar
h)\left[\frac{\Gamma(-i\bar u-\bar s)}{\Gamma(-i\bar u+\bar s)}\right]^N.
\label{Wronskian-anti}
\ee
For $u=i(s+\epsilon)$ and $\bar u=-i(\bar s+\epsilon)$ we find from the
Wronskians \re{Wronskian} and \re{Wronskian-anti} that the functions
$\Phi(\epsilon)$ and $\bar\Phi(\epsilon)$, defined in \re{phi-1}, satisfy the
relations
\be
\Phi(\epsilon)-\Phi(\epsilon+k)={\cal O}(\epsilon^N)\,,\qquad
\bar\Phi(\epsilon)-\bar\Phi(\epsilon+k)={\cal O}(\epsilon^N)\,,
\label{phi-phi}
\ee
with $k$ being positive integer. Here, we used the fact that
$Q(i(s+\epsilon);h,q)$ and $Q(i(s+\epsilon);1-h,q)$ are finite for $\epsilon\to
0$.

\subsection*{Series representation}

We can obtain a series representation for the block $Q(u;h,q)$ in the different
regions on the complex $u-$plane, by replacing the function $Q_1(z)$ in
\re{blocks-def} by its expressions in terms of the fundamental solutions,
Eqs.~\re{C-choice} and \re{global-Q}, defined in \re{Q-0-h} and \re{set-1}. For
$z\to 1$ the function $Q_1(z)$ is given by
\be
Q_1(z)=z^{1-s} (1-z)^{Ns-h-1}\sum_{n=0}^\infty v_n (1-z)^n
\label{Q1-series-1}
\ee
with $v_0=1$ and the expansion coefficients $v_n\equiv v^{(1)}_{n}(-q)$ defined
in \re{set-1} and \re{v-series}. In this way, one gets from \re{blocks-def}
\be
Q(u;h,q)=\frac{\Gamma(1-s+iu)}{\Gamma(-h+Ns)}
\sum_{n=0}^\infty v_n\,\frac{\Gamma(n-h+Ns)}{\Gamma(n+1-h-s+iu+Ns)}\,.
\label{Q-series-1}
\ee
For $z\to 0$ the function $Q_1(z)$ is given by
\be
Q_1(z)=z^{1-s}\sum_{k=0}^{N-1}(\ln z)^k \sum_{n=0}^\infty w^{(k)}_{n} z^n\,,
\label{Q1-series-2}
\ee
with $w^{(k)}_{n}=\sum_{b=k+1}^N [\Omega^{-1}(q)]_{1b}\,c_{b-1}^k
u^{(b-k)}_{n}(-q)$ and the expansion coefficients $u^{(m)}_{n}$ defined in
\re{Q-0-h} and \re{power-series-0}. This leads to
\be
Q(u;h,q)=\frac{1}{\Gamma(Ns-h)}\sum_{n=0}^\infty\sum_{k=0}^{N-1}\frac{(-1)^k\,
w^{(k)}_{n}\, k!}{(iu-s+n+1)^{k+1}}\,,
\label{Q-series-2}
\ee
where the sum over $n$ goes over the $k-$th order poles located at $u=i(n+1-s)$.

We notice that \re{Q-series-1} reproduces correctly the asymptotic behaviour of
$Q(u;h,q)$ at large $u$, Eq.~\re{block-asym}, and the position of its poles on
the $u-$plane, Eq.~\re{poles-block}, but not their order. The reason for this is
that the series \re{Q-series-1} is convergent only for $\Re(1-s+iu)\geq 0$.
Indeed, the large-order behaviour of the expansion coefficients $v_n$ in
\re{Q1-series-1} is determined by the asymptotics of the function $Q_1(z)$ at
$z=0$, Eq.~\re{Q1-series-2}
\be
v_n=\frac{1}{2\pi i}\oint_{|1-z|<\epsilon} dz\,
\frac{z^{s-1}Q_1(z)}{(1-z)^{n-h+Ns}}\sim
\int^0_{-\infty} dz
%\frac{1}{2\pi i}\oint dz
\, \frac{\ln^{N-1}z}{(1-z)^{n+1}}\sim\frac{\ln^{N-1} n}{n}\,.
\ee
Therefore, for $\Re(1-s+iu)< 0$ the series \re{Q-series-1} diverges at large $n$
as $\sum_n v_n/n^{1-s+iu}\sim\sum_n\ln^{N-1} n/n^{2-s+iu}$.

\section{Appendix: Contour integral representation}\label{App-contour}

Let us demonstrate that the two-dimensional integral \re{Q-2dim} can be
decomposed into the sum of products of simple contour integrals, Eq.~\re{Q-1dim}.
The derivation is based on the technique developed in \cite{M} for calculation of
the correlation functions in CFT. To simplify notations, we rewrite the integral
over \re{Q-2dim} as
\be
Q=\int d^2z\, \sum_{n,k=1}^N q_n(z) C_{nk} \bar q_k(\bar z) \equiv \int d^2z\,
q^T(z)\cdot \mathbf{C} \cdot\bar q(\bar z)\,.
\label{App-Q}
\ee
Here, the functions  $q_n(z)=z^{-iu-1}Q_n(z)$ and $\bar q_k(\bar z)=\bar
z^{-i\bar u-1}\bar Q_k(\bar z)$ have three isolated singular points located at
$z_i=\bar z_i=0$, $1$ and $\infty$. Around these points, they have a nontrivial
monodromy
\be
q_n(z) \stackrel{z \circlearrowleft z_i}{\longmapsto} [M_i]_{nk} \,
q_k(z)\,,\qquad
\bar q_n(\bar z)\stackrel{\bar z \circlearrowright
\bar z_i}{\longmapsto} [\widebar M_i]_{nk} \, \bar q_k(\bar z)\,,
\ee
with $M_i$ and $\widebar M_i$ being the corresponding monodromy matrices such
that $M_0 M_1 M_\infty=\II$ and $\widebar M_0\widebar M_1\widebar M_\infty=\II$.
Here, $z$ and $\bar z$ encircle the singular points on the complex plane in the
counterclockwise and clockwise directions, respectively. For the integrand in
\re{App-Q} to be a single-valued function on the two-dimensional plane, the
mixing matrix in Eq.~\re{App-Q} has to satisfy the condition
\be
M_i^T\, C\,\widebar M_i = C\,,\qquad (i=0,1,\infty).
\label{App-MCM}
\ee

%%%%%%%%%%%%%%%%%%%%%%%%%%%%%%%%%%%%%%%%%%%%%%%%%%%%%%%%%%%%%%%%%%%%
\begin{figure}[t]
\centerline{{\epsfxsize10.0cm\epsfbox{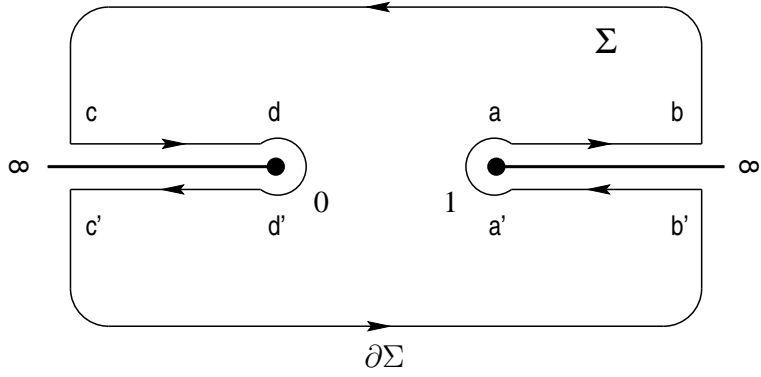}}}
\centerline{$\partial\Sigma$}
%\vspace*{0.5cm}
\caption[]{The integration contour $\partial\Sigma$ in Eq.~\re{Stokes}.}
\label{fig:cuts}
\end{figure}
%%%%%%%%%%%%%%%%%%%%%%%%%%%%%%%%%%%%%%%%%%%%%%%%%%%%%%%%%%%%%%%%%%%%

Since $q(z)$ and $\bar q(\bar z)$ are analytical functions on the complex plane
with two cuts running from the singular points $0$ and $1$ to infinity, as shown
in Figure~\ref{fig:cuts}, one can apply the Stokes's theorem
\be
Q=\int_\Sigma d^2z\, \frac{\partial}{\partial \bar z}\left[q^T(z)\cdot \mathbf{C}
\cdot\int^{\bar z}_{\bar z_{\rm aux}} d\bar z'\, \bar q(\bar z')\right]
=\frac1{2i}\int_{\partial\Sigma}dz\,q^T(z)\cdot \mathbf{C}
\cdot\int^{\bar z}_{\bar z_{\rm aux}} d\bar z'\, \bar q(\bar z')\,.
\label{Stokes}
\ee
Here, $\bar z_{\rm aux}$ is an arbitrary reference point and the $\bar
z'-$integration goes along the contour that does not cross the cut. It is
convenient to choose $\bar z_{\rm aux}=0$ and split the integral over (infinite)
contour $\partial\Sigma$ into four integrals along the different edges of two
cuts. Their contribution to \re{Stokes} can be calculated as follows
\ba
Q_{[b'a']}&=&-\frac1{2i}\int_{a'}^{b'} dz\,q^T(z)\cdot \mathbf{C}
\cdot\left[\int^{\bar z}_1 d\bar z'\, \bar q(\bar z')+\int^1_0 d\bar z'\, \bar q(\bar
z')\right],
\nonumber
\\
Q_{[ab]}&=&\frac1{2i}\int_a^b dz\,q^T(z)\cdot \mathbf{C}
\cdot\left[\int^{\bar z}_1 d\bar z'\, \bar q(\bar z')
+\int^1_0 d\bar z'\, \bar q(\bar z')\right].
\label{App-Q-part1}
\ea
In these relations the integration goes along two different edges of the same
cut. In spite of the fact that $q(z)$ and $\bar q(\bar z)$ are discontinuous
across the cut, their bilinear combination remains continuous due to \re{App-MCM}
$$
\int_{a'}^{b'} dz\,q^T(z)\,\mathbf{C}
\int^{\bar z}_1 d\bar z'\, \bar q(\bar z')=\int_{a}^{b} dz\,[M_1q(z)]^T\,\mathbf{C}
\int^{\bar z}_1 d\bar z'\, \bar M_1\bar q(\bar z')
=\int_{a}^{b} dz\,q^T(z)\,\mathbf{C}
\int^{\bar z}_1 d\bar z'\, \bar q(\bar z')\,.
$$
This leads to a partial cancellation of terms in the sum $Q_{[b'a']}+Q_{[ab]}$.
Then, one calculates the contribution of the second cut
\ba
Q_{[cd]}&=&-\frac1{2i}\int_{d}^{c} dz\,q^T(z)\cdot \mathbf{C}
\cdot\int^{\bar z}_0 d\bar z'\, \bar q(\bar z')\,,
\nonumber
\\
 Q_{[d'c']}&=&\frac1{2i}\int_{d'}^{c'} dz\,q^T(z)\cdot \mathbf{C}
\cdot\int^{\bar z}_0 d\bar z'\, \bar q(\bar z')\,.
\label{App-Q-part2}
\ea
and finds that the same property leads to $Q_{[cd]}+Q_{[d'c']}=0$. Combining
together \re{App-Q-part1} and \re{App-Q-part2}, we obtain the following
expression for the two-dimensional integral \re{Stokes}
\ba
Q&=&\frac1{2i}\left[\int_a^b dz\,q^T(z)-\int_{a'}^{b'} dz\,q^T(z)
\right]\cdot\mathbf{C}\cdot\int^1_0 d\bar z'\, \bar q(\bar z')
\nonumber
\\
&=&\frac1{2i}\int_1^\infty dz\,q^T(z)\cdot (1-M_1^T)\mathbf{C}\cdot\int^1_0 d\bar
z'\,\bar q(\bar z')\,.
\label{App-Q-dec}
\ea
Finally, one replaces the functions $q(z)$ and $\bar q(\bar z)$ by their actual
expressions, Eq.~\re{App-Q}, and arrives at \re{Q-1dim}.

One can obtain another (though equivalent) representation for $Q$ if one starts
with
\be
Q=\int_\Sigma d^2z\, \frac{\partial}{\partial z}\left[\int^{z}_{0} d
z'\,q^T(z')\cdot \mathbf{C}
\cdot \bar q(\bar z)\right]
=-\frac1{2i}\int_{\partial\Sigma}d\bar z\,\int^z_0 dz'\, q^T(z')\cdot
\mathbf{C}
\cdot\bar q(\bar z)
\label{Stokes-1}
\ee
instead of \re{Stokes}. Repeating the same analysis one gets
\be
Q=-\frac1{2i}\int_0^1 dz'\,q^T(z')\cdot
\mathbf{C}(1-\widebar M_1)\cdot\int_1^\infty d\bar z\,\bar q(\bar z)\,.
\label{App-Q-dec-1}
\ee
Let us now take into account that $q(z)$ is analytical inside $\Sigma$ and,
therefore, $\int_{\partial\Sigma}dz\,q(z)=0$. As before, splitting the
integration contour into four pieces and taking into account that $ M_1\int_a^b
dz\,q(z)=\int_{a'}^{b'} dz\,q(z)$ and $\int_c^d dz\,q(z)=M_0\int_{c'}^{d'}
dz\,q(z)$ one obtains
\be
-(1-M_0^{-1})\int_0^\infty dz\,q(z)+(1-M_1)\int_1^\infty dz\,q(z)=0\,.
\label{q-iden}
\ee
Here, two terms in the l.h.s.\ correspond to the contribution of two cuts. We
conclude from \re{q-iden} that
\be
-(1-M_0)\int_0^1 dz\,q(z)=(1-M_0M_1)\int_1^\infty dz\,q(z)\,.
\label{two-int}
\ee
Obviously, similar property holds for the function $\bar q(\bar z)$ in the
antiholomorphic sector. Recalling the definition of the function $q(z)$,
Eq.~\re{App-Q}, one obtains
\be
\int_0^1 dz\,q_n(z)=\int_0^1 dz\,z^{-iu-1}Q_n(z)\,,\qquad
\int_1^\infty dz\,q_n(z)=\int_0^1 dz\,z^{iu-1}Q_n(1/z)\,.
\ee
Eq.~\re{two-int} allows us to rewrite \re{App-Q-dec} and \re{App-Q-dec-1} in the
form
\be
Q=\frac1{2i}\int_0^1 dz\,q^T(z) \cdot
\lr{1-\lr{1-M_0^T}^{-1}-\lr{1-M_1^T}^{-1}}^{-1}
\mathbf{C}\cdot\int_0^1 d\bar z\,\bar q(\bar z)\,,
\ee
in which the symmetry between $z-$ and $\bar z-$sectors becomes manifest.

\section{Appendix: Degenerate $Q-$blocks}\label{App-6}

In our analysis of the quantization conditions performed in Section~4.3, we have
tacitly assumed that the blocks $Q(u;h,q)$ are well-defined for arbitrary spins
$h=(1+n_h)/2+i\nu_h$ and, in addition, $Q(u;h,q)$ is finite at $u=i(s+n-1)$ with
$n\ge 1$. As was already mentioned in Section~4.2, the first condition is not
satisfied at $\nu_h=0$. The second condition does not hold for (half)integer
spins $s$ since, by the definition, the block $Q(u;h,q)$ has poles at
$u_m^-=-i(s-m)$ with $m\ge 1$ and for positive integer $n$ and $m$, such that
$2s-1=m-n$, the point $u=i(s+n-1)$ coincides with the pole $u_m^-$. In this
Section, we will work out the quantization conditions for (half)integer spins
$h=(1+n_h)/2$ and $s$. It worth mentioning that one has to deal with these two
cases calculating the ground state of the Schr\"odinger equation \re{Sch} at
$h=1/2$ and $s=0$.

To start with, let us examine the series representation \re{Q1-series-1} for the
block $Q(u;h,q)$ at $h=(1+n_h)/2+i\nu_h$ in the limit $\nu_h\to 0$ and $n_h>0$.
The expansion coefficients $v_n$ entering \re{Q1-series-1} satisfy the $N-$term
recurrence relations, which lead to $v_n\sim 1/\nu_h$ as $\nu_h\to 0$ for $n\ge
n_h$. The resulting expression for $Q(u;h,q)$ can be written as
\be
Q(u;h,q)=\frac{A_{n_h}(q)}{\nu_h} Q(u;1-h,q)+\widetilde Q(u;n_h,q)+{\cal
O}(\nu_h)\,,
\label{tilde-Q}
\ee
where $\widetilde Q(u;h,q)$ and $Q(u;1-h,q)$ are finite for $\nu_h\to 0$ and
$A_{n_h}(q)=\lim_{\nu_h\to 0} [v_{n_h}(q)\nu_h]$. Taking into account
\re{block-asym}, we find that the function $\widetilde Q(u;n_h,q)$ defined in
this way has the following asymptotic behaviour at large $u$
\ba
\widetilde Q(u;n_h,q)&\sim& (iu)^{-Ns+(1+n_h)/2}\,\left[1+{\cal O}(1/u)\right]
\label{tilde-u}
\\[2mm]
&& +A_{n_h}(q)\,(iu)^{-Ns+(1-n_h)/2} \,\ln u\,\left[1+{\cal O}(1/u)\right]\,.
\nonumber
\ea
According to the definition \re{tilde-Q}, the function $\widetilde Q(u;n_h,q)$ is
a linear combination of two degenerate blocks and, therefore, it satisfies the
holomorphic Baxter equation \re{Bax-eq}. Eqs.~\re{tilde-Q} and \re{tilde-u} are
valid only for $n_h>0$. At $n_h=0$, or equivalently $h=1/2$, the function $\tilde
Q(u;0,q)$ is defined as
\be
\widetilde Q(u;0,q)=\partial_{\nu_h}Q(u;1/2+i\nu_h,q)\bigg|_{\nu_h=0}
\stackrel{u\to \infty}{\sim }(iu)^{-Ns+1/2} \,\ln u\,\left[1+{\cal
O}(1/u)\right]\,.
\ee
It is straightforward to verify that for $n_h\ge 0$ the function $\widetilde
Q(u;n_h,q)$ satisfies the chiral Baxter equation \re{Bax-eq}.

Let us now insert \re{tilde-Q} into the quantization conditions \re{const-2} and
examine the limit $\nu_h\to 0$. It follows from \re{phi-1} that
\be
\Phi(\epsilon)=\frac{A_{n_h}(q)}{\nu_h}
+\frac{\widetilde Q(i(s+\epsilon);n_h,q)}{Q(i(s+\epsilon);(1-n_h)/2,q)} +{\cal
O}(\nu_h)\,.
\ee
Then, comparing the coefficients in front of different powers of $\nu_h$ we find
{}from \re{const-2} the set of $N-1$ quantization conditions on the charges
$q_3$, $...$, $q_N$
\be
\frac{\partial^n}{\partial\epsilon^n}
\Im\left[\lr{A_{n_h}(q)}^*\frac{\widetilde Q(i(s+\epsilon);n_h,q)
}{Q(i(s+\epsilon);(1-n_h)/2,q)}\right]\Bigg|_{\epsilon=0} =0\,,
\label{deg-qc}
\ee
with $n=1,...,N-1$ and $h=(1+n_h)/2$. We recall that the additional set of $N$
quantization conditions follows from \re{ell-bar} in the antiholomorphic sector.
At $h=1/2$ the relation \re{deg-qc} leads to
\be
\frac{\partial^n}{\partial\epsilon^n}\frac{\partial}{\partial\nu_h}
\Im\,\ln Q\bigg(i(s+\epsilon);1/2+i\nu_h,q\bigg)\bigg|_{\epsilon=\nu_h=0} =0\,.
\ee

Let us now consider the quantization conditions for (half) integer spins $s$. In
distinction with the previous case, the blocks $Q_s(u;h,q)$ as well as the
eigenvalues of the Baxter operator $Q(u,\bar u)$ remain finite in the limit
$s=(1+n_s)/2$ and possess a correct analytical properties as functions of $u$ and
$\bar u$. Nevertheless, the quantization conditions have to be modified because
the two sets of points \re{spur-poles-new} overlap. Repeating the analysis of
Section 4.4, we find that the function $Q(u,\bar u)$ has correct analytical
properties  for (half)integer $s$ provided that the following conditions are
satisfied
\be
{\rm
arg}\left[\frac{Q(i\lr{1+|n_s|}/2+\epsilon;h,q)}{Q(i\lr{1+|n_s|}/2+\epsilon;1-h,q)}\right]
=\pi\lr{\frac{n_h}2+\ell}-\Theta_{q,\bar q}+ {\cal O}(\epsilon^{2N})\,.
\label{ell-new}
\ee
Here, in distinction with \re{ell}, the $Q-$blocks are calculated in the vicinity
of the point $u=i(1+|n_s|)/2$ that belongs to the both sets \re{spur-poles-new}
simultaneously. In addition, the small $\epsilon-$expansion in the r.h.s.\ starts
with the terms $\sim\epsilon^{2N}$.

\end{document}